\documentclass[aps,prd,twocolumn,showpacs,superscriptaddress,nofootinbib,preprintnumbers]{revtex4-1}
\pdfoutput=1

\usepackage{amssymb,amsmath,latexsym,mathrsfs}
\usepackage{graphicx,subfigure}
\usepackage{epsfig}
\usepackage{varioref,xr-hyper}
\usepackage{multirow}
\usepackage{tikz}
\usepackage{array}
\usepackage{hyperref}
\usepackage{wasysym}
\usepackage{float}
\usepackage{soul}
\usepackage{color}
\usepackage[utf8]{inputenc}
\usepackage[T1]{fontenc}
\hypersetup{
     colorlinks   = true,
     citecolor    = blue
}

\begin{document}

\title{Concerns regarding the use of black hole shadows as standard rulers}

\author{Sunny Vagnozzi}
\email{sunny.vagnozzi@ast.cam.ac.uk}
\affiliation{Kavli Institute for Cosmology (KICC) and Institute of Astronomy, University of Cambridge, Madingley Road, Cambridge CB3 0HA, United Kingdom}

\author{Cosimo Bambi}
\email{bambi@fudan.edu.cn}
\affiliation{Center for Field Theory and Particle Physics and Department of Physics, Fudan University, 200438 Shanghai, China}

\author{Luca Visinelli}
\email{l.visinelli@uva.nl}
\affiliation{Gravitation Astroparticle Physics Amsterdam (GRAPPA), University of Amsterdam, Science Park 904, 1098 XH Amsterdam, The Netherlands}

\date{\today}

\begin{abstract}
\noindent Recently, Tsupko \textit{et al.} have put forward the very interesting proposal to use the shadows of high-redshift supermassive black holes (SMBHs) as standard rulers. This would in principle allow us to probe the expansion history within a redshift range which would otherwise be challenging to access. In this note, we critically examine this proposal, and identify a number of important issues which had been previously overlooked. These include difficulties in obtaining reliable SMBH mass estimates and reaching the required angular resolution, and an insufficient knowledge of the accretion dynamics of high-redshift SMBHs. While these issues currently appear to prevent high-redshift SMBH shadows from being used as robust standard rulers, we hope that our flagging them early will help in making this probe theoretically mature by the time it will be experimentally feasible.
\end{abstract}
\maketitle

\section{Introduction}
\label{sec:intro}

One of the most important breakthroughs in 21st century cosmology has been the ability to probe the expansion history of the universe and the relation between distance and redshift far beyond our local neighbourhood. These determinations usually rely on objects (or classes of objects) with well-known intrinsic properties, such as so-called \textit{standard candles}~\cite{Perlmutter:1998np,Riess:1998cb}, \textit{standard sirens}~\cite{Schutz:1986gp, Holz:2005df}, \textit{standard rulers}~\cite{Eisenstein:1998tu,Eisenstein:2005su}, and \textit{standard clocks}~\cite{Jimenez:2001gg,Heavens:2014rja}. Here, we shall mostly be concerned with the concept of a standard ruler (SR), an object of known intrinsic size. The distance to a SR can be then determined by comparing its observed angular size to its known physical size. The archetype of SRs (which is more precisely a statistical SR) is represented by the scale imprinted by Baryon Acoustic Oscillations (BAOs)~\cite{Eisenstein:2005su} which are set up by the interplay between radiation pressure and gravity of the strongly coupled photon-baryon fluid in the early universe. BAOs imprint a scale corresponding to the sound horizon at baryon drag in the distribution of matter, resulting in a preferred clustering scale for tracers of the large-scale structure. A statistical analysis of a given large-scale structure tracer at a given redshift allows one to extract this preferred scale, and hence the distance to the redshift in question.

The use of BAOs as SRs has revolutionised our understanding of dark energy and cosmic acceleration and has been instrumental in establishing the $\Lambda$CDM concordance cosmological model~\cite{Aubourg:2014yra}. Nonetheless, there are plenty of theoretical and observational reasons to believe that $\Lambda$CDM might not be the end of the story, ranging from considerations over the theoretical implausibility of a cosmological constant of the observed magnitude~\cite{Weinberg:1988cp}, to mismatches between cosmological parameters estimated from independent probes (such as the ``$H_0$ tension'', see e.g.~\cite{Giusarma:2016phn,DiValentino:2016hlg,Bernal:2016gxb,Vagnozzi:2017ovm,
Renk:2017rzu,Mortsell:2018mfj,Vagnozzi:2018jhn,Nunes:2018xbm,Yang:2018euj,Guo:2018ans,
Aylor:2018drw,Poulin:2018cxd,DiValentino:2019exe,Pan:2019gop,Vagnozzi:2019ezj,
Visinelli:2019qqu,Cai:2019bdh,Pan:2019hac,DiValentino:2019ffd,Escudero:2019gvw,
DiValentino:2019jae}) suggesting that the $\Lambda$CDM description of the dark sectors of the Universe might be incomplete. Anticipated improvements in BAO measurements from future surveys such as DESI~\cite{Aghamousa:2016zmz} and Euclid~\cite{Laureijs:2011gra} will be crucial towards either further strengthening the case for $\Lambda$CDM, or conclusively finding evidence for new physics.

Regardless of the success of BAOs in mapping the late-time expansion history, it is desirable to find novel and independent standard rulers, which might be used to either cross-validate existing BAO distance measurements or, more intriguingly, allow us to probe a new redshift window otherwise not accessible to BAOs. A wide variety of novel standard rulers have been proposed in the literature, including (but not limited to): double-lobed radio sources~\cite{Buchalter:1997vz,Carlberg:1998rk}, X-ray gas mass fractions from galaxy clusters~\cite{Allen:2002sr,Mantz:2007qh}, ultra-compact radio sources~\cite{1993Natur.361..134K,1994ApJ...425..442G}, Minkowski functionals of the large-scale structure density field~\cite{Park:2009ja,Blake:2013noa}, dust time lags~\cite{Hoenig:2014jca,Honig:2016oyn}, strongly-lensed systems~\cite{Paraficz:2009xj,Agnello:2015ala}, the cosmic homogeneity scale~\cite{Ntelis:2018ctq,Nesseris:2019mlr}, velocity-induced acoustic oscillations at Cosmic Dawn~\cite{Munoz:2019fkt,Munoz:2019rhi}, and light echos~\cite{Kervella:2008ne,Bond:2008ax}. However, it is fair to say that none has (yet) even gone close to achieving the same level of maturity and reliability of BAOs, both exploiting the sound-horizon standard ruler as well as the so-called linear point standard ruler~\cite{Anselmi:2015dha,Anselmi:2018hdn,Anselmi:2017zss,Anselmi:2018vjz,
ODwyer:2019rvi}.

Recently, a very interesting possibility for a new SR making use of black hole (BH) shadows has been proposed by Tsupko \textit{et al.} in~\cite{Tsupko:2019pzg}. A BH shadow is the apparent (\textit{i.e.} gravitationally lensed) image of the photon sphere, the region in the vicinity of the BH along which photons travel in unstable circular orbits. More precisely, the proposal advocated by~\cite{Tsupko:2019pzg} makes us of measurements of the angular sizes of supermassive black hole (SMBH) shadows (whose evolution as a function of redshift is in principle known, if the SMBH mass is known) for SMBHs located at cosmological distances. A very interesting follow-up in~\cite{Qi:2019zdk} examined the cosmological implications of this SR, finding that such a probe can potentially lead to exquisite constraints on the expansion history at very high redshift ($z \gtrsim 10$), as well as on cosmological parameters such as $\Omega_m$. On the other hand, at low redshifts SMBH shadows might allow for precise constraints on the Hubble constant $H_0$, thus possibly providing more insight into the $H_0$ tension. Therefore, it appears that the use of SMBH shadows as standard rulers can provide an extremely successful cosmological probe.

In this note, we wish to advocate a more cautious approach on the subject, despite the promising results of Tsupko \textit{et al.}~\cite{Tsupko:2019pzg} being formally correct. In particular, our goal is to point out a number of rather important practical issues and difficulties overlooked by~\cite{Tsupko:2019pzg}, which render the use of SMBH shadows as standard rulers more problematic than what has been originally thought. While we certainly do not want to discourage astrophysicists and cosmologists from thinking about using SMBH shadows as standard rulers, given the huge potential therein, we believe that at the same time it is important to point out the associated difficulties as early in the process as possible, in order to allow such a probe to reach a high level of theoretical maturity by the time it will be experimentally feasible.

The rest of this paper is organized as follows. In Sec.~\ref{sec:shadow} we review of the concept of a BH shadow, how its angular size evolves with redshift, and how it can be used as a standard ruler in an expanding universe. In Sec.~\ref{sec:issue} we discuss why we find such a probe to be problematic, identifying six independent concerns. We provide concluding remarks in Sec.~\ref{sec:conclusions}. Throughout the paper, we work in Planck units with $G = c = \hbar = 1$.

\section{Black hole shadows as standard rulers}
\label{sec:shadow}

Black holes are unique regions of space-time, and might hold the key towards the unification of Quantum Mechanics and General Relativity (GR)~\cite{Hawking:1976ra,Mathur:2005ai,Dvali:2011aa,Giddings:2017jts,Giddings:2019jwy}. They represent the final state of continuous gravitational collapse of matter and are defined by their event horizon, a one-way causal space-time boundary from which nothing can escape~\cite{Einstein:1916vd,Schwarzschild:1916uq,Penrose:1964wq}. Observationally speaking, BHs are ubiquitous in a wide range of environments (for a recent review on astrophysical BHs see~\cite{Bambi:2019xzp}). Of particular interest are so-called supermassive BHs (SMBHs), with masses in the range $ \left ( 10^{5}-10^{10} \right ) \,M_{\odot}$. It is believed that most sufficiently massive galaxies harbor SMBHs at their centres~\cite{LyndenBell:1969yx,Kormendy:1995er}.

The so-called BH shadow is an important feature resulting from the combination of an event horizon (or, more precisely, of a photon sphere, around which photons orbit the BH on unstable circular orbits) and the strong gravitational lensing in the vicinity of a BH. More formally, the BH shadow constitutes a closed curved on the sky which separates capture orbits from scattering orbits, see~\cite{Dokuchaev:2019jqq} for a review. In particular, for a BH surrounded by a geometrically thick, optically thin emission region, the shadow should be visible as a dark region on the sky, surrounded by a bright emission ring (see e.g.~\cite{Luminet:1979nyg,Lu:2014zja,Cunha:2018acu,Gralla:2019xty,Narayan:2019imo}). For a Schwarzschild BH, the radius of the shadow $r_{\rm sh}=3\sqrt{3}M \approx 5.2M$ is equal neither to the Schwarzschild radius $r_{\rm s}=2M$ nor to the photon sphere radius $r_{\rm ph}=3M$, but is actually slightly larger than both due to the fact that the shadow is the gravitationally lensed image of the photon sphere~\cite{Luminet:1979nyg}.

Very-long-baseline interferometry (VLBI) has been argued to be a promising technique to image the shadows of SMBHs~\cite{Falcke:1999pj}. A very successful example is represented by the Event Horizon Telescope (EHT)~\cite{Fish:2016jil}, a global network of radio telescopes which in 2019 imaged the shadow of the SMBH M87*~\cite{Akiyama:2019cqa,Akiyama:2019brx,Akiyama:2019sww,Akiyama:2019bqs,Akiyama:2019fyp,Akiyama:2019eap}. The shadow of M87* appears to be broadly consistent with that of GR Kerr BH~\cite{Kerr:1963ud}, although the possibility that M87* might be a more complex object (either a non-Kerr BH or a BH mimicker) cannot yet be excluded. In fact, a number of works have examined the possibility of using M87*'s shadow as a probe of fundamental physics, and possibly of deviations from GR~\cite{Moffat:2019uxp,Nokhrina:2019sxv,Abdikamalov:2019ztb,Held:2019xde,Wei:2019pjf,Shaikh:2019fpu,Tamburini:2019vrf,Davoudiasl:2019nlo,Ovgun:2019yor,Bambi:2019tjh,Nemmen:2019idv,Churilova:2019jqx,Safarzadeh:2019imq,Firouzjaee:2019aij,Konoplya:2019nzp,Kawashima:2019ljv,Contreras:2019nih,Bar:2019pnz,Jusufi:2019nrn,Vagnozzi:2019apd,Banerjee:2019cjk,Roy:2019esk,Ali:2019khp,Long:2019nox,Zhu:2019ura,Contreras:2019cmf,Dokuchaev:2019pcx,Wang:2019tto,Konoplya:2019goy,Roy:2019hqf,Pavlovic:2019rim,Biswas:2019gia,Wang:2019skw,Nalewajko:2019mxh,Tian:2019yhn,Cunha:2019ikd,Banerjee:2019nnj,Shaikh:2019hbm,Vrba:2019vqh,Kumar:2019pjp,Allahyari:2019jqz,Li:2019lsm,Jusufi:2019ltj,Rummel:2019ads,Kumar:2020hgm}.

The proposal put forward by Tsupko \textit{et al.} in~\cite{Tsupko:2019pzg} is to use SMBH shadows as standard rulers, by computing the angular size $\alpha_{\rm sh}(z)$ of the shadow of a Schwarzschild BH at arbitrary redshift. The issue of computing the size of a BH shadow at cosmological distances is actually highly non-trivial. The main difficulties in performing an analytical calculation are first of all that of finding an adequate description of a BH embedded in an expanding universe, and next that of computing light ray trajectories in the strong gravity regime. Usually the problem is approached by exploiting constants of motion which are either conserved or approximately conserved. However, the Friedmann-Lema\^{i}tre-Robertson-Walker (FLRW) space-time does not possess a time-like Killing vector, implying that energy is not conserved, which complicates the analytical computation of BH shadows therein. The issue of embedding a BH solution in an expanding universe has been tackled in recent years, for instance within the so-called Einstein-Straus model~\cite{Einstein:1945id,Einstein:1946zz} or within the McVittie metric~\cite{McVittie:1933zz,Nolan:1998xs,Nolan:1999kk,Nolan:1999wf} (see also~\cite{Carrera:2008pi,Gibbons:2009dr,Nandra:2011ug,Nandra:2011ui}). Other works focused on computing the shadow of a Schwarzschild BH embedded in a de Sitter universe~\cite{Stuchlik:1999qk,Bakala:2007pw,Stuchlik:2018qyz}. More progress was made in~\cite{Perlick:2018iye}, where the authors computed the size of a Schwarzschild BH shadow as seen by a comoving observer in an expanding universe with a cosmological constant.

A later study in~\cite{Bisnovatyi-Kogan:2018vxl} proposes an approximate method for computing the size of the shadows of Schwarzschild BHs in an expanding FLRW universe as seen by a comoving observer. The key observation made in~\cite{Bisnovatyi-Kogan:2018vxl} (see also~\cite{Bisnovatyi-Kogan:2019wdd}) is that for BHs located at cosmological distances (\textit{i.e.} well within the Hubble flow) the observer is typically very far from the BH event horizon, and the expansion of the universe is slow enough that it can be neglected near the BH. Within these approximations, one can compute the size of the BH shadow in a FLRW universe with arbitrary energy content, by first neglecting the expansion of the universe as light rays propagate near the BH, and then neglecting the strong BH gravity as light rays propagate towards the distant observer. Under these approximations, which are most certainly reasonable for SMBHs located at cosmological distances (but not applicable to SMBHs situated in the local universe, such as M87*), the expression for the angular size of a Schwarzschild BH shadow at redshift $z$ is~\cite{Bisnovatyi-Kogan:2018vxl}:
\begin{eqnarray}
	\alpha_{\rm sh}(z) \simeq \frac{3\sqrt{3}M}{D_A(z)}\,.
	\label{eq:ash}
\end{eqnarray}
Here, $D_A(z)$ is the angular diameter distance to redshift $z$, which depends on the energy content of the universe (photons, baryons, dark matter, dark energy, and neutrinos) as a function of time through the Hubble expansion rate $H(z)$ at redshift $z$, as:
\begin{eqnarray}
	D_A(z) = \frac{1}{(1+z)}\int_{0}^{z}\frac{dz'}{H(z')}\,.
\label{eq:da}
\end{eqnarray}
For sufficiently small $z \ll 1$ one finds that $D_A(z) \approx z/H_0$, with $H_0$ the Hubble constant (but in this regime neglecting the strong BH gravity might not be justified, and peculiar velocities become important).

The expression in Eq.~\eqref{eq:ash} reflects the fact that the shadow of a Schwarzschild BH is an object of known intrinsic physical size, so that the influence of gravity on the propagation of photons can be neglected. The apparent angular size of the BH is related to its intrinsic physical size through the angular diameter distance at redshift $z$. It is worth remaking once more that this approximation is valid only for observers sufficiently far from the BH~\cite{Bisnovatyi-Kogan:2018vxl}. Within this regime, the validity of Eq.~(\ref{eq:ash}) has been checked in~\cite{Bisnovatyi-Kogan:2018vxl} against the full computation performed in~\cite{Perlick:2018iye}.

\begin{figure}[!ht]
	\includegraphics[width=0.5\textwidth]{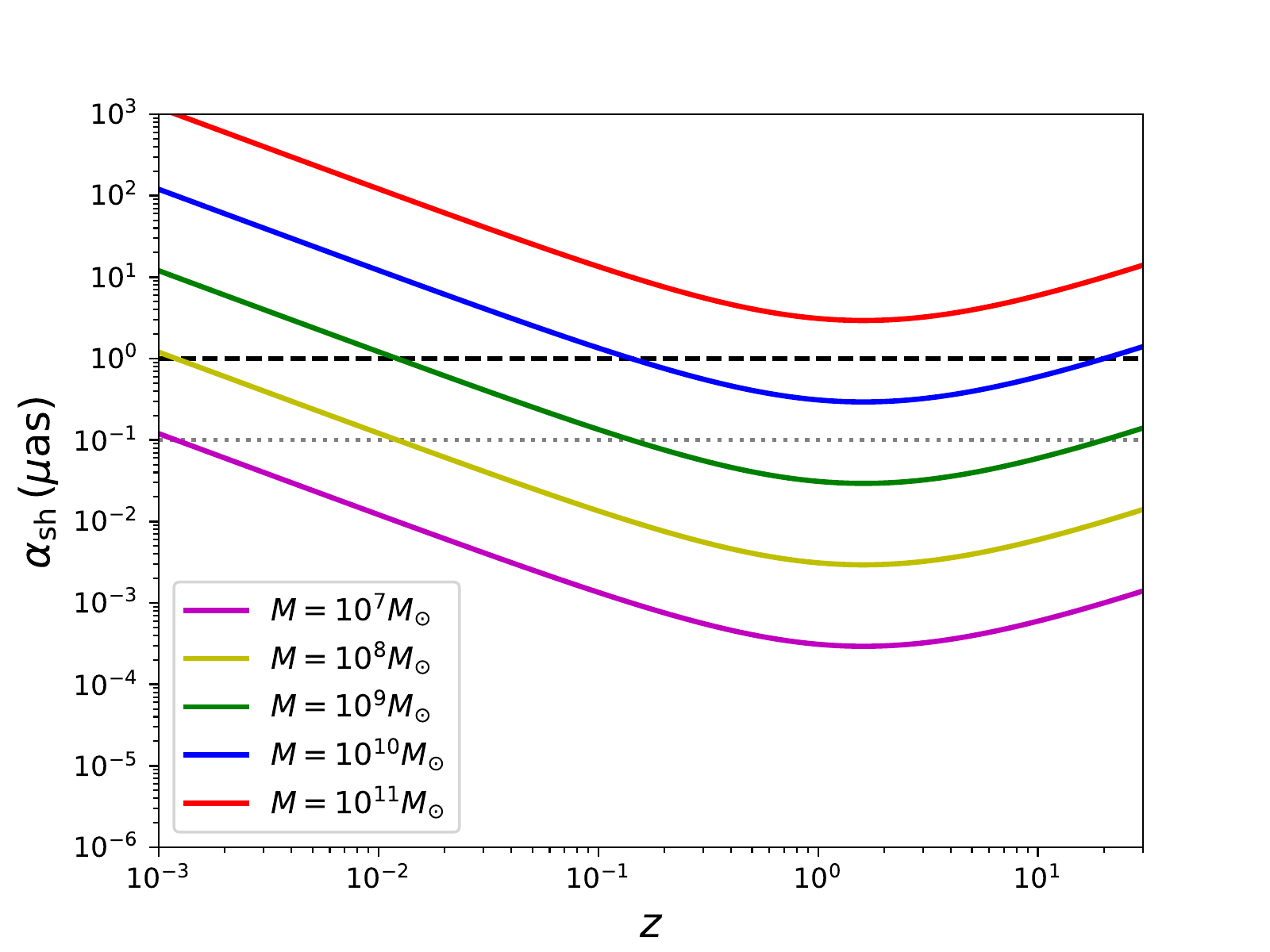}
	\caption{Angular size of supermassive black hole shadows $\alpha_{\rm sh}$ (in $\mu$as) as a function of redshift $z$, given by Eq.~(\ref{eq:ash}). The angular sizes are computed for various SMBH masses: $M=10^7\,M_{\odot}$ (magenta curve), $10^8\,M_{\odot}$ (yellow curve), $10^9\,M_{\odot}$ (green curve), $10^{10}\,M_{\odot}$ (blue curve), and $10^{11}\,M_{\odot}$ (red curve). The horizontal black dashed line denotes the angular resolution of $1\,\mu{\rm as}$, which approximately corresponds to the current sensitivity of the Event Horizon Telescope. We have furthermore also included a very optimistic forecast sensitivity for an angular resolution of $0.1\,\mu{\rm as}$ (gray dotted line). $10^{11} M_{\odot}$ represents a more-than-optimistic rough optimistic upper limit to the heaviest SMBH existing in Nature, and hence the red curve gives a rough upper limit to the size of a SMBH shadow one might ever hope to observe at any given redshift.}
	\label{fig:alphash}
\end{figure}

Interestingly, given the well-known fact that in an Universe with a cosmological constant the angular diameter distance $D_A(z)$ reaches a maximum at $z_{\max} \approx 1.5$ before continuously decreasing, the angular size of SMBH shadows dramatically increases for $z \gg z_{\rm max}$. We show this in Fig.~\ref{fig:alphash}, where we plot the angular size of SMBH shadows as a function of redshift, for various values of the SMBH mass as reported in the caption. The angular size is reported in $\mu$as, and we consider SMBHs with masses up to $10^{11}M_{\odot}$, with heavier SMBHs leading to larger shadows as is obvious from Eq.~(\ref{eq:ash}). The heaviest SMBHs known to us fall just short of the $10^{11}M_{\odot}$ threshold. For example, TON618 is the heaviest SMBH known, and weighs about $6.6 \times 10^{10}M_{\odot}$~\cite{Shemmer:2004ph}. The next-to-heaviest SMBHs known are Homberg 15A~\cite{2019ApJ...887..195M}, IC 1101~\cite{2017MNRAS.471.2321D}, and S5 0014+81~\cite{2009MNRAS.399L..24G,2010MNRAS.405..387G}, all with masses $\approx 4 \times 10^{10}M_{\odot}$. Therefore, $10^{11}M_{\odot}$ can be considered to be a loose more-than-optimistic upper limit for the heaviest SMBHs existing in Nature, and by extension the red curve in Fig.~\ref{fig:alphash} gives a rough upper limit to the size of how large the size of a SMBH shadow can be at any given redshift.

The proposal put forward by Tsupko \textit{et al.} in~\cite{Tsupko:2019pzg} is that an independent determination of the mass $M$ and redshift $z$ of SMBHs at cosmological distances, whose shadow angular size has been measured, leads to an indirect measurement of the angular diameter distance $D_A(z)$ through Eq.~\eqref{eq:ash}. As we see from Fig.~\ref{fig:alphash}, given that the current angular resolution of the EHT is of ${\cal O}(\mu{\rm as})$, such a technique could in principle allow us to probe the distance-redshift relation within the redshift window $z \gtrsim \mathcal{O}(10)$ for SMBHs with masses $M \gtrsim 10^{10}\,M_{\odot}$, corresponding to the heaviest SMBHs known. This redshift window is extremely intriguing, and is well beyond the region that is currently accessible by conventional distance ladder methods, for example by the use of Supernovae Type Ia (SNeIa) or BAOs, which in the most optimistic cases can reach redshifts $z \lesssim 2-3$. Future 21-cm measurements might instead probe the same redshift window as SMBH shadows (see e.g.~\cite{Sprenger:2018tdb,Brinckmann:2018owf,Munoz:2019fkt,Munoz:2019rhi,Munoz:2019hjh}). As shown in~\cite{Qi:2019zdk}, assuming that SMBH shadows could be used as a standard ruler, a combination of SMBH shadows and SNeIa measurements would lead to exquisite constraints on $\Omega_m$ and $H_0$~\cite{Qi:2019zdk}. In the next Section, we will advocate a more cautious approach towards the problem, highlighting a number of issues which were overlooked in the original proposal of Tsupko \textit{et al.}~\cite{Tsupko:2019pzg}, and which appear to prevent SMBH shadows from becoming, at least at present, a reliable standard ruler.

\section{Issues with the use of black hole shadows as standard rulers}
\label{sec:issue}

In this Section, we discuss in more detail the difficulties which have been overlooked on the road towards using SMBH shadows as standard rulers.

\subsection{Reliably determining black hole masses}
\label{subsec:mass}

Assuming that Eq.~(\ref{eq:ash}) is valid (see the later Sec.~\ref{subsec:modeldependence} for concerns on the matter), it is clear that in order to obtain a reliable distance measurement, an equally precise determination of the mass of the SMBH in question is required. Ideally, independent determinations of the SMBH mass should agree between each other. Unfortunately, this is far from being the case even with current SMBH mass determinations.

For instance, aside from the EHT-based determination of M87*'s mass, there are essentially two main ways to determine this quantity: either using stellar dynamics measurements (e.g.~\cite{Gebhardt:2011yw}) or gas dynamics observations (e.g.~\cite{Walsh:2013uua}). These two methods to determine M87*'s mass disagree by about a factor of $2$, and a similar level of disagreement is present for most SMBH mass estimates at low redshift. There are preliminary indications that incorporating non-Keplerian components in the modelling of the gas orbits might solve this discrepancy~\cite{Jeter:2018eoh}, however the situation is extremely far from being settled. Overall, it is clear that current SMBH mass determinations come with a significant ($\gtrsim 100\%$) systematic uncertainty budget, which directly translates into an equally large uncertainty budget on the inferred distance if SMBH shadows are used as standard rulers. It is impossible to do precision cosmology with such a large systematic uncertainty budget floating around.

Another possibility, especially useful at high redshifts, is reverberation mapping~\cite{1982ApJ...255..419B,2001sac..conf....3P}. However, present uncertainties obtained through this method are huge, again $\gtrsim 100\%$. Moreover, the uncertainty is dominated by systematics in our understanding of the so-called broad emission-line region form factor (see e.g.~\cite{Denney:2008gk,Shen:2013pea,Campitiello:2019otf}). Until these broad emission-line regions are better understood, it will not be possible to improve this uncertainty budget, thus calling into question whether it will even be feasible to obtain precise measurements for SMBH masses at high redshift. The problem of reliably determining SMBH masses does not depend on the SMBH redshift, and we therefore expect it to be a significant limitation over the whole redshift window.

\subsection{Reaching the required angular sensitivity}
\label{subsec:sensitivity}

From Fig.~\ref{fig:alphash}, we see that in order to realistically resolve high-redshift SMBH shadows, a better than $0.1\,\mu{\rm as}$ angular resolution is required. Note that the red curve in Fig.~\ref{fig:alphash} is very optimistic, since we do not know of any SMBH as heavy as $10^{11}M_{\odot}$, whereas only a handful of SMBHs with masses of order $10^{10}M_{\odot}$ are known. Most known SMBHs have masses of order $10^9M_{\odot}$ (for instance, M87* has a mass of about $6.5 \times 10^9M_{\odot}$).

An angular resolution of better than $0.1\,\mu{\rm as}$ requires an improvement of over an order of magnitude compared to the current angular resolution of the EHT. While the EHT (as well as planned surveys/space observatories) do plan to improve their sensitivity by both including multiple space-based telescopes, as well as moving to different frequencies, even the most optimistic setup does not seem to be able to achieve the required sensitivity of $0.1\,\mu{\rm as}$ or better (see e.g.~\cite{Kardashev:2015xua,2019arXiv190309539F,2019ApJ...881...62P}). While we cannot exclude that future VLBI technology will be able to reach such a sensitivity, this target appears very futuristic at present.

In~\cite{Tsupko:2019pzg}, it was suggested that the target resolution might be reached by using VLBI technology in the optical band (recall that the EHT is currently observing at $1.3\,{\rm mm}$). However, there are reasons to be skeptical about high-redshift optical VLBI. In fact, the presence of dust in galactic nuclei strongly limits the capabilities of optical observations, which thus do not appear to be a plausible solution to the issue of increasing the angular sensitivity.

A perhaps more plausible alternative is that of using X-ray interferometry (XRI) techniques which, employing a constellation of satellites, may reach the necessary resolution in a relatively distant future~\cite{Uttley:2019ngm} (the build and launch of constellation sub-$\mu$as XRI facilities can be expected indicatively no earlier than 2060). However, XRI facilities aim at observing the direct image of SMHBs with optically thick disks, in which case the shadow does not correspond to the apparent image of the photon sphere, but to the inner edge of the accretion disk, which should also strongly depend on the black hole spin parameter (see further discussions below in Sec.~\ref{subsec:highredshift}). More generally, the issue of what is the most appropriate electromagnetic wavelength to use is closely related to the emission mechanisms of the accreting material, which are far from being well understood, as we will discuss below in Sec.~\ref{subsec:highredshift}. In addition, XRI projects are expected to be able to image the shadows of SMBHs located near us, not at cosmological distances. In summary, the issue of reaching a sensitivity of $0.1\,\mu{\rm as}$ or better appears to be a severe limitation for most of the redshift range under consideration, unless a substantial population of high-redshift SMBHs with masses $>10^{10}M_{\odot}$ exists and can be observed.

\subsection{Do we understand high-redshift black holes well enough?}
\label{subsec:highredshift}

Another possible concern is that the key expression for $\alpha_{\rm sh}(z)$, Eq.~(\ref{eq:ash}), might be modified in the presence of accretion flow which inevitably surrounds the SMBH. One might in fact worry that the observed size of the shadow would depend strongly on the shape and inclination of the accretion disk. More generally, the observed shadow might depend on the details of the accretion flow themselves (in fact, such a concern was recently raised in~\cite{Gralla:2019xty}, see also a partial response in~\cite{Narayan:2019imo}), making SMBH shadows unsuitable for cosmological studies unless the accretion details were sufficiently understood. Fortunately, it is known that for advection dominated accretion flow (ADAF)~\cite{Narayan:1994et,Narayan:1994is,Yuan:2014gma}, the BH shadow is indeed the apparent image of the photon sphere, whose size is thus insensitive to the details of the accretion flow (see e.g.~\cite{Narayan:2019imo}). The ADAF model is believed to be a valid description of the accretion flow around M87* and SgrA*, and in fact for several low-redshift SMBHs.

Is this still the case at high redshift? Unfortunately, things appear to be significantly more complicated. In fact, observations of SMBHs at redshifts as high as $z \sim 7-8$ (see e.g.~\cite{Mortlock:2011va,DeRosa:2013iia,2015Natur.518..512W}) suggest that objects as massive as $M \approx  \left ( 10^{9}-10^{10} \right ) \,M_{\odot}$ were in place less than $1\,{\rm Gyr}$ after the Big Bang~\cite{Fan:2005eq}. This challenges the conventional picture of SMBH growth~\cite{Volonteri:2010wz}, which would require significantly longer timescales to build up so massive objects. It is not clear what the solution to this conundrum is, although a possibility very seriously considered in the literature is that the process of accretion around SMBHs at high redshift is significantly modified (see e.g.~\cite{Volonteri:2005fj,Madau:2014pta,Alexander:2014noa}). In several of the scenarios advocated to explain the anomalously large population of high-redshift SMBHs, the details of the accretion flow are substantially different from the standard ADAF scenario, see for instance~\cite{Haehnelt:1997js,Nulsen:1999mt,Barausse:2012fy,Volonteri:2014lja,Pacucci:2015efa,Pacucci:2015rwa,Pacucci:2015wea}. This implies that the resulting shadows of high-redshift SMBHs might be significantly affected by the details of the accretion flow, making them unsuitable for cosmological studies until the details of accretion onto high-redshift SMBHs is better understood.

On completely general grounds, one would in fact expect much higher accretion rates around high-redshift SMBHs, which would lead to an optically thick accretion flow. In this case, we expect the shadow to corresponds to the apparent image of the inner edge of the accretion disk, ranging from the innermost stable circular orbit (ISCO) for sources accreting at $\sim 10$\% of the Eddington limit to the marginally bound orbit near the Eddington limit, and this clearly modifies Eq.~(\ref{eq:ash}). While this can in principle be imaged by XRI as we discussed in Sec.~\ref{subsec:sensitivity}, the main issue is that the resulting angular size is extremely sensitive to both the SMBH spin and its inclination angle, and can vary by up to a factor of $\approx 10$. Thus, a reliable use of the angular sizes of high-redshift SMBHs with optically thick accretion flow requires a simultaneous precise measurement of both the BH spin and inclination angle, which appears to be extremely challenging at present.


Overall, it is more than fair to state that there is yet no general consensus regarding the formation and accretion dynamics of high-redshift SMBHs. This is of course a very active field of research, and there is all the reason to hope that improvements in future surveys will shed significantly more light on these issues (see for instance~\cite{Paliya:2019oyn}). Only once the picture becomes clearer may we seriously start investigating realistic shadows of SMBHs at high-redshift (to the best of our knowledge, no such study exists in the literature). This issue makes it very premature to even consider using the shadows of SMBHs at redshift $z \gtrsim 7$ (even assuming they can be detected).

\subsection{Model-dependence}
\label{subsec:modeldependence}

One more potential concern regarding the use of SMBH shadows as standard rulers is the model-dependence of the shadow angular size, or more precisely the model-dependence of Eq.~(\ref{eq:ash}). In fact, a reliable standard ruler (or standard candle/siren/clock for that matter) should be as model-independent as possible, \textit{i.e.} the interpretation of the resulting measurement should not depend (or only depend weakly) on the assumption of any specific model. The expression for $\alpha_{\rm sh}(z)$ in Eq.~(\ref{eq:ash}) is valid only for Schwarzschild (\textit{i.e.} non-rotating) BHs in GR. The extension to rotating (Kerr) BHs in GR is in principle not too problematic, since for Kerr BHs the main difference with respect to Schwarzschild BHs is the fact that the shadow becomes less circular (particularly at high observation angles, see for instance Fig.~1 in~\cite{Bambi:2019tjh}), whereas its angular size remains roughly unchanged (although it does shrink slightly). The main problem appears when one considers theories of gravity beyond GR, many of which have been invoked in the literature to address the issues of cosmic acceleration, cosmic inflation, or dark matter (see e.g.~\cite{Hu:2007nk,Boehmer:2007kx,Saridakis:2009bv,Clifton:2011jh,Capozziello:2011et,
Chamseddine:2013kea,Myrzakulov:2015qaa,Myrzakulov:2015kda,Cai:2015emx,Rinaldi:2016oqp,
Sebastiani:2016ras,Capozziello:2017rvz,Nojiri:2017ncd,Vagnozzi:2017ilo,Dutta:2017fjw,
Casalino:2018tcd}). While the Kerr solution persists as a solution to various theories beyond GR, in several other well-motivated theories this is not the case. As a result, the sizes of the shadows of beyond-GR BHs can deviate from the $r_{\rm sh}=3\sqrt{3}M$ predicted from GR. Essentially the same problem can occur when moving beyond BHs and considering so-called ``BH mimickers'' (including for instance horizon-less compact objects).

The literature on the shadows of BHs beyond GR and BH mimickers is too vast to be summarized here. Important works in this direction (studying for instance BH shadows in theories such as Chern-Simons gravity, brane-world gravity, dilaton gravity, scalar-vector-tensor gravity, or Einstein-Born-Infeld gravity, and shadows of BH mimickers such as superspinars, gravastars, and so on) can be found in e.g.~\cite{Bambi:2008jg,Amarilla:2010zq,Amarilla:2011fx,Amarilla:2013sj,Nedkova:2013msa,Tinchev:2013nba,Wei:2013kza,Grenzebach:2014fha,Papnoi:2014aaa,Sakai:2014pga,Wei:2015dua,Moffat:2015kva,Ghasemi-Nodehi:2015raa,Atamurotov:2015xfa,Cunha:2015yba,Amir:2016cen,Dastan:2016vhb,Tretyakova:2016ale,Mureika:2016efo,Sharif:2016znp,Alhamzawi:2017iyn,Cunha:2016wzk,Singh:2017vfr,Tsukamoto:2017fxq,Eiroa:2017uuq,Kumar:2017vuh,Hennigar:2018hza,Vetsov:2018mld,Shaikh:2018lcc,Shaikh:2018kfv,Mizuno:2018lxz,Amir:2018pcu,Ovgun:2018tua,Ayzenberg:2018jip,Okounkova:2018abo,Wang:2018prk,Haroon:2019new,Kumar:2019ohr,Ovgun:2019jdo,Das:2019sty} (see for instance~\cite{Amarilla:2015pgp} for a review). For many of the solutions studied, the size of the BH shadow can deviate appreciably from $3\sqrt{3}M$. Even in the highly idealized case where we are able to measure a SMBH mass to high accuracy (an issue which is in itself problematic as per our earlier discussion, see Sec.~\ref{subsec:mass}), if the true underlying model of gravity is such that the angular size of the SMBH shadow is not given by Eq.~(\ref{eq:ash}) but by the same equation rescaled by a factor of $\beta$, incorrectly interpreting the observed angular size as being that of a GR BH directly translates into a biased determination of the angular diameter distance by the same factor of $\beta$ (note that $\beta$ can be both $\gtrsim 1$ or $\lesssim 1$). Another potentially important concern is that, at least for certain models of dark matter, the details of the dark matter halo by which BHs are surrounded could significantly affect the size of the shadow, as shown in a few studies (see e.g.~\cite{Jusufi:2019nrn,Haroon:2019new,Xu:2018wow,Hou:2018bar,Haroon:2018ryd,Hou:2018avu,Konoplya:2019sns,Jusufi:2019ltj}).

It should be remarked that in most alternative theories the BH shadow size does not deviate too much from the GR predictions. Theories where such a deviation is substantial ($\gtrsim 100\%$) are few and arguably more exotic. Still, this model-dependence underlying the use of SMBH shadows as standard rulers should be kept in mind, and is potentially an important concern. Assuming that we will be able to detect the shadows of high-redshift SMBHs, a possible way to address this concern would be to independently show that such SMBHs are indeed GR SMBHs. We leave open the question as to what would be the best way to do so.

\subsection{Decrease in flux and surface brightness at high redshift}
\label{subsec:flux}

Another concern is that detecting the shadows of high-redshift SMBHs, despite their angular size increasing with respect to their low-redshift counterparts, might be more challenging than na\"{i}vely expected. In fact, the SMBH shadow angular size increasing at high redshift is not the only relevant factor. What's perhaps more important is the fact that the observed flux decreases dramatically as $(1+z)^2$, \textit{i.e.} a factor of ${\cal O}(100)$ at $z \sim 10$. When doing imaging or interferometry, an even more relevant quantity is that of surface brightness, which actually decreases even more dramatically as $(1+z)^4$, \textit{i.e.} a factor of ${\cal O}(10000)$ at $z \sim 10$.

These effects could in principle be counteracted if the luminosity function of active galactic nuclei (AGNs, extremely luminous objects resulting from the accretion of matter onto SMBHs at the centers of galaxies) peaked at a higher luminosity as one moves up in redshift. However, the exact opposite occurs in reality, as one could expect~\cite{Boyle:1988zz}. In fact, at high redshift the AGN luminosity function first peaks at $z \approx 1.5$ before declining rapidly~\cite{Wyithe:2002ij,Silverman:2007qa,Fiore:2011iv,Georgakakis:2015rfa,
Kulkarni:2018ebj}. Moreover, the bright-end slope also steepens, meaning that high-luminosity AGNs become increasingly rarer.

Together, the two effects [evolution of the AGN luminosity function at high redshift, and surface brightness decreasing as $(1+z)^4$] conspire to seriously complicate the detection prospects of SMBHs at high redshift, in spite of the fact that their angular size increases at sufficiently high redshift. This difficulty should be taken into account in realistic forecasts for the use of SMBH shadows as standard rulers.

\subsection{Is weak lensing an issue?}
\label{subsec:weaklensing}

One final potentially important concern is that of weak lensing (WL), the deflection of photons by intervening matter along the line-of-sight in the limit where the deflection only causes small modifications to the photon's path but not visually striking phenomena such as multiple images (see for instance~\cite{Bartelmann:1999yn,Refregier:2003ct,Mandelbaum:2017jpr} for reviews). Photons coming from SMBHs at high redshift will inevitably encounter several lenses (or, more precisely, gravitational potentials) along the line-of-sight to us. To understand whether WL is a concern we need to estimate both the typical angular deflections of photons coming from high-redshift SMBHs as well as the angular coherence scale of the potentials responsible for these deflections, and compare these numbers to the typical shadow angular sizes plotted in Fig.~\ref{fig:alphash}.

Consider a photon belonging to the boundary of a high-redshift SMBH shadow travelling to us and encountering several gravitational potentials along its way, and let us focus on one chosen gravitational potential. Denoting by $\Psi_i$ the depth of the gravitational potential at the point of closest approach on the non-deflected path, General Relativity predicts that the photon will be deflected by an angle $\delta_i \sim 4\Psi_i$. Typical gravitational potentials have a depth of $\Psi_i \sim 2 \times 10^{-5}$, leading to a typical deflection angle of $\delta_i \sim 10^{-4}$. How many such potentials does a high-redshift photon encounter on its path to us? The typical comoving size of gravitational potentials is $\sim 300\,{\rm Mpc}$ (twice the BAO scale), whereas the comoving distance to $z \sim 10$ is $\chi \approx 10000\,{\rm Mpc}$. This leads us to expect that a typical photon from a high-redshift SMBH will encounter about 30 gravitational potentials along its path to us. Assuming uncorrelated potentials, this gives a total rms deflection angle of about $\delta_{\rm tot} \sim \sqrt{30}\delta_i \approx 5 \times 10^{-4}$, or approximately $2\,{\rm arcmin}$.

Therefore, we expect a typical photon coming from the boundary of a high-redshift SMBH shadow to experience WL deflections of ${\cal O}({\rm arcmin})$, 8 orders of magnitude larger than the typical high-redshift SMBH shadow. This could na\"{i}vely suggest that WL is a severe limitation towards the use of SMBH shadows as standard rulers. However, what one should really be concerned about is not so much the overall weak lensing deflection, but rather the differential deflection experienced by photons coming from different points across the SMBH shadow boundary: in other words, what is the net shear experienced by these photons. We expect this shear to be small, at the percent level or smaller, consistently with what we see for high-redshift galaxies.

A more insightful way of understanding that WL is not a severe issue is to consider the angular scale across which the potentials responsible for weak lensing are coherent. We expect a coherence scale of $\sim 300\,{\rm Mpc}/10000\,{\rm Mpc} \approx 1\,{\rm deg}$. Therefore, the potentials which are responsible for WL are coherent over degree-scale patches, much larger than the ${\rm O}(\mu{\rm as})$ size of the SMBH shadow. This implies that photons coming from different parts of the shadow boundary are expect to experience on average the same amount of deflection, meaning that overall the weakly lensed image of the SMBH shadow is not distorted nor blurred, but simply offset by ${\cal O}({\rm arcmin})$ with respect to its original angular position.

However, the use of SMBH shadows as angular probes does not require knowledge of the original shadow angular position, but only of its angular size, which is expected to be preserved by WL given the large coherence scale of the potentials. Of course, this conclusion is contingent on the WL approximation holding. Overall, we find that unlike what one might na\"{i}vely conclude, WL does not appear to be a severe contaminant towards the use of high-redshift SMBH shadows as standard rulers.

\section{Conclusions}
\label{sec:conclusions}

In this note, we have critically examined the intriguing proposal put forward by Tsupko \textit{et al.}~\cite{Tsupko:2019pzg} of using the shadows of high-redshift supermassive black holes as standard rulers. This is a very interesting proposal which rests upon the fact that for sufficiently high redshift in an expanding dark energy-dominated universe, the angular sizes of SMBH shadows increase with increasing redshift (see Fig.~\ref{fig:alphash}). If feasible, such a probe could potentially lead to exquisite constraints on cosmological parameters~\cite{Qi:2019zdk}, potentially also shedding light on the persisting $H_0$ tension.

We have critically examined the feasibility of such a proposal, finding several limitations and concerns regarding the use of high-redshift SMBH shadows as a cosmological probe. These issues identified include: difficulties in obtaining reliable measurements of SMBH masses, currently limited by $>100\%$ systematics, and the determination of which is crucial for the proposal in question, see Eq.~(\ref{eq:ash}); reaching an angular sensitivity of $0.1\,\mu{\rm as}$ or better, which remains challenging even when considering alternative techniques such as X-ray interferometry; an insufficient knowledge of the accretion dynamics of high-redshift SMBHs, and consequently of our understanding of how the shadows of the latter should appear; the model-dependence of the key equation for the angular size of SMBH shadows at high redshift, Eq.~(\ref{eq:ash}), which can be modified if the underlying theory of gravity is not General Relativity; and finally the fact that the flux and surface brightness of high-redshift SMBHs decrease dramatically compared to their low-redshift counterparts. While weak lensing by gravitational potentials along the line-of-sight might na\"{i}vely also appear to be a limitation, given the typical ${\cal O}({\rm arcmin})$ deviations induced on the path of high-redshift photons, we have argued that it is actually not an issue because of the coherence of typical gravitational potentials across angular scales of ${\cal O}({\rm deg})$.

In conclusion, we have found a number of critical issues which appear to undermine the very interesting possibility put forward by Tsupko \textit{et al.} of using SMBH shadows as standard rulers~\cite{Tsupko:2019pzg}, at least at present. While we of course do not want to discourage astrophysicists and cosmologists from further considering this probe, we believe it is important to highlight any shortcomings thereof as early as possible, in order for these to be thoroughly addressed by the time the proposal will be experimentally feasible. We leave the issue of proposing possible solutions to the issues identified to future work.

\begin{acknowledgments}
We thank Stefano Anselmi, Edvard M\"{o}rtsell, and David Poletti for useful remarks. S.V. acknowledges support from the Isaac Newton Trust and the Kavli Foundation through a Newton-Kavli Fellowship, and acknowledges a College Research Associateship at Homerton College, University of Cambridge. C.B. acknowledges support by the Innovation Program of the Shanghai Municipal Education Commission, Grant No.~2019-01-07-00-07-E00035, and the National Natural Science Foundation of China (NSFC), Grant No.~11973019. L.V. is supported through the research program ``The Hidden Universe of Weakly Interacting Particles'' with project number 680.92.18.03 (NWO Vrije Programma), which is partly financed by the Dutch Research Council.
\end{acknowledgments}

\bibliography{BH.bib}

\begin{thebibliography}{255}%
\makeatletter
\providecommand \@ifxundefined [1]{%
 \@ifx{#1\undefined}
}%
\providecommand \@ifnum [1]{%
 \ifnum #1\expandafter \@firstoftwo
 \else \expandafter \@secondoftwo
 \fi
}%
\providecommand \@ifx [1]{%
 \ifx #1\expandafter \@firstoftwo
 \else \expandafter \@secondoftwo
 \fi
}%
\providecommand \natexlab [1]{#1}%
\providecommand \enquote  [1]{``#1''}%
\providecommand \bibnamefont  [1]{#1}%
\providecommand \bibfnamefont [1]{#1}%
\providecommand \citenamefont [1]{#1}%
\providecommand \href@noop [0]{\@secondoftwo}%
\providecommand \href [0]{\begingroup \@sanitize@url \@href}%
\providecommand \@href[1]{\@@startlink{#1}\@@href}%
\providecommand \@@href[1]{\endgroup#1\@@endlink}%
\providecommand \@sanitize@url [0]{\catcode `\\12\catcode `\$12\catcode
  `\&12\catcode `\#12\catcode `\^12\catcode `\_12\catcode `\%12\relax}%
\providecommand \@@startlink[1]{}%
\providecommand \@@endlink[0]{}%
\providecommand \url  [0]{\begingroup\@sanitize@url \@url }%
\providecommand \@url [1]{\endgroup\@href {#1}{\urlprefix }}%
\providecommand \urlprefix  [0]{URL }%
\providecommand \Eprint [0]{\href }%
\providecommand \doibase [0]{http://dx.doi.org/}%
\providecommand \selectlanguage [0]{\@gobble}%
\providecommand \bibinfo  [0]{\@secondoftwo}%
\providecommand \bibfield  [0]{\@secondoftwo}%
\providecommand \translation [1]{[#1]}%
\providecommand \BibitemOpen [0]{}%
\providecommand \bibitemStop [0]{}%
\providecommand \bibitemNoStop [0]{.\EOS\space}%
\providecommand \EOS [0]{\spacefactor3000\relax}%
\providecommand \BibitemShut  [1]{\csname bibitem#1\endcsname}%
\let\auto@bib@innerbib\@empty
\bibitem [{\citenamefont {Perlmutter}\ \emph {et~al.}(1999)\citenamefont
  {Perlmutter} \emph {et~al.}}]{Perlmutter:1998np}%
  \BibitemOpen
  \bibfield  {author} {\bibinfo {author} {\bibfnamefont {S.}~\bibnamefont
  {Perlmutter}} \emph {et~al.} (\bibinfo {collaboration} {Supernova Cosmology
  Project}),\ }\href {\doibase 10.1086/307221} {\bibfield  {journal} {\bibinfo
  {journal} {Astrophys. J.}\ }\textbf {\bibinfo {volume} {517}},\ \bibinfo
  {pages} {565} (\bibinfo {year} {1999})},\ \Eprint
  {http://arxiv.org/abs/astro-ph/9812133} {arXiv:astro-ph/9812133 [astro-ph]}
  \BibitemShut {NoStop}%
\bibitem [{\citenamefont {Riess}\ \emph {et~al.}(1998)\citenamefont {Riess}
  \emph {et~al.}}]{Riess:1998cb}%
  \BibitemOpen
  \bibfield  {author} {\bibinfo {author} {\bibfnamefont {A.~G.}\ \bibnamefont
  {Riess}} \emph {et~al.} (\bibinfo {collaboration} {Supernova Search Team}),\
  }\href {\doibase 10.1086/300499} {\bibfield  {journal} {\bibinfo  {journal}
  {Astron. J.}\ }\textbf {\bibinfo {volume} {116}},\ \bibinfo {pages} {1009}
  (\bibinfo {year} {1998})},\ \Eprint {http://arxiv.org/abs/astro-ph/9805201}
  {arXiv:astro-ph/9805201 [astro-ph]} \BibitemShut {NoStop}%
\bibitem [{\citenamefont {Schutz}(1986)}]{Schutz:1986gp}%
  \BibitemOpen
  \bibfield  {author} {\bibinfo {author} {\bibfnamefont {B.~F.}\ \bibnamefont
  {Schutz}},\ }\href {\doibase 10.1038/323310a0} {\bibfield  {journal}
  {\bibinfo  {journal} {Nature}\ }\textbf {\bibinfo {volume} {323}},\ \bibinfo
  {pages} {310} (\bibinfo {year} {1986})}\BibitemShut {NoStop}%
\bibitem [{\citenamefont {Holz}\ and\ \citenamefont
  {Hughes}(2005)}]{Holz:2005df}%
  \BibitemOpen
  \bibfield  {author} {\bibinfo {author} {\bibfnamefont {D.~E.}\ \bibnamefont
  {Holz}}\ and\ \bibinfo {author} {\bibfnamefont {S.~A.}\ \bibnamefont
  {Hughes}},\ }\href {\doibase 10.1086/431341} {\bibfield  {journal} {\bibinfo
  {journal} {Astrophys. J.}\ }\textbf {\bibinfo {volume} {629}},\ \bibinfo
  {pages} {15} (\bibinfo {year} {2005})},\ \Eprint
  {http://arxiv.org/abs/astro-ph/0504616} {arXiv:astro-ph/0504616 [astro-ph]}
  \BibitemShut {NoStop}%
\bibitem [{\citenamefont {Eisenstein}\ \emph {et~al.}(1998)\citenamefont
  {Eisenstein}, \citenamefont {Hu},\ and\ \citenamefont
  {Tegmark}}]{Eisenstein:1998tu}%
  \BibitemOpen
  \bibfield  {author} {\bibinfo {author} {\bibfnamefont {D.~J.}\ \bibnamefont
  {Eisenstein}}, \bibinfo {author} {\bibfnamefont {W.}~\bibnamefont {Hu}}, \
  and\ \bibinfo {author} {\bibfnamefont {M.}~\bibnamefont {Tegmark}},\ }\href
  {\doibase 10.1086/311582} {\bibfield  {journal} {\bibinfo  {journal}
  {Astrophys. J.}\ }\textbf {\bibinfo {volume} {504}},\ \bibinfo {pages} {L57}
  (\bibinfo {year} {1998})},\ \Eprint {http://arxiv.org/abs/astro-ph/9805239}
  {arXiv:astro-ph/9805239 [astro-ph]} \BibitemShut {NoStop}%
\bibitem [{\citenamefont {Eisenstein}\ \emph {et~al.}(2005)\citenamefont
  {Eisenstein} \emph {et~al.}}]{Eisenstein:2005su}%
  \BibitemOpen
  \bibfield  {author} {\bibinfo {author} {\bibfnamefont {D.~J.}\ \bibnamefont
  {Eisenstein}} \emph {et~al.} (\bibinfo {collaboration} {SDSS}),\ }\href
  {\doibase 10.1086/466512} {\bibfield  {journal} {\bibinfo  {journal}
  {Astrophys. J.}\ }\textbf {\bibinfo {volume} {633}},\ \bibinfo {pages} {560}
  (\bibinfo {year} {2005})},\ \Eprint {http://arxiv.org/abs/astro-ph/0501171}
  {arXiv:astro-ph/0501171 [astro-ph]} \BibitemShut {NoStop}%
\bibitem [{\citenamefont {Jimenez}\ and\ \citenamefont
  {Loeb}(2002)}]{Jimenez:2001gg}%
  \BibitemOpen
  \bibfield  {author} {\bibinfo {author} {\bibfnamefont {R.}~\bibnamefont
  {Jimenez}}\ and\ \bibinfo {author} {\bibfnamefont {A.}~\bibnamefont {Loeb}},\
  }\href {\doibase 10.1086/340549} {\bibfield  {journal} {\bibinfo  {journal}
  {Astrophys. J.}\ }\textbf {\bibinfo {volume} {573}},\ \bibinfo {pages} {37}
  (\bibinfo {year} {2002})},\ \Eprint {http://arxiv.org/abs/astro-ph/0106145}
  {arXiv:astro-ph/0106145 [astro-ph]} \BibitemShut {NoStop}%
\bibitem [{\citenamefont {Heavens}\ \emph {et~al.}(2014)\citenamefont
  {Heavens}, \citenamefont {Jimenez},\ and\ \citenamefont
  {Verde}}]{Heavens:2014rja}%
  \BibitemOpen
  \bibfield  {author} {\bibinfo {author} {\bibfnamefont {A.}~\bibnamefont
  {Heavens}}, \bibinfo {author} {\bibfnamefont {R.}~\bibnamefont {Jimenez}}, \
  and\ \bibinfo {author} {\bibfnamefont {L.}~\bibnamefont {Verde}},\ }\href
  {\doibase 10.1103/PhysRevLett.113.241302} {\bibfield  {journal} {\bibinfo
  {journal} {Phys. Rev. Lett.}\ }\textbf {\bibinfo {volume} {113}},\ \bibinfo
  {pages} {241302} (\bibinfo {year} {2014})},\ \Eprint
  {http://arxiv.org/abs/1409.6217} {arXiv:1409.6217 [astro-ph.CO]} \BibitemShut
  {NoStop}%
\bibitem [{\citenamefont {Aubourg}\ \emph {et~al.}(2015)\citenamefont {Aubourg}
  \emph {et~al.}}]{Aubourg:2014yra}%
  \BibitemOpen
  \bibfield  {author} {\bibinfo {author} {\bibfnamefont {E.}~\bibnamefont
  {Aubourg}} \emph {et~al.},\ }\href {\doibase 10.1103/PhysRevD.92.123516}
  {\bibfield  {journal} {\bibinfo  {journal} {Phys. Rev.}\ }\textbf {\bibinfo
  {volume} {D92}},\ \bibinfo {pages} {123516} (\bibinfo {year} {2015})},\
  \Eprint {http://arxiv.org/abs/1411.1074} {arXiv:1411.1074 [astro-ph.CO]}
  \BibitemShut {NoStop}%
\bibitem [{\citenamefont {Weinberg}(1989)}]{Weinberg:1988cp}%
  \BibitemOpen
  \bibfield  {author} {\bibinfo {author} {\bibfnamefont {S.}~\bibnamefont
  {Weinberg}},\ }\href {\doibase 10.1103/RevModPhys.61.1} {\bibfield  {journal}
  {\bibinfo  {journal} {Rev. Mod. Phys.}\ }\textbf {\bibinfo {volume} {61}},\
  \bibinfo {pages} {1} (\bibinfo {year} {1989})},\ \bibinfo {note}
  {[,569(1988)]}\BibitemShut {NoStop}%
\bibitem [{\citenamefont {Giusarma}\ \emph {et~al.}(2016)\citenamefont
  {Giusarma}, \citenamefont {Gerbino}, \citenamefont {Mena}, \citenamefont
  {Vagnozzi}, \citenamefont {Ho},\ and\ \citenamefont
  {Freese}}]{Giusarma:2016phn}%
  \BibitemOpen
  \bibfield  {author} {\bibinfo {author} {\bibfnamefont {E.}~\bibnamefont
  {Giusarma}}, \bibinfo {author} {\bibfnamefont {M.}~\bibnamefont {Gerbino}},
  \bibinfo {author} {\bibfnamefont {O.}~\bibnamefont {Mena}}, \bibinfo {author}
  {\bibfnamefont {S.}~\bibnamefont {Vagnozzi}}, \bibinfo {author}
  {\bibfnamefont {S.}~\bibnamefont {Ho}}, \ and\ \bibinfo {author}
  {\bibfnamefont {K.}~\bibnamefont {Freese}},\ }\href {\doibase
  10.1103/PhysRevD.94.083522} {\bibfield  {journal} {\bibinfo  {journal} {Phys.
  Rev.}\ }\textbf {\bibinfo {volume} {D94}},\ \bibinfo {pages} {083522}
  (\bibinfo {year} {2016})},\ \Eprint {http://arxiv.org/abs/1605.04320}
  {arXiv:1605.04320 [astro-ph.CO]} \BibitemShut {NoStop}%
\bibitem [{\citenamefont {Di~Valentino}\ \emph {et~al.}(2016)\citenamefont
  {Di~Valentino}, \citenamefont {Melchiorri},\ and\ \citenamefont
  {Silk}}]{DiValentino:2016hlg}%
  \BibitemOpen
  \bibfield  {author} {\bibinfo {author} {\bibfnamefont {E.}~\bibnamefont
  {Di~Valentino}}, \bibinfo {author} {\bibfnamefont {A.}~\bibnamefont
  {Melchiorri}}, \ and\ \bibinfo {author} {\bibfnamefont {J.}~\bibnamefont
  {Silk}},\ }\href {\doibase 10.1016/j.physletb.2016.08.043} {\bibfield
  {journal} {\bibinfo  {journal} {Phys. Lett.}\ }\textbf {\bibinfo {volume}
  {B761}},\ \bibinfo {pages} {242} (\bibinfo {year} {2016})},\ \Eprint
  {http://arxiv.org/abs/1606.00634} {arXiv:1606.00634 [astro-ph.CO]}
  \BibitemShut {NoStop}%
\bibitem [{\citenamefont {Bernal}\ \emph {et~al.}(2016)\citenamefont {Bernal},
  \citenamefont {Verde},\ and\ \citenamefont {Riess}}]{Bernal:2016gxb}%
  \BibitemOpen
  \bibfield  {author} {\bibinfo {author} {\bibfnamefont {J.~L.}\ \bibnamefont
  {Bernal}}, \bibinfo {author} {\bibfnamefont {L.}~\bibnamefont {Verde}}, \
  and\ \bibinfo {author} {\bibfnamefont {A.~G.}\ \bibnamefont {Riess}},\ }\href
  {\doibase 10.1088/1475-7516/2016/10/019} {\bibfield  {journal} {\bibinfo
  {journal} {JCAP}\ }\textbf {\bibinfo {volume} {1610}},\ \bibinfo {pages}
  {019} (\bibinfo {year} {2016})},\ \Eprint {http://arxiv.org/abs/1607.05617}
  {arXiv:1607.05617 [astro-ph.CO]} \BibitemShut {NoStop}%
\bibitem [{\citenamefont {Vagnozzi}\ \emph {et~al.}(2017)\citenamefont
  {Vagnozzi}, \citenamefont {Giusarma}, \citenamefont {Mena}, \citenamefont
  {Freese}, \citenamefont {Gerbino}, \citenamefont {Ho},\ and\ \citenamefont
  {Lattanzi}}]{Vagnozzi:2017ovm}%
  \BibitemOpen
  \bibfield  {author} {\bibinfo {author} {\bibfnamefont {S.}~\bibnamefont
  {Vagnozzi}}, \bibinfo {author} {\bibfnamefont {E.}~\bibnamefont {Giusarma}},
  \bibinfo {author} {\bibfnamefont {O.}~\bibnamefont {Mena}}, \bibinfo {author}
  {\bibfnamefont {K.}~\bibnamefont {Freese}}, \bibinfo {author} {\bibfnamefont
  {M.}~\bibnamefont {Gerbino}}, \bibinfo {author} {\bibfnamefont
  {S.}~\bibnamefont {Ho}}, \ and\ \bibinfo {author} {\bibfnamefont
  {M.}~\bibnamefont {Lattanzi}},\ }\href {\doibase 10.1103/PhysRevD.96.123503}
  {\bibfield  {journal} {\bibinfo  {journal} {Phys. Rev.}\ }\textbf {\bibinfo
  {volume} {D96}},\ \bibinfo {pages} {123503} (\bibinfo {year} {2017})},\
  \Eprint {http://arxiv.org/abs/1701.08172} {arXiv:1701.08172 [astro-ph.CO]}
  \BibitemShut {NoStop}%
\bibitem [{\citenamefont {Renk}\ \emph {et~al.}(2017)\citenamefont {Renk},
  \citenamefont {Zumalacárregui}, \citenamefont {Montanari},\ and\
  \citenamefont {Barreira}}]{Renk:2017rzu}%
  \BibitemOpen
  \bibfield  {author} {\bibinfo {author} {\bibfnamefont {J.}~\bibnamefont
  {Renk}}, \bibinfo {author} {\bibfnamefont {M.}~\bibnamefont
  {Zumalacárregui}}, \bibinfo {author} {\bibfnamefont {F.}~\bibnamefont
  {Montanari}}, \ and\ \bibinfo {author} {\bibfnamefont {A.}~\bibnamefont
  {Barreira}},\ }\href {\doibase 10.1088/1475-7516/2017/10/020} {\bibfield
  {journal} {\bibinfo  {journal} {JCAP}\ }\textbf {\bibinfo {volume} {1710}},\
  \bibinfo {pages} {020} (\bibinfo {year} {2017})},\ \Eprint
  {http://arxiv.org/abs/1707.02263} {arXiv:1707.02263 [astro-ph.CO]}
  \BibitemShut {NoStop}%
\bibitem [{\citenamefont {Mörtsell}\ and\ \citenamefont
  {Dhawan}(2018)}]{Mortsell:2018mfj}%
  \BibitemOpen
  \bibfield  {author} {\bibinfo {author} {\bibfnamefont {E.}~\bibnamefont
  {Mörtsell}}\ and\ \bibinfo {author} {\bibfnamefont {S.}~\bibnamefont
  {Dhawan}},\ }\href {\doibase 10.1088/1475-7516/2018/09/025} {\bibfield
  {journal} {\bibinfo  {journal} {JCAP}\ }\textbf {\bibinfo {volume} {1809}},\
  \bibinfo {pages} {025} (\bibinfo {year} {2018})},\ \Eprint
  {http://arxiv.org/abs/1801.07260} {arXiv:1801.07260 [astro-ph.CO]}
  \BibitemShut {NoStop}%
\bibitem [{\citenamefont {Vagnozzi}\ \emph {et~al.}(2018)\citenamefont
  {Vagnozzi}, \citenamefont {Dhawan}, \citenamefont {Gerbino}, \citenamefont
  {Freese}, \citenamefont {Goobar},\ and\ \citenamefont
  {Mena}}]{Vagnozzi:2018jhn}%
  \BibitemOpen
  \bibfield  {author} {\bibinfo {author} {\bibfnamefont {S.}~\bibnamefont
  {Vagnozzi}}, \bibinfo {author} {\bibfnamefont {S.}~\bibnamefont {Dhawan}},
  \bibinfo {author} {\bibfnamefont {M.}~\bibnamefont {Gerbino}}, \bibinfo
  {author} {\bibfnamefont {K.}~\bibnamefont {Freese}}, \bibinfo {author}
  {\bibfnamefont {A.}~\bibnamefont {Goobar}}, \ and\ \bibinfo {author}
  {\bibfnamefont {O.}~\bibnamefont {Mena}},\ }\href {\doibase
  10.1103/PhysRevD.98.083501} {\bibfield  {journal} {\bibinfo  {journal} {Phys.
  Rev.}\ }\textbf {\bibinfo {volume} {D98}},\ \bibinfo {pages} {083501}
  (\bibinfo {year} {2018})},\ \Eprint {http://arxiv.org/abs/1801.08553}
  {arXiv:1801.08553 [astro-ph.CO]} \BibitemShut {NoStop}%
\bibitem [{\citenamefont {Nunes}(2018)}]{Nunes:2018xbm}%
  \BibitemOpen
  \bibfield  {author} {\bibinfo {author} {\bibfnamefont {R.~C.}\ \bibnamefont
  {Nunes}},\ }\href {\doibase 10.1088/1475-7516/2018/05/052} {\bibfield
  {journal} {\bibinfo  {journal} {JCAP}\ }\textbf {\bibinfo {volume} {1805}},\
  \bibinfo {pages} {052} (\bibinfo {year} {2018})},\ \Eprint
  {http://arxiv.org/abs/1802.02281} {arXiv:1802.02281 [gr-qc]} \BibitemShut
  {NoStop}%
\bibitem [{\citenamefont {Yang}\ \emph {et~al.}(2018)\citenamefont {Yang},
  \citenamefont {Pan}, \citenamefont {Di~Valentino}, \citenamefont {Nunes},
  \citenamefont {Vagnozzi},\ and\ \citenamefont {Mota}}]{Yang:2018euj}%
  \BibitemOpen
  \bibfield  {author} {\bibinfo {author} {\bibfnamefont {W.}~\bibnamefont
  {Yang}}, \bibinfo {author} {\bibfnamefont {S.}~\bibnamefont {Pan}}, \bibinfo
  {author} {\bibfnamefont {E.}~\bibnamefont {Di~Valentino}}, \bibinfo {author}
  {\bibfnamefont {R.~C.}\ \bibnamefont {Nunes}}, \bibinfo {author}
  {\bibfnamefont {S.}~\bibnamefont {Vagnozzi}}, \ and\ \bibinfo {author}
  {\bibfnamefont {D.~F.}\ \bibnamefont {Mota}},\ }\href {\doibase
  10.1088/1475-7516/2018/09/019} {\bibfield  {journal} {\bibinfo  {journal}
  {JCAP}\ }\textbf {\bibinfo {volume} {1809}},\ \bibinfo {pages} {019}
  (\bibinfo {year} {2018})},\ \Eprint {http://arxiv.org/abs/1805.08252}
  {arXiv:1805.08252 [astro-ph.CO]} \BibitemShut {NoStop}%
\bibitem [{\citenamefont {Guo}\ \emph {et~al.}(2019)\citenamefont {Guo},
  \citenamefont {Zhang},\ and\ \citenamefont {Zhang}}]{Guo:2018ans}%
  \BibitemOpen
  \bibfield  {author} {\bibinfo {author} {\bibfnamefont {R.-Y.}\ \bibnamefont
  {Guo}}, \bibinfo {author} {\bibfnamefont {J.-F.}\ \bibnamefont {Zhang}}, \
  and\ \bibinfo {author} {\bibfnamefont {X.}~\bibnamefont {Zhang}},\ }\href
  {\doibase 10.1088/1475-7516/2019/02/054} {\bibfield  {journal} {\bibinfo
  {journal} {JCAP}\ }\textbf {\bibinfo {volume} {1902}},\ \bibinfo {pages}
  {054} (\bibinfo {year} {2019})},\ \Eprint {http://arxiv.org/abs/1809.02340}
  {arXiv:1809.02340 [astro-ph.CO]} \BibitemShut {NoStop}%
\bibitem [{\citenamefont {Aylor}\ \emph {et~al.}(2019)\citenamefont {Aylor},
  \citenamefont {Joy}, \citenamefont {Knox}, \citenamefont {Millea},
  \citenamefont {Raghunathan},\ and\ \citenamefont {Wu}}]{Aylor:2018drw}%
  \BibitemOpen
  \bibfield  {author} {\bibinfo {author} {\bibfnamefont {K.}~\bibnamefont
  {Aylor}}, \bibinfo {author} {\bibfnamefont {M.}~\bibnamefont {Joy}}, \bibinfo
  {author} {\bibfnamefont {L.}~\bibnamefont {Knox}}, \bibinfo {author}
  {\bibfnamefont {M.}~\bibnamefont {Millea}}, \bibinfo {author} {\bibfnamefont
  {S.}~\bibnamefont {Raghunathan}}, \ and\ \bibinfo {author} {\bibfnamefont
  {W.~L.~K.}\ \bibnamefont {Wu}},\ }\href {\doibase 10.3847/1538-4357/ab0898}
  {\bibfield  {journal} {\bibinfo  {journal} {Astrophys. J.}\ }\textbf
  {\bibinfo {volume} {874}},\ \bibinfo {pages} {4} (\bibinfo {year} {2019})},\
  \Eprint {http://arxiv.org/abs/1811.00537} {arXiv:1811.00537 [astro-ph.CO]}
  \BibitemShut {NoStop}%
\bibitem [{\citenamefont {Poulin}\ \emph {et~al.}(2019)\citenamefont {Poulin},
  \citenamefont {Smith}, \citenamefont {Karwal},\ and\ \citenamefont
  {Kamionkowski}}]{Poulin:2018cxd}%
  \BibitemOpen
  \bibfield  {author} {\bibinfo {author} {\bibfnamefont {V.}~\bibnamefont
  {Poulin}}, \bibinfo {author} {\bibfnamefont {T.~L.}\ \bibnamefont {Smith}},
  \bibinfo {author} {\bibfnamefont {T.}~\bibnamefont {Karwal}}, \ and\ \bibinfo
  {author} {\bibfnamefont {M.}~\bibnamefont {Kamionkowski}},\ }\href {\doibase
  10.1103/PhysRevLett.122.221301} {\bibfield  {journal} {\bibinfo  {journal}
  {Phys. Rev. Lett.}\ }\textbf {\bibinfo {volume} {122}},\ \bibinfo {pages}
  {221301} (\bibinfo {year} {2019})},\ \Eprint
  {http://arxiv.org/abs/1811.04083} {arXiv:1811.04083 [astro-ph.CO]}
  \BibitemShut {NoStop}%
\bibitem [{\citenamefont {Di~Valentino}\ \emph
  {et~al.}(2019{\natexlab{a}})\citenamefont {Di~Valentino}, \citenamefont
  {Ferreira}, \citenamefont {Visinelli},\ and\ \citenamefont
  {Danielsson}}]{DiValentino:2019exe}%
  \BibitemOpen
  \bibfield  {author} {\bibinfo {author} {\bibfnamefont {E.}~\bibnamefont
  {Di~Valentino}}, \bibinfo {author} {\bibfnamefont {R.~Z.}\ \bibnamefont
  {Ferreira}}, \bibinfo {author} {\bibfnamefont {L.}~\bibnamefont {Visinelli}},
  \ and\ \bibinfo {author} {\bibfnamefont {U.}~\bibnamefont {Danielsson}},\
  }\href {\doibase 10.1016/j.dark.2019.100385} {\bibfield  {journal} {\bibinfo
  {journal} {Phys. Dark Univ.}\ }\textbf {\bibinfo {volume} {26}},\ \bibinfo
  {pages} {100385} (\bibinfo {year} {2019}{\natexlab{a}})},\ \Eprint
  {http://arxiv.org/abs/1906.11255} {arXiv:1906.11255 [astro-ph.CO]}
  \BibitemShut {NoStop}%
\bibitem [{\citenamefont {Pan}\ \emph {et~al.}(2019{\natexlab{a}})\citenamefont
  {Pan}, \citenamefont {Yang}, \citenamefont {Di~Valentino}, \citenamefont
  {Saridakis},\ and\ \citenamefont {Chakraborty}}]{Pan:2019gop}%
  \BibitemOpen
  \bibfield  {author} {\bibinfo {author} {\bibfnamefont {S.}~\bibnamefont
  {Pan}}, \bibinfo {author} {\bibfnamefont {W.}~\bibnamefont {Yang}}, \bibinfo
  {author} {\bibfnamefont {E.}~\bibnamefont {Di~Valentino}}, \bibinfo {author}
  {\bibfnamefont {E.~N.}\ \bibnamefont {Saridakis}}, \ and\ \bibinfo {author}
  {\bibfnamefont {S.}~\bibnamefont {Chakraborty}},\ }\href {\doibase
  10.1103/PhysRevD.100.103520} {\bibfield  {journal} {\bibinfo  {journal}
  {Phys. Rev.}\ }\textbf {\bibinfo {volume} {D100}},\ \bibinfo {pages} {103520}
  (\bibinfo {year} {2019}{\natexlab{a}})},\ \Eprint
  {http://arxiv.org/abs/1907.07540} {arXiv:1907.07540 [astro-ph.CO]}
  \BibitemShut {NoStop}%
\bibitem [{\citenamefont {Vagnozzi}(2019)}]{Vagnozzi:2019ezj}%
  \BibitemOpen
  \bibfield  {author} {\bibinfo {author} {\bibfnamefont {S.}~\bibnamefont
  {Vagnozzi}},\ }\href@noop {} {\  (\bibinfo {year} {2019})},\ \Eprint
  {http://arxiv.org/abs/1907.07569} {arXiv:1907.07569 [astro-ph.CO]}
  \BibitemShut {NoStop}%
\bibitem [{\citenamefont {Visinelli}\ \emph {et~al.}(2019)\citenamefont
  {Visinelli}, \citenamefont {Vagnozzi},\ and\ \citenamefont
  {Danielsson}}]{Visinelli:2019qqu}%
  \BibitemOpen
  \bibfield  {author} {\bibinfo {author} {\bibfnamefont {L.}~\bibnamefont
  {Visinelli}}, \bibinfo {author} {\bibfnamefont {S.}~\bibnamefont {Vagnozzi}},
  \ and\ \bibinfo {author} {\bibfnamefont {U.}~\bibnamefont {Danielsson}},\
  }\href {\doibase 10.3390/sym11081035} {\bibfield  {journal} {\bibinfo
  {journal} {Symmetry}\ }\textbf {\bibinfo {volume} {11}},\ \bibinfo {pages}
  {1035} (\bibinfo {year} {2019})},\ \Eprint {http://arxiv.org/abs/1907.07953}
  {arXiv:1907.07953 [astro-ph.CO]} \BibitemShut {NoStop}%
\bibitem [{\citenamefont {Cai}\ \emph {et~al.}(2019)\citenamefont {Cai},
  \citenamefont {Khurshudyan},\ and\ \citenamefont {Saridakis}}]{Cai:2019bdh}%
  \BibitemOpen
  \bibfield  {author} {\bibinfo {author} {\bibfnamefont {Y.-F.}\ \bibnamefont
  {Cai}}, \bibinfo {author} {\bibfnamefont {M.}~\bibnamefont {Khurshudyan}}, \
  and\ \bibinfo {author} {\bibfnamefont {E.~N.}\ \bibnamefont {Saridakis}},\
  }\href@noop {} {\  (\bibinfo {year} {2019})},\ \Eprint
  {http://arxiv.org/abs/1907.10813} {arXiv:1907.10813 [astro-ph.CO]}
  \BibitemShut {NoStop}%
\bibitem [{\citenamefont {Pan}\ \emph {et~al.}(2019{\natexlab{b}})\citenamefont
  {Pan}, \citenamefont {Yang}, \citenamefont {Di~Valentino}, \citenamefont
  {Shafieloo},\ and\ \citenamefont {Chakraborty}}]{Pan:2019hac}%
  \BibitemOpen
  \bibfield  {author} {\bibinfo {author} {\bibfnamefont {S.}~\bibnamefont
  {Pan}}, \bibinfo {author} {\bibfnamefont {W.}~\bibnamefont {Yang}}, \bibinfo
  {author} {\bibfnamefont {E.}~\bibnamefont {Di~Valentino}}, \bibinfo {author}
  {\bibfnamefont {A.}~\bibnamefont {Shafieloo}}, \ and\ \bibinfo {author}
  {\bibfnamefont {S.}~\bibnamefont {Chakraborty}},\ }\href@noop {} {\
  (\bibinfo {year} {2019}{\natexlab{b}})},\ \Eprint
  {http://arxiv.org/abs/1907.12551} {arXiv:1907.12551 [astro-ph.CO]}
  \BibitemShut {NoStop}%
\bibitem [{\citenamefont {Di~Valentino}\ \emph
  {et~al.}(2019{\natexlab{b}})\citenamefont {Di~Valentino}, \citenamefont
  {Melchiorri}, \citenamefont {Mena},\ and\ \citenamefont
  {Vagnozzi}}]{DiValentino:2019ffd}%
  \BibitemOpen
  \bibfield  {author} {\bibinfo {author} {\bibfnamefont {E.}~\bibnamefont
  {Di~Valentino}}, \bibinfo {author} {\bibfnamefont {A.}~\bibnamefont
  {Melchiorri}}, \bibinfo {author} {\bibfnamefont {O.}~\bibnamefont {Mena}}, \
  and\ \bibinfo {author} {\bibfnamefont {S.}~\bibnamefont {Vagnozzi}},\
  }\href@noop {} {\  (\bibinfo {year} {2019}{\natexlab{b}})},\ \Eprint
  {http://arxiv.org/abs/1908.04281} {arXiv:1908.04281 [astro-ph.CO]}
  \BibitemShut {NoStop}%
\bibitem [{\citenamefont {Escudero}\ and\ \citenamefont
  {Witte}(2019)}]{Escudero:2019gvw}%
  \BibitemOpen
  \bibfield  {author} {\bibinfo {author} {\bibfnamefont {M.}~\bibnamefont
  {Escudero}}\ and\ \bibinfo {author} {\bibfnamefont {S.~J.}\ \bibnamefont
  {Witte}},\ }\href@noop {} {\  (\bibinfo {year} {2019})},\ \Eprint
  {http://arxiv.org/abs/1909.04044} {arXiv:1909.04044 [astro-ph.CO]}
  \BibitemShut {NoStop}%
\bibitem [{\citenamefont {Di~Valentino}\ \emph
  {et~al.}(2019{\natexlab{c}})\citenamefont {Di~Valentino}, \citenamefont
  {Melchiorri}, \citenamefont {Mena},\ and\ \citenamefont
  {Vagnozzi}}]{DiValentino:2019jae}%
  \BibitemOpen
  \bibfield  {author} {\bibinfo {author} {\bibfnamefont {E.}~\bibnamefont
  {Di~Valentino}}, \bibinfo {author} {\bibfnamefont {A.}~\bibnamefont
  {Melchiorri}}, \bibinfo {author} {\bibfnamefont {O.}~\bibnamefont {Mena}}, \
  and\ \bibinfo {author} {\bibfnamefont {S.}~\bibnamefont {Vagnozzi}},\
  }\href@noop {} {\  (\bibinfo {year} {2019}{\natexlab{c}})},\ \Eprint
  {http://arxiv.org/abs/1910.09853} {arXiv:1910.09853 [astro-ph.CO]}
  \BibitemShut {NoStop}%
\bibitem [{\citenamefont {Aghamousa}\ \emph {et~al.}(2016)\citenamefont
  {Aghamousa} \emph {et~al.}}]{Aghamousa:2016zmz}%
  \BibitemOpen
  \bibfield  {author} {\bibinfo {author} {\bibfnamefont {A.}~\bibnamefont
  {Aghamousa}} \emph {et~al.} (\bibinfo {collaboration} {DESI}),\ }\href@noop
  {} {\  (\bibinfo {year} {2016})},\ \Eprint {http://arxiv.org/abs/1611.00036}
  {arXiv:1611.00036 [astro-ph.IM]} \BibitemShut {NoStop}%
\bibitem [{\citenamefont {Laureijs}\ \emph {et~al.}(2011)\citenamefont
  {Laureijs} \emph {et~al.}}]{Laureijs:2011gra}%
  \BibitemOpen
  \bibfield  {author} {\bibinfo {author} {\bibfnamefont {R.}~\bibnamefont
  {Laureijs}} \emph {et~al.} (\bibinfo {collaboration} {EUCLID}),\ }\href@noop
  {} {\  (\bibinfo {year} {2011})},\ \Eprint {http://arxiv.org/abs/1110.3193}
  {arXiv:1110.3193 [astro-ph.CO]} \BibitemShut {NoStop}%
\bibitem [{\citenamefont {Buchalter}\ \emph {et~al.}(1998)\citenamefont
  {Buchalter}, \citenamefont {Helfand}, \citenamefont {Becker},\ and\
  \citenamefont {White}}]{Buchalter:1997vz}%
  \BibitemOpen
  \bibfield  {author} {\bibinfo {author} {\bibfnamefont {A.}~\bibnamefont
  {Buchalter}}, \bibinfo {author} {\bibfnamefont {D.~J.}\ \bibnamefont
  {Helfand}}, \bibinfo {author} {\bibfnamefont {R.~H.}\ \bibnamefont {Becker}},
  \ and\ \bibinfo {author} {\bibfnamefont {R.~L.}\ \bibnamefont {White}},\
  }\href {\doibase 10.1086/305236} {\bibfield  {journal} {\bibinfo  {journal}
  {Astrophys. J.}\ }\textbf {\bibinfo {volume} {494}},\ \bibinfo {pages} {503}
  (\bibinfo {year} {1998})},\ \Eprint {http://arxiv.org/abs/astro-ph/9709174}
  {arXiv:astro-ph/9709174 [astro-ph]} \BibitemShut {NoStop}%
\bibitem [{\citenamefont {Carlberg}\ \emph {et~al.}(1998)\citenamefont
  {Carlberg} \emph {et~al.}}]{Carlberg:1998rk}%
  \BibitemOpen
  \bibfield  {author} {\bibinfo {author} {\bibfnamefont {R.~G.}\ \bibnamefont
  {Carlberg}} \emph {et~al.},\ }in\ \href@noop {} {\emph {\bibinfo {booktitle}
  {{Proceedings, 33rd Rencontres de Moriond fundamental parameters in
  cosmology: Les Arcs, France, Jan 17-24, 1998}}}}\ (\bibinfo {year} {1998})\
  pp.\ \bibinfo {pages} {279--282},\ \Eprint
  {http://arxiv.org/abs/astro-ph/9804312} {arXiv:astro-ph/9804312 [astro-ph]}
  \BibitemShut {NoStop}%
\bibitem [{\citenamefont {Allen}\ \emph {et~al.}(2002)\citenamefont {Allen},
  \citenamefont {Schmidt},\ and\ \citenamefont {Fabian}}]{Allen:2002sr}%
  \BibitemOpen
  \bibfield  {author} {\bibinfo {author} {\bibfnamefont {S.~W.}\ \bibnamefont
  {Allen}}, \bibinfo {author} {\bibfnamefont {R.~W.}\ \bibnamefont {Schmidt}},
  \ and\ \bibinfo {author} {\bibfnamefont {A.~C.}\ \bibnamefont {Fabian}},\
  }\href {\doibase 10.1046/j.1365-8711.2002.05601.x} {\bibfield  {journal}
  {\bibinfo  {journal} {Mon. Not. Roy. Astron. Soc.}\ }\textbf {\bibinfo
  {volume} {334}},\ \bibinfo {pages} {L11} (\bibinfo {year} {2002})},\ \Eprint
  {http://arxiv.org/abs/astro-ph/0205007} {arXiv:astro-ph/0205007 [astro-ph]}
  \BibitemShut {NoStop}%
\bibitem [{\citenamefont {Mantz}\ \emph {et~al.}(2008)\citenamefont {Mantz},
  \citenamefont {Allen}, \citenamefont {Ebeling},\ and\ \citenamefont
  {Rapetti}}]{Mantz:2007qh}%
  \BibitemOpen
  \bibfield  {author} {\bibinfo {author} {\bibfnamefont {A.}~\bibnamefont
  {Mantz}}, \bibinfo {author} {\bibfnamefont {S.~W.}\ \bibnamefont {Allen}},
  \bibinfo {author} {\bibfnamefont {H.}~\bibnamefont {Ebeling}}, \ and\
  \bibinfo {author} {\bibfnamefont {D.}~\bibnamefont {Rapetti}},\ }\href
  {\doibase 10.1111/j.1365-2966.2008.13311.x} {\bibfield  {journal} {\bibinfo
  {journal} {Mon. Not. Roy. Astron. Soc.}\ }\textbf {\bibinfo {volume} {387}},\
  \bibinfo {pages} {1179} (\bibinfo {year} {2008})},\ \Eprint
  {http://arxiv.org/abs/0709.4294} {arXiv:0709.4294 [astro-ph]} \BibitemShut
  {NoStop}%
\bibitem [{\citenamefont {{Kellermann}}(1993)}]{1993Natur.361..134K}%
  \BibitemOpen
  \bibfield  {author} {\bibinfo {author} {\bibfnamefont {K.~I.}\ \bibnamefont
  {{Kellermann}}},\ }\href {\doibase 10.1038/361134a0} {\bibfield  {journal}
  {\bibinfo  {journal} {Nature}\ }\textbf {\bibinfo {volume} {361}},\ \bibinfo
  {pages} {134} (\bibinfo {year} {1993})}\BibitemShut {NoStop}%
\bibitem [{\citenamefont {{Gurvits}}(1994)}]{1994ApJ...425..442G}%
  \BibitemOpen
  \bibfield  {author} {\bibinfo {author} {\bibfnamefont {L.~I.}\ \bibnamefont
  {{Gurvits}}},\ }\href {\doibase 10.1086/173999} {\bibfield  {journal}
  {\bibinfo  {journal} {Astrophys. J.}\ }\textbf {\bibinfo {volume} {425}},\
  \bibinfo {pages} {442} (\bibinfo {year} {1994})}\BibitemShut {NoStop}%
\bibitem [{\citenamefont {Park}\ and\ \citenamefont {Kim}(2010)}]{Park:2009ja}%
  \BibitemOpen
  \bibfield  {author} {\bibinfo {author} {\bibfnamefont {C.}~\bibnamefont
  {Park}}\ and\ \bibinfo {author} {\bibfnamefont {Y.-R.}\ \bibnamefont {Kim}},\
  }\href {\doibase 10.1088/2041-8205/715/2/L185} {\bibfield  {journal}
  {\bibinfo  {journal} {Astrophys. J.}\ }\textbf {\bibinfo {volume} {715}},\
  \bibinfo {pages} {L185} (\bibinfo {year} {2010})},\ \Eprint
  {http://arxiv.org/abs/0905.2268} {arXiv:0905.2268 [astro-ph.CO]} \BibitemShut
  {NoStop}%
\bibitem [{\citenamefont {Blake}\ \emph {et~al.}(2014)\citenamefont {Blake},
  \citenamefont {James},\ and\ \citenamefont {Poole}}]{Blake:2013noa}%
  \BibitemOpen
  \bibfield  {author} {\bibinfo {author} {\bibfnamefont {C.}~\bibnamefont
  {Blake}}, \bibinfo {author} {\bibfnamefont {J.~B.}\ \bibnamefont {James}}, \
  and\ \bibinfo {author} {\bibfnamefont {G.~B.}\ \bibnamefont {Poole}},\ }\href
  {\doibase 10.1093/mnras/stt2062} {\bibfield  {journal} {\bibinfo  {journal}
  {Mon. Not. Roy. Astron. Soc.}\ }\textbf {\bibinfo {volume} {437}},\ \bibinfo
  {pages} {2488} (\bibinfo {year} {2014})},\ \Eprint
  {http://arxiv.org/abs/1310.6810} {arXiv:1310.6810 [astro-ph.CO]} \BibitemShut
  {NoStop}%
\bibitem [{\citenamefont {H{\"o}nig}(2014)}]{Hoenig:2014jca}%
  \BibitemOpen
  \bibfield  {author} {\bibinfo {author} {\bibfnamefont {S.~F.}\ \bibnamefont
  {H{\"o}nig}},\ }\href {\doibase 10.1088/2041-8205/784/1/L4} {\bibfield
  {journal} {\bibinfo  {journal} {Astrophys. J.}\ }\textbf {\bibinfo {volume}
  {784}},\ \bibinfo {pages} {L4} (\bibinfo {year} {2014})},\ \Eprint
  {http://arxiv.org/abs/1401.2999} {arXiv:1401.2999 [astro-ph.GA]} \BibitemShut
  {NoStop}%
\bibitem [{\citenamefont {H{\"o}nig}\ \emph {et~al.}(2017)\citenamefont
  {H{\"o}nig} \emph {et~al.}}]{Honig:2016oyn}%
  \BibitemOpen
  \bibfield  {author} {\bibinfo {author} {\bibfnamefont {S.~F.}\ \bibnamefont
  {H{\"o}nig}} \emph {et~al.},\ }\href {\doibase 10.1093/mnras/stw2484}
  {\bibfield  {journal} {\bibinfo  {journal} {Mon. Not. Roy. Astron. Soc.}\
  }\textbf {\bibinfo {volume} {464}},\ \bibinfo {pages} {1693} (\bibinfo {year}
  {2017})},\ \Eprint {http://arxiv.org/abs/1609.09091} {arXiv:1609.09091
  [astro-ph.CO]} \BibitemShut {NoStop}%
\bibitem [{\citenamefont {Paraficz}\ and\ \citenamefont
  {Hjorth}(2009)}]{Paraficz:2009xj}%
  \BibitemOpen
  \bibfield  {author} {\bibinfo {author} {\bibfnamefont {D.}~\bibnamefont
  {Paraficz}}\ and\ \bibinfo {author} {\bibfnamefont {J.}~\bibnamefont
  {Hjorth}},\ }\href {\doibase 10.1051/0004-6361/200913307} {\bibfield
  {journal} {\bibinfo  {journal} {Astron. Astrophys.}\ }\textbf {\bibinfo
  {volume} {507}},\ \bibinfo {pages} {L49} (\bibinfo {year} {2009})},\ \Eprint
  {http://arxiv.org/abs/0910.5823} {arXiv:0910.5823 [astro-ph.CO]} \BibitemShut
  {NoStop}%
\bibitem [{\citenamefont {Agnello}\ \emph {et~al.}(2016)\citenamefont
  {Agnello}, \citenamefont {Sonnenfeld}, \citenamefont {Suyu}, \citenamefont
  {Treu}, \citenamefont {Fassnacht}, \citenamefont {Mason}, \citenamefont
  {Brada{\v c}},\ and\ \citenamefont {Auger}}]{Agnello:2015ala}%
  \BibitemOpen
  \bibfield  {author} {\bibinfo {author} {\bibfnamefont {A.}~\bibnamefont
  {Agnello}}, \bibinfo {author} {\bibfnamefont {A.}~\bibnamefont {Sonnenfeld}},
  \bibinfo {author} {\bibfnamefont {S.~H.}\ \bibnamefont {Suyu}}, \bibinfo
  {author} {\bibfnamefont {T.}~\bibnamefont {Treu}}, \bibinfo {author}
  {\bibfnamefont {C.~D.}\ \bibnamefont {Fassnacht}}, \bibinfo {author}
  {\bibfnamefont {C.}~\bibnamefont {Mason}}, \bibinfo {author} {\bibfnamefont
  {M.}~\bibnamefont {Brada{\v c}}}, \ and\ \bibinfo {author} {\bibfnamefont
  {M.~W.}\ \bibnamefont {Auger}},\ }\href {\doibase 10.1093/mnras/stw529}
  {\bibfield  {journal} {\bibinfo  {journal} {Mon. Not. Roy. Astron. Soc.}\
  }\textbf {\bibinfo {volume} {458}},\ \bibinfo {pages} {3830} (\bibinfo {year}
  {2016})},\ \Eprint {http://arxiv.org/abs/1506.02720} {arXiv:1506.02720
  [astro-ph.CO]} \BibitemShut {NoStop}%
\bibitem [{\citenamefont {Ntelis}\ \emph {et~al.}(2018)\citenamefont {Ntelis},
  \citenamefont {Ealet}, \citenamefont {Escoffier}, \citenamefont {Hamilton},
  \citenamefont {Hawken}, \citenamefont {Le~Goff}, \citenamefont {Rich},\ and\
  \citenamefont {Tilquin}}]{Ntelis:2018ctq}%
  \BibitemOpen
  \bibfield  {author} {\bibinfo {author} {\bibfnamefont {P.}~\bibnamefont
  {Ntelis}}, \bibinfo {author} {\bibfnamefont {A.}~\bibnamefont {Ealet}},
  \bibinfo {author} {\bibfnamefont {S.}~\bibnamefont {Escoffier}}, \bibinfo
  {author} {\bibfnamefont {J.-C.}\ \bibnamefont {Hamilton}}, \bibinfo {author}
  {\bibfnamefont {A.~J.}\ \bibnamefont {Hawken}}, \bibinfo {author}
  {\bibfnamefont {J.-M.}\ \bibnamefont {Le~Goff}}, \bibinfo {author}
  {\bibfnamefont {J.}~\bibnamefont {Rich}}, \ and\ \bibinfo {author}
  {\bibfnamefont {A.}~\bibnamefont {Tilquin}},\ }\href {\doibase
  10.1088/1475-7516/2018/12/014} {\bibfield  {journal} {\bibinfo  {journal}
  {JCAP}\ }\textbf {\bibinfo {volume} {1812}},\ \bibinfo {pages} {014}
  (\bibinfo {year} {2018})},\ \Eprint {http://arxiv.org/abs/1810.09362}
  {arXiv:1810.09362 [astro-ph.CO]} \BibitemShut {NoStop}%
\bibitem [{\citenamefont {Nesseris}\ and\ \citenamefont
  {Trashorras}(2019)}]{Nesseris:2019mlr}%
  \BibitemOpen
  \bibfield  {author} {\bibinfo {author} {\bibfnamefont {S.}~\bibnamefont
  {Nesseris}}\ and\ \bibinfo {author} {\bibfnamefont {M.}~\bibnamefont
  {Trashorras}},\ }\href {\doibase 10.1103/PhysRevD.99.063539} {\bibfield
  {journal} {\bibinfo  {journal} {Phys. Rev.}\ }\textbf {\bibinfo {volume}
  {D99}},\ \bibinfo {pages} {063539} (\bibinfo {year} {2019})},\ \Eprint
  {http://arxiv.org/abs/1901.02400} {arXiv:1901.02400 [astro-ph.CO]}
  \BibitemShut {NoStop}%
\bibitem [{\citenamefont {Mu{\~n}oz}(2019{\natexlab{a}})}]{Munoz:2019fkt}%
  \BibitemOpen
  \bibfield  {author} {\bibinfo {author} {\bibfnamefont {J.~B.}\ \bibnamefont
  {Mu{\~n}oz}},\ }\href {\doibase 10.1103/PhysRevLett.123.131301} {\bibfield
  {journal} {\bibinfo  {journal} {Phys. Rev. Lett.}\ }\textbf {\bibinfo
  {volume} {123}},\ \bibinfo {pages} {131301} (\bibinfo {year}
  {2019}{\natexlab{a}})},\ \Eprint {http://arxiv.org/abs/1904.07868}
  {arXiv:1904.07868 [astro-ph.CO]} \BibitemShut {NoStop}%
\bibitem [{\citenamefont {Mu{\~n}oz}(2019{\natexlab{b}})}]{Munoz:2019rhi}%
  \BibitemOpen
  \bibfield  {author} {\bibinfo {author} {\bibfnamefont {J.~B.}\ \bibnamefont
  {Mu{\~n}oz}},\ }\href {\doibase 10.1103/PhysRevD.100.063538} {\bibfield
  {journal} {\bibinfo  {journal} {Phys. Rev.}\ }\textbf {\bibinfo {volume}
  {D100}},\ \bibinfo {pages} {063538} (\bibinfo {year} {2019}{\natexlab{b}})},\
  \Eprint {http://arxiv.org/abs/1904.07881} {arXiv:1904.07881 [astro-ph.CO]}
  \BibitemShut {NoStop}%
\bibitem [{\citenamefont {Kervella}\ \emph {et~al.}(2008)\citenamefont
  {Kervella}, \citenamefont {Merand}, \citenamefont {Szabados}, \citenamefont
  {Fouque}, \citenamefont {Bersier}, \citenamefont {Pompei},\ and\
  \citenamefont {Perrin}}]{Kervella:2008ne}%
  \BibitemOpen
  \bibfield  {author} {\bibinfo {author} {\bibfnamefont {P.}~\bibnamefont
  {Kervella}}, \bibinfo {author} {\bibfnamefont {A.}~\bibnamefont {Merand}},
  \bibinfo {author} {\bibfnamefont {L.}~\bibnamefont {Szabados}}, \bibinfo
  {author} {\bibfnamefont {P.}~\bibnamefont {Fouque}}, \bibinfo {author}
  {\bibfnamefont {D.}~\bibnamefont {Bersier}}, \bibinfo {author} {\bibfnamefont
  {E.}~\bibnamefont {Pompei}}, \ and\ \bibinfo {author} {\bibfnamefont
  {G.}~\bibnamefont {Perrin}},\ }\href {\doibase 10.1051/0004-6361:20078961}
  {\bibfield  {journal} {\bibinfo  {journal} {Astron. Astrophys.}\ }\textbf
  {\bibinfo {volume} {480}},\ \bibinfo {pages} {167} (\bibinfo {year}
  {2008})},\ \Eprint {http://arxiv.org/abs/0802.1501} {arXiv:0802.1501
  [astro-ph]} \BibitemShut {NoStop}%
\bibitem [{\citenamefont {Bond}\ and\ \citenamefont
  {Sparks}(2009)}]{Bond:2008ax}%
  \BibitemOpen
  \bibfield  {author} {\bibinfo {author} {\bibfnamefont {H.~E.}\ \bibnamefont
  {Bond}}\ and\ \bibinfo {author} {\bibfnamefont {W.~B.}\ \bibnamefont
  {Sparks}},\ }\href {\doibase 10.1051/0004-6361:200810280} {\bibfield
  {journal} {\bibinfo  {journal} {Astron. Astrophys.}\ }\textbf {\bibinfo
  {volume} {495}},\ \bibinfo {pages} {371} (\bibinfo {year} {2009})},\ \Eprint
  {http://arxiv.org/abs/0811.2943} {arXiv:0811.2943 [astro-ph]} \BibitemShut
  {NoStop}%
\bibitem [{\citenamefont {Anselmi}\ \emph {et~al.}(2016)\citenamefont
  {Anselmi}, \citenamefont {Starkman},\ and\ \citenamefont
  {Sheth}}]{Anselmi:2015dha}%
  \BibitemOpen
  \bibfield  {author} {\bibinfo {author} {\bibfnamefont {S.}~\bibnamefont
  {Anselmi}}, \bibinfo {author} {\bibfnamefont {G.~D.}\ \bibnamefont
  {Starkman}}, \ and\ \bibinfo {author} {\bibfnamefont {R.~K.}\ \bibnamefont
  {Sheth}},\ }\href {\doibase 10.1093/mnras/stv2436} {\bibfield  {journal}
  {\bibinfo  {journal} {Mon. Not. Roy. Astron. Soc.}\ }\textbf {\bibinfo
  {volume} {455}},\ \bibinfo {pages} {2474} (\bibinfo {year} {2016})},\ \Eprint
  {http://arxiv.org/abs/1508.01170} {arXiv:1508.01170 [astro-ph.CO]}
  \BibitemShut {NoStop}%
\bibitem [{\citenamefont {Anselmi}\ \emph
  {et~al.}(2018{\natexlab{a}})\citenamefont {Anselmi}, \citenamefont
  {Starkman}, \citenamefont {Corasaniti}, \citenamefont {Sheth},\ and\
  \citenamefont {Zehavi}}]{Anselmi:2018hdn}%
  \BibitemOpen
  \bibfield  {author} {\bibinfo {author} {\bibfnamefont {S.}~\bibnamefont
  {Anselmi}}, \bibinfo {author} {\bibfnamefont {G.~D.}\ \bibnamefont
  {Starkman}}, \bibinfo {author} {\bibfnamefont {P.-S.}\ \bibnamefont
  {Corasaniti}}, \bibinfo {author} {\bibfnamefont {R.~K.}\ \bibnamefont
  {Sheth}}, \ and\ \bibinfo {author} {\bibfnamefont {I.}~\bibnamefont
  {Zehavi}},\ }\href {\doibase 10.1103/PhysRevLett.121.021302} {\bibfield
  {journal} {\bibinfo  {journal} {Phys. Rev. Lett.}\ }\textbf {\bibinfo
  {volume} {121}},\ \bibinfo {pages} {021302} (\bibinfo {year}
  {2018}{\natexlab{a}})},\ \Eprint {http://arxiv.org/abs/1703.01275}
  {arXiv:1703.01275 [astro-ph.CO]} \BibitemShut {NoStop}%
\bibitem [{\citenamefont {Anselmi}\ \emph
  {et~al.}(2018{\natexlab{b}})\citenamefont {Anselmi}, \citenamefont
  {Corasaniti}, \citenamefont {Starkman}, \citenamefont {Sheth},\ and\
  \citenamefont {Zehavi}}]{Anselmi:2017zss}%
  \BibitemOpen
  \bibfield  {author} {\bibinfo {author} {\bibfnamefont {S.}~\bibnamefont
  {Anselmi}}, \bibinfo {author} {\bibfnamefont {P.-S.}\ \bibnamefont
  {Corasaniti}}, \bibinfo {author} {\bibfnamefont {G.~D.}\ \bibnamefont
  {Starkman}}, \bibinfo {author} {\bibfnamefont {R.~K.}\ \bibnamefont {Sheth}},
  \ and\ \bibinfo {author} {\bibfnamefont {I.}~\bibnamefont {Zehavi}},\ }\href
  {\doibase 10.1103/PhysRevD.98.023527} {\bibfield  {journal} {\bibinfo
  {journal} {Phys. Rev.}\ }\textbf {\bibinfo {volume} {D98}},\ \bibinfo {pages}
  {023527} (\bibinfo {year} {2018}{\natexlab{b}})},\ \Eprint
  {http://arxiv.org/abs/1711.09063} {arXiv:1711.09063 [astro-ph.CO]}
  \BibitemShut {NoStop}%
\bibitem [{\citenamefont {Anselmi}\ \emph {et~al.}(2019)\citenamefont
  {Anselmi}, \citenamefont {Corasaniti}, \citenamefont {Sanchez}, \citenamefont
  {Starkman}, \citenamefont {Sheth},\ and\ \citenamefont
  {Zehavi}}]{Anselmi:2018vjz}%
  \BibitemOpen
  \bibfield  {author} {\bibinfo {author} {\bibfnamefont {S.}~\bibnamefont
  {Anselmi}}, \bibinfo {author} {\bibfnamefont {P.-S.}\ \bibnamefont
  {Corasaniti}}, \bibinfo {author} {\bibfnamefont {A.~G.}\ \bibnamefont
  {Sanchez}}, \bibinfo {author} {\bibfnamefont {G.~D.}\ \bibnamefont
  {Starkman}}, \bibinfo {author} {\bibfnamefont {R.~K.}\ \bibnamefont {Sheth}},
  \ and\ \bibinfo {author} {\bibfnamefont {I.}~\bibnamefont {Zehavi}},\ }\href
  {\doibase 10.1103/PhysRevD.99.123515} {\bibfield  {journal} {\bibinfo
  {journal} {Phys. Rev.}\ }\textbf {\bibinfo {volume} {D99}},\ \bibinfo {pages}
  {123515} (\bibinfo {year} {2019})},\ \Eprint
  {http://arxiv.org/abs/1811.12312} {arXiv:1811.12312 [astro-ph.CO]}
  \BibitemShut {NoStop}%
\bibitem [{\citenamefont {O'Dwyer}\ \emph {et~al.}(2019)\citenamefont
  {O'Dwyer}, \citenamefont {Anselmi}, \citenamefont {Starkman}, \citenamefont
  {Corasaniti}, \citenamefont {Sheth},\ and\ \citenamefont
  {Zehavi}}]{ODwyer:2019rvi}%
  \BibitemOpen
  \bibfield  {author} {\bibinfo {author} {\bibfnamefont {M.}~\bibnamefont
  {O'Dwyer}}, \bibinfo {author} {\bibfnamefont {S.}~\bibnamefont {Anselmi}},
  \bibinfo {author} {\bibfnamefont {G.~D.}\ \bibnamefont {Starkman}}, \bibinfo
  {author} {\bibfnamefont {P.-S.}\ \bibnamefont {Corasaniti}}, \bibinfo
  {author} {\bibfnamefont {R.~K.}\ \bibnamefont {Sheth}}, \ and\ \bibinfo
  {author} {\bibfnamefont {I.}~\bibnamefont {Zehavi}},\ }\href@noop {} {\
  (\bibinfo {year} {2019})},\ \Eprint {http://arxiv.org/abs/1910.10698}
  {arXiv:1910.10698 [astro-ph.CO]} \BibitemShut {NoStop}%
\bibitem [{\citenamefont {Tsupko}\ \emph {et~al.}(2019)\citenamefont {Tsupko},
  \citenamefont {Fan},\ and\ \citenamefont
  {Bisnovatyi-Kogan}}]{Tsupko:2019pzg}%
  \BibitemOpen
  \bibfield  {author} {\bibinfo {author} {\bibfnamefont {O.~{\relax Yu}.}\
  \bibnamefont {Tsupko}}, \bibinfo {author} {\bibfnamefont {Z.}~\bibnamefont
  {Fan}}, \ and\ \bibinfo {author} {\bibfnamefont {G.~S.}\ \bibnamefont
  {Bisnovatyi-Kogan}},\ }\href@noop {} {\  (\bibinfo {year} {2019})},\ \Eprint
  {http://arxiv.org/abs/1905.10509} {arXiv:1905.10509 [gr-qc]} \BibitemShut
  {NoStop}%
\bibitem [{\citenamefont {Qi}\ and\ \citenamefont {Zhang}(2019)}]{Qi:2019zdk}%
  \BibitemOpen
  \bibfield  {author} {\bibinfo {author} {\bibfnamefont {J.-Z.}\ \bibnamefont
  {Qi}}\ and\ \bibinfo {author} {\bibfnamefont {X.}~\bibnamefont {Zhang}},\
  }\href@noop {} {\  (\bibinfo {year} {2019})},\ \Eprint
  {http://arxiv.org/abs/1906.10825} {arXiv:1906.10825 [astro-ph.CO]}
  \BibitemShut {NoStop}%
\bibitem [{\citenamefont {Hawking}(1976)}]{Hawking:1976ra}%
  \BibitemOpen
  \bibfield  {author} {\bibinfo {author} {\bibfnamefont {S.~W.}\ \bibnamefont
  {Hawking}},\ }\href {\doibase 10.1103/PhysRevD.14.2460} {\bibfield  {journal}
  {\bibinfo  {journal} {Phys. Rev.}\ }\textbf {\bibinfo {volume} {D14}},\
  \bibinfo {pages} {2460} (\bibinfo {year} {1976})}\BibitemShut {NoStop}%
\bibitem [{\citenamefont {Mathur}(2006)}]{Mathur:2005ai}%
  \BibitemOpen
  \bibfield  {author} {\bibinfo {author} {\bibfnamefont {S.~D.}\ \bibnamefont
  {Mathur}},\ }\href {\doibase 10.1088/0264-9381/23/11/R01} {\bibfield
  {journal} {\bibinfo  {journal} {Class. Quant. Grav.}\ }\textbf {\bibinfo
  {volume} {23}},\ \bibinfo {pages} {R115} (\bibinfo {year} {2006})},\ \Eprint
  {http://arxiv.org/abs/hep-th/0510180} {arXiv:hep-th/0510180 [hep-th]}
  \BibitemShut {NoStop}%
\bibitem [{\citenamefont {Dvali}\ and\ \citenamefont
  {Gomez}(2013)}]{Dvali:2011aa}%
  \BibitemOpen
  \bibfield  {author} {\bibinfo {author} {\bibfnamefont {G.}~\bibnamefont
  {Dvali}}\ and\ \bibinfo {author} {\bibfnamefont {C.}~\bibnamefont {Gomez}},\
  }\href {\doibase 10.1002/prop.201300001} {\bibfield  {journal} {\bibinfo
  {journal} {Fortsch. Phys.}\ }\textbf {\bibinfo {volume} {61}},\ \bibinfo
  {pages} {742} (\bibinfo {year} {2013})},\ \Eprint
  {http://arxiv.org/abs/1112.3359} {arXiv:1112.3359 [hep-th]} \BibitemShut
  {NoStop}%
\bibitem [{\citenamefont {Giddings}(2017)}]{Giddings:2017jts}%
  \BibitemOpen
  \bibfield  {author} {\bibinfo {author} {\bibfnamefont {S.~B.}\ \bibnamefont
  {Giddings}},\ }\href {\doibase 10.1038/s41550-017-0067} {\bibfield  {journal}
  {\bibinfo  {journal} {Nat. Astron.}\ }\textbf {\bibinfo {volume} {1}},\
  \bibinfo {pages} {0067} (\bibinfo {year} {2017})},\ \Eprint
  {http://arxiv.org/abs/1703.03387} {arXiv:1703.03387 [gr-qc]} \BibitemShut
  {NoStop}%
\bibitem [{\citenamefont {Giddings}(2019)}]{Giddings:2019jwy}%
  \BibitemOpen
  \bibfield  {author} {\bibinfo {author} {\bibfnamefont {S.~B.}\ \bibnamefont
  {Giddings}},\ }\href {\doibase 10.3390/universe5090201} {\bibfield  {journal}
  {\bibinfo  {journal} {Universe}\ }\textbf {\bibinfo {volume} {5}},\ \bibinfo
  {pages} {201} (\bibinfo {year} {2019})},\ \Eprint
  {http://arxiv.org/abs/1904.05287} {arXiv:1904.05287 [gr-qc]} \BibitemShut
  {NoStop}%
\bibitem [{\citenamefont {Einstein}(1916)}]{Einstein:1916vd}%
  \BibitemOpen
  \bibfield  {author} {\bibinfo {author} {\bibfnamefont {A.}~\bibnamefont
  {Einstein}},\ }\href {\doibase 10.1002/andp.200590044,
  10.1002/andp.19163540702} {\bibfield  {journal} {\bibinfo  {journal} {Annalen
  Phys.}\ }\textbf {\bibinfo {volume} {49}},\ \bibinfo {pages} {769} (\bibinfo
  {year} {1916})},\ \bibinfo {note} {[Annalen
  Phys.354,no.7,769(1916)]}\BibitemShut {NoStop}%
\bibitem [{\citenamefont {Schwarzschild}(1916)}]{Schwarzschild:1916uq}%
  \BibitemOpen
  \bibfield  {author} {\bibinfo {author} {\bibfnamefont {K.}~\bibnamefont
  {Schwarzschild}},\ }\href@noop {} {\bibfield  {journal} {\bibinfo  {journal}
  {Sitzungsber. Preuss. Akad. Wiss. Berlin (Math. Phys.)}\ }\textbf {\bibinfo
  {volume} {1916}},\ \bibinfo {pages} {189} (\bibinfo {year} {1916})},\ \Eprint
  {http://arxiv.org/abs/physics/9905030} {arXiv:physics/9905030 [physics]}
  \BibitemShut {NoStop}%
\bibitem [{\citenamefont {Penrose}(1965)}]{Penrose:1964wq}%
  \BibitemOpen
  \bibfield  {author} {\bibinfo {author} {\bibfnamefont {R.}~\bibnamefont
  {Penrose}},\ }\href {\doibase 10.1103/PhysRevLett.14.57} {\bibfield
  {journal} {\bibinfo  {journal} {Phys. Rev. Lett.}\ }\textbf {\bibinfo
  {volume} {14}},\ \bibinfo {pages} {57} (\bibinfo {year} {1965})}\BibitemShut
  {NoStop}%
\bibitem [{\citenamefont {Bambi}(2019)}]{Bambi:2019xzp}%
  \BibitemOpen
  \bibfield  {author} {\bibinfo {author} {\bibfnamefont {C.}~\bibnamefont
  {Bambi}}\ }(\bibinfo {year} {2019})\ \Eprint
  {http://arxiv.org/abs/1906.03871} {arXiv:1906.03871 [astro-ph.HE]}
  \BibitemShut {NoStop}%
\bibitem [{\citenamefont {Lynden-Bell}(1969)}]{LyndenBell:1969yx}%
  \BibitemOpen
  \bibfield  {author} {\bibinfo {author} {\bibfnamefont {D.}~\bibnamefont
  {Lynden-Bell}},\ }\href {\doibase 10.1038/223690a0} {\bibfield  {journal}
  {\bibinfo  {journal} {Nature}\ }\textbf {\bibinfo {volume} {223}},\ \bibinfo
  {pages} {690} (\bibinfo {year} {1969})}\BibitemShut {NoStop}%
\bibitem [{\citenamefont {Kormendy}\ and\ \citenamefont
  {Richstone}(1995)}]{Kormendy:1995er}%
  \BibitemOpen
  \bibfield  {author} {\bibinfo {author} {\bibfnamefont {J.}~\bibnamefont
  {Kormendy}}\ and\ \bibinfo {author} {\bibfnamefont {D.}~\bibnamefont
  {Richstone}},\ }\href {\doibase 10.1146/annurev.aa.33.090195.003053}
  {\bibfield  {journal} {\bibinfo  {journal} {Ann. Rev. Astron. Astrophys.}\
  }\textbf {\bibinfo {volume} {33}},\ \bibinfo {pages} {581} (\bibinfo {year}
  {1995})}\BibitemShut {NoStop}%
\bibitem [{\citenamefont {Dokuchaev}\ and\ \citenamefont
  {Nazarova}(2019{\natexlab{a}})}]{Dokuchaev:2019jqq}%
  \BibitemOpen
  \bibfield  {author} {\bibinfo {author} {\bibfnamefont {V.~I.}\ \bibnamefont
  {Dokuchaev}}\ and\ \bibinfo {author} {\bibfnamefont {N.~O.}\ \bibnamefont
  {Nazarova}},\ }\href@noop {} {\  (\bibinfo {year} {2019}{\natexlab{a}})},\
  \Eprint {http://arxiv.org/abs/1911.07695} {arXiv:1911.07695 [gr-qc]}
  \BibitemShut {NoStop}%
\bibitem [{\citenamefont {Luminet}(1979)}]{Luminet:1979nyg}%
  \BibitemOpen
  \bibfield  {author} {\bibinfo {author} {\bibfnamefont {J.~P.}\ \bibnamefont
  {Luminet}},\ }\href@noop {} {\bibfield  {journal} {\bibinfo  {journal}
  {Astron. Astrophys.}\ }\textbf {\bibinfo {volume} {75}},\ \bibinfo {pages}
  {228} (\bibinfo {year} {1979})}\BibitemShut {NoStop}%
\bibitem [{\citenamefont {Lu}\ \emph {et~al.}(2014)\citenamefont {Lu},
  \citenamefont {Broderick}, \citenamefont {Baron}, \citenamefont {Monnier},
  \citenamefont {Fish}, \citenamefont {Doeleman},\ and\ \citenamefont
  {Pankratius}}]{Lu:2014zja}%
  \BibitemOpen
  \bibfield  {author} {\bibinfo {author} {\bibfnamefont {R.-S.}\ \bibnamefont
  {Lu}}, \bibinfo {author} {\bibfnamefont {A.~E.}\ \bibnamefont {Broderick}},
  \bibinfo {author} {\bibfnamefont {F.}~\bibnamefont {Baron}}, \bibinfo
  {author} {\bibfnamefont {J.~D.}\ \bibnamefont {Monnier}}, \bibinfo {author}
  {\bibfnamefont {V.~L.}\ \bibnamefont {Fish}}, \bibinfo {author}
  {\bibfnamefont {S.~S.}\ \bibnamefont {Doeleman}}, \ and\ \bibinfo {author}
  {\bibfnamefont {V.}~\bibnamefont {Pankratius}},\ }\href {\doibase
  10.1088/0004-637X/788/2/120} {\bibfield  {journal} {\bibinfo  {journal}
  {Astrophys. J.}\ }\textbf {\bibinfo {volume} {788}},\ \bibinfo {pages} {120}
  (\bibinfo {year} {2014})},\ \Eprint {http://arxiv.org/abs/1404.7095}
  {arXiv:1404.7095 [astro-ph.IM]} \BibitemShut {NoStop}%
\bibitem [{\citenamefont {Cunha}\ and\ \citenamefont
  {Herdeiro}(2018)}]{Cunha:2018acu}%
  \BibitemOpen
  \bibfield  {author} {\bibinfo {author} {\bibfnamefont {P.~V.~P.}\
  \bibnamefont {Cunha}}\ and\ \bibinfo {author} {\bibfnamefont {C.~A.~R.}\
  \bibnamefont {Herdeiro}},\ }\href {\doibase 10.1007/s10714-018-2361-9}
  {\bibfield  {journal} {\bibinfo  {journal} {Gen. Rel. Grav.}\ }\textbf
  {\bibinfo {volume} {50}},\ \bibinfo {pages} {42} (\bibinfo {year} {2018})},\
  \Eprint {http://arxiv.org/abs/1801.00860} {arXiv:1801.00860 [gr-qc]}
  \BibitemShut {NoStop}%
\bibitem [{\citenamefont {Gralla}\ \emph {et~al.}(2019)\citenamefont {Gralla},
  \citenamefont {Holz},\ and\ \citenamefont {Wald}}]{Gralla:2019xty}%
  \BibitemOpen
  \bibfield  {author} {\bibinfo {author} {\bibfnamefont {S.~E.}\ \bibnamefont
  {Gralla}}, \bibinfo {author} {\bibfnamefont {D.~E.}\ \bibnamefont {Holz}}, \
  and\ \bibinfo {author} {\bibfnamefont {R.~M.}\ \bibnamefont {Wald}},\ }\href
  {\doibase 10.1103/PhysRevD.100.024018} {\bibfield  {journal} {\bibinfo
  {journal} {Phys. Rev.}\ }\textbf {\bibinfo {volume} {D100}},\ \bibinfo
  {pages} {024018} (\bibinfo {year} {2019})},\ \Eprint
  {http://arxiv.org/abs/1906.00873} {arXiv:1906.00873 [astro-ph.HE]}
  \BibitemShut {NoStop}%
\bibitem [{\citenamefont {Narayan}\ \emph {et~al.}(2019)\citenamefont
  {Narayan}, \citenamefont {Johnson},\ and\ \citenamefont
  {Gammie}}]{Narayan:2019imo}%
  \BibitemOpen
  \bibfield  {author} {\bibinfo {author} {\bibfnamefont {R.}~\bibnamefont
  {Narayan}}, \bibinfo {author} {\bibfnamefont {M.~D.}\ \bibnamefont
  {Johnson}}, \ and\ \bibinfo {author} {\bibfnamefont {C.~F.}\ \bibnamefont
  {Gammie}},\ }\href {\doibase 10.3847/2041-8213/ab518c} {\bibfield  {journal}
  {\bibinfo  {journal} {Astrophys. J.}\ }\textbf {\bibinfo {volume} {885}},\
  \bibinfo {pages} {L33} (\bibinfo {year} {2019})},\ \Eprint
  {http://arxiv.org/abs/1910.02957} {arXiv:1910.02957 [astro-ph.HE]}
  \BibitemShut {NoStop}%
\bibitem [{\citenamefont {Falcke}\ \emph {et~al.}(2000)\citenamefont {Falcke},
  \citenamefont {Melia},\ and\ \citenamefont {Agol}}]{Falcke:1999pj}%
  \BibitemOpen
  \bibfield  {author} {\bibinfo {author} {\bibfnamefont {H.}~\bibnamefont
  {Falcke}}, \bibinfo {author} {\bibfnamefont {F.}~\bibnamefont {Melia}}, \
  and\ \bibinfo {author} {\bibfnamefont {E.}~\bibnamefont {Agol}},\ }\href
  {\doibase 10.1086/312423} {\bibfield  {journal} {\bibinfo  {journal}
  {Astrophys. J.}\ }\textbf {\bibinfo {volume} {528}},\ \bibinfo {pages} {L13}
  (\bibinfo {year} {2000})},\ \Eprint {http://arxiv.org/abs/astro-ph/9912263}
  {arXiv:astro-ph/9912263 [astro-ph]} \BibitemShut {NoStop}%
\bibitem [{\citenamefont {Fish}\ \emph {et~al.}(2016)\citenamefont {Fish},
  \citenamefont {Akiyama}, \citenamefont {Bouman}, \citenamefont {Chael},
  \citenamefont {Johnson}, \citenamefont {Doeleman}, \citenamefont {Blackburn},
  \citenamefont {Wardle},\ and\ \citenamefont {Freeman}}]{Fish:2016jil}%
  \BibitemOpen
  \bibfield  {author} {\bibinfo {author} {\bibfnamefont {V.~L.}\ \bibnamefont
  {Fish}}, \bibinfo {author} {\bibfnamefont {K.}~\bibnamefont {Akiyama}},
  \bibinfo {author} {\bibfnamefont {K.~L.}\ \bibnamefont {Bouman}}, \bibinfo
  {author} {\bibfnamefont {A.~A.}\ \bibnamefont {Chael}}, \bibinfo {author}
  {\bibfnamefont {M.~D.}\ \bibnamefont {Johnson}}, \bibinfo {author}
  {\bibfnamefont {S.~S.}\ \bibnamefont {Doeleman}}, \bibinfo {author}
  {\bibfnamefont {L.}~\bibnamefont {Blackburn}}, \bibinfo {author}
  {\bibfnamefont {J.~F.~C.}\ \bibnamefont {Wardle}}, \ and\ \bibinfo {author}
  {\bibfnamefont {W.~T.}\ \bibnamefont {Freeman}} (\bibinfo {collaboration}
  {Event Horizon Telescope}),\ }\bibfield  {booktitle} {\emph {\bibinfo
  {booktitle} {{Proceedings, Blazars through Sharp Multi-wavelength Eyes:
  Malaga, Spain, May 30-June 3, 2016}}},\ }\href {\doibase
  10.3390/galaxies4040054} {\bibfield  {journal} {\bibinfo  {journal}
  {Galaxies}\ }\textbf {\bibinfo {volume} {4}},\ \bibinfo {pages} {54}
  (\bibinfo {year} {2016})},\ \Eprint {http://arxiv.org/abs/1607.03034}
  {arXiv:1607.03034 [astro-ph.IM]} \BibitemShut {NoStop}%
\bibitem [{\citenamefont {Akiyama}\ \emph
  {et~al.}(2019{\natexlab{a}})\citenamefont {Akiyama} \emph
  {et~al.}}]{Akiyama:2019cqa}%
  \BibitemOpen
  \bibfield  {author} {\bibinfo {author} {\bibfnamefont {K.}~\bibnamefont
  {Akiyama}} \emph {et~al.} (\bibinfo {collaboration} {Event Horizon
  Telescope}),\ }\href {\doibase 10.3847/2041-8213/ab0ec7} {\bibfield
  {journal} {\bibinfo  {journal} {Astrophys. J.}\ }\textbf {\bibinfo {volume}
  {875}},\ \bibinfo {pages} {L1} (\bibinfo {year} {2019}{\natexlab{a}})},\
  \Eprint {http://arxiv.org/abs/1906.11238} {arXiv:1906.11238 [astro-ph.GA]}
  \BibitemShut {NoStop}%
\bibitem [{\citenamefont {Akiyama}\ \emph
  {et~al.}(2019{\natexlab{b}})\citenamefont {Akiyama} \emph
  {et~al.}}]{Akiyama:2019brx}%
  \BibitemOpen
  \bibfield  {author} {\bibinfo {author} {\bibfnamefont {K.}~\bibnamefont
  {Akiyama}} \emph {et~al.} (\bibinfo {collaboration} {Event Horizon
  Telescope}),\ }\href {\doibase 10.3847/2041-8213/ab0c96} {\bibfield
  {journal} {\bibinfo  {journal} {Astrophys. J.}\ }\textbf {\bibinfo {volume}
  {875}},\ \bibinfo {pages} {L2} (\bibinfo {year} {2019}{\natexlab{b}})},\
  \Eprint {http://arxiv.org/abs/1906.11239} {arXiv:1906.11239 [astro-ph.IM]}
  \BibitemShut {NoStop}%
\bibitem [{\citenamefont {Akiyama}\ \emph
  {et~al.}(2019{\natexlab{c}})\citenamefont {Akiyama} \emph
  {et~al.}}]{Akiyama:2019sww}%
  \BibitemOpen
  \bibfield  {author} {\bibinfo {author} {\bibfnamefont {K.}~\bibnamefont
  {Akiyama}} \emph {et~al.} (\bibinfo {collaboration} {Event Horizon
  Telescope}),\ }\href {\doibase 10.3847/2041-8213/ab0c57} {\bibfield
  {journal} {\bibinfo  {journal} {Astrophys. J.}\ }\textbf {\bibinfo {volume}
  {875}},\ \bibinfo {pages} {L3} (\bibinfo {year} {2019}{\natexlab{c}})},\
  \Eprint {http://arxiv.org/abs/1906.11240} {arXiv:1906.11240 [astro-ph.GA]}
  \BibitemShut {NoStop}%
\bibitem [{\citenamefont {Akiyama}\ \emph
  {et~al.}(2019{\natexlab{d}})\citenamefont {Akiyama} \emph
  {et~al.}}]{Akiyama:2019bqs}%
  \BibitemOpen
  \bibfield  {author} {\bibinfo {author} {\bibfnamefont {K.}~\bibnamefont
  {Akiyama}} \emph {et~al.} (\bibinfo {collaboration} {Event Horizon
  Telescope}),\ }\href {\doibase 10.3847/2041-8213/ab0e85} {\bibfield
  {journal} {\bibinfo  {journal} {Astrophys. J.}\ }\textbf {\bibinfo {volume}
  {875}},\ \bibinfo {pages} {L4} (\bibinfo {year} {2019}{\natexlab{d}})},\
  \Eprint {http://arxiv.org/abs/1906.11241} {arXiv:1906.11241 [astro-ph.GA]}
  \BibitemShut {NoStop}%
\bibitem [{\citenamefont {Akiyama}\ \emph
  {et~al.}(2019{\natexlab{e}})\citenamefont {Akiyama} \emph
  {et~al.}}]{Akiyama:2019fyp}%
  \BibitemOpen
  \bibfield  {author} {\bibinfo {author} {\bibfnamefont {K.}~\bibnamefont
  {Akiyama}} \emph {et~al.} (\bibinfo {collaboration} {Event Horizon
  Telescope}),\ }\href {\doibase 10.3847/2041-8213/ab0f43} {\bibfield
  {journal} {\bibinfo  {journal} {Astrophys. J.}\ }\textbf {\bibinfo {volume}
  {875}},\ \bibinfo {pages} {L5} (\bibinfo {year} {2019}{\natexlab{e}})},\
  \Eprint {http://arxiv.org/abs/1906.11242} {arXiv:1906.11242 [astro-ph.GA]}
  \BibitemShut {NoStop}%
\bibitem [{\citenamefont {Akiyama}\ \emph
  {et~al.}(2019{\natexlab{f}})\citenamefont {Akiyama} \emph
  {et~al.}}]{Akiyama:2019eap}%
  \BibitemOpen
  \bibfield  {author} {\bibinfo {author} {\bibfnamefont {K.}~\bibnamefont
  {Akiyama}} \emph {et~al.} (\bibinfo {collaboration} {Event Horizon
  Telescope}),\ }\href {\doibase 10.3847/2041-8213/ab1141} {\bibfield
  {journal} {\bibinfo  {journal} {Astrophys. J.}\ }\textbf {\bibinfo {volume}
  {875}},\ \bibinfo {pages} {L6} (\bibinfo {year} {2019}{\natexlab{f}})},\
  \Eprint {http://arxiv.org/abs/1906.11243} {arXiv:1906.11243 [astro-ph.GA]}
  \BibitemShut {NoStop}%
\bibitem [{\citenamefont {Kerr}(1963)}]{Kerr:1963ud}%
  \BibitemOpen
  \bibfield  {author} {\bibinfo {author} {\bibfnamefont {R.~P.}\ \bibnamefont
  {Kerr}},\ }\href {\doibase 10.1103/PhysRevLett.11.237} {\bibfield  {journal}
  {\bibinfo  {journal} {Phys. Rev. Lett.}\ }\textbf {\bibinfo {volume} {11}},\
  \bibinfo {pages} {237} (\bibinfo {year} {1963})}\BibitemShut {NoStop}%
\bibitem [{\citenamefont {Moffat}\ and\ \citenamefont
  {Toth}(2019)}]{Moffat:2019uxp}%
  \BibitemOpen
  \bibfield  {author} {\bibinfo {author} {\bibfnamefont {J.~W.}\ \bibnamefont
  {Moffat}}\ and\ \bibinfo {author} {\bibfnamefont {V.~T.}\ \bibnamefont
  {Toth}},\ }\href@noop {} {\  (\bibinfo {year} {2019})},\ \Eprint
  {http://arxiv.org/abs/1904.04142} {arXiv:1904.04142 [gr-qc]} \BibitemShut
  {NoStop}%
\bibitem [{\citenamefont {Nokhrina}\ \emph {et~al.}(2019)\citenamefont
  {Nokhrina}, \citenamefont {Gurvits}, \citenamefont {Beskin}, \citenamefont
  {Nakamura}, \citenamefont {Asada},\ and\ \citenamefont
  {Hada}}]{Nokhrina:2019sxv}%
  \BibitemOpen
  \bibfield  {author} {\bibinfo {author} {\bibfnamefont {E.~E.}\ \bibnamefont
  {Nokhrina}}, \bibinfo {author} {\bibfnamefont {L.~I.}\ \bibnamefont
  {Gurvits}}, \bibinfo {author} {\bibfnamefont {V.~S.}\ \bibnamefont {Beskin}},
  \bibinfo {author} {\bibfnamefont {M.}~\bibnamefont {Nakamura}}, \bibinfo
  {author} {\bibfnamefont {K.}~\bibnamefont {Asada}}, \ and\ \bibinfo {author}
  {\bibfnamefont {K.}~\bibnamefont {Hada}},\ }\href {\doibase
  10.1093/mnras/stz2116} {\bibfield  {journal} {\bibinfo  {journal} {Mon. Not.
  Roy. Astron. Soc.}\ }\textbf {\bibinfo {volume} {489}},\ \bibinfo {pages}
  {1197} (\bibinfo {year} {2019})},\ \Eprint {http://arxiv.org/abs/1904.05665}
  {arXiv:1904.05665 [astro-ph.HE]} \BibitemShut {NoStop}%
\bibitem [{\citenamefont {Abdikamalov}\ \emph {et~al.}(2019)\citenamefont
  {Abdikamalov}, \citenamefont {Abdujabbarov}, \citenamefont {Ayzenberg},
  \citenamefont {Malafarina}, \citenamefont {Bambi},\ and\ \citenamefont
  {Ahmedov}}]{Abdikamalov:2019ztb}%
  \BibitemOpen
  \bibfield  {author} {\bibinfo {author} {\bibfnamefont {A.~B.}\ \bibnamefont
  {Abdikamalov}}, \bibinfo {author} {\bibfnamefont {A.~A.}\ \bibnamefont
  {Abdujabbarov}}, \bibinfo {author} {\bibfnamefont {D.}~\bibnamefont
  {Ayzenberg}}, \bibinfo {author} {\bibfnamefont {D.}~\bibnamefont
  {Malafarina}}, \bibinfo {author} {\bibfnamefont {C.}~\bibnamefont {Bambi}}, \
  and\ \bibinfo {author} {\bibfnamefont {B.}~\bibnamefont {Ahmedov}},\ }\href
  {\doibase 10.1103/PhysRevD.100.024014} {\bibfield  {journal} {\bibinfo
  {journal} {Phys. Rev.}\ }\textbf {\bibinfo {volume} {D100}},\ \bibinfo
  {pages} {024014} (\bibinfo {year} {2019})},\ \Eprint
  {http://arxiv.org/abs/1904.06207} {arXiv:1904.06207 [gr-qc]} \BibitemShut
  {NoStop}%
\bibitem [{\citenamefont {Held}\ \emph {et~al.}(2019)\citenamefont {Held},
  \citenamefont {Gold},\ and\ \citenamefont {Eichhorn}}]{Held:2019xde}%
  \BibitemOpen
  \bibfield  {author} {\bibinfo {author} {\bibfnamefont {A.}~\bibnamefont
  {Held}}, \bibinfo {author} {\bibfnamefont {R.}~\bibnamefont {Gold}}, \ and\
  \bibinfo {author} {\bibfnamefont {A.}~\bibnamefont {Eichhorn}},\ }\href
  {\doibase 10.1088/1475-7516/2019/06/029} {\bibfield  {journal} {\bibinfo
  {journal} {JCAP}\ }\textbf {\bibinfo {volume} {1906}},\ \bibinfo {pages}
  {029} (\bibinfo {year} {2019})},\ \Eprint {http://arxiv.org/abs/1904.07133}
  {arXiv:1904.07133 [gr-qc]} \BibitemShut {NoStop}%
\bibitem [{\citenamefont {Wei}\ \emph {et~al.}(2019)\citenamefont {Wei},
  \citenamefont {Zou}, \citenamefont {Liu},\ and\ \citenamefont
  {Mann}}]{Wei:2019pjf}%
  \BibitemOpen
  \bibfield  {author} {\bibinfo {author} {\bibfnamefont {S.-W.}\ \bibnamefont
  {Wei}}, \bibinfo {author} {\bibfnamefont {Y.-C.}\ \bibnamefont {Zou}},
  \bibinfo {author} {\bibfnamefont {Y.-X.}\ \bibnamefont {Liu}}, \ and\
  \bibinfo {author} {\bibfnamefont {R.~B.}\ \bibnamefont {Mann}},\ }\href
  {\doibase 10.1088/1475-7516/2019/08/030} {\bibfield  {journal} {\bibinfo
  {journal} {JCAP}\ }\textbf {\bibinfo {volume} {1908}},\ \bibinfo {pages}
  {030} (\bibinfo {year} {2019})},\ \Eprint {http://arxiv.org/abs/1904.07710}
  {arXiv:1904.07710 [gr-qc]} \BibitemShut {NoStop}%
\bibitem [{\citenamefont {Shaikh}(2019)}]{Shaikh:2019fpu}%
  \BibitemOpen
  \bibfield  {author} {\bibinfo {author} {\bibfnamefont {R.}~\bibnamefont
  {Shaikh}},\ }\href {\doibase 10.1103/PhysRevD.100.024028} {\bibfield
  {journal} {\bibinfo  {journal} {Phys. Rev.}\ }\textbf {\bibinfo {volume}
  {D100}},\ \bibinfo {pages} {024028} (\bibinfo {year} {2019})},\ \Eprint
  {http://arxiv.org/abs/1904.08322} {arXiv:1904.08322 [gr-qc]} \BibitemShut
  {NoStop}%
\bibitem [{\citenamefont {Tamburini}\ \emph {et~al.}(2019)\citenamefont
  {Tamburini}, \citenamefont {Thid{\'e}},\ and\ \citenamefont
  {Della~Valle}}]{Tamburini:2019vrf}%
  \BibitemOpen
  \bibfield  {author} {\bibinfo {author} {\bibfnamefont {F.}~\bibnamefont
  {Tamburini}}, \bibinfo {author} {\bibfnamefont {B.}~\bibnamefont
  {Thid{\'e}}}, \ and\ \bibinfo {author} {\bibfnamefont {M.}~\bibnamefont
  {Della~Valle}},\ }\href@noop {} {\  (\bibinfo {year} {2019})},\ \Eprint
  {http://arxiv.org/abs/1904.07923} {arXiv:1904.07923 [astro-ph.HE]}
  \BibitemShut {NoStop}%
\bibitem [{\citenamefont {Davoudiasl}\ and\ \citenamefont
  {Denton}(2019)}]{Davoudiasl:2019nlo}%
  \BibitemOpen
  \bibfield  {author} {\bibinfo {author} {\bibfnamefont {H.}~\bibnamefont
  {Davoudiasl}}\ and\ \bibinfo {author} {\bibfnamefont {P.~B.}\ \bibnamefont
  {Denton}},\ }\href {\doibase 10.1103/PhysRevLett.123.021102} {\bibfield
  {journal} {\bibinfo  {journal} {Phys. Rev. Lett.}\ }\textbf {\bibinfo
  {volume} {123}},\ \bibinfo {pages} {021102} (\bibinfo {year} {2019})},\
  \Eprint {http://arxiv.org/abs/1904.09242} {arXiv:1904.09242 [astro-ph.CO]}
  \BibitemShut {NoStop}%
\bibitem [{\citenamefont {{\"O}vg{\"u}n}\ \emph {et~al.}(2019)\citenamefont
  {{\"O}vg{\"u}n}, \citenamefont {Sakalli},\ and\ \citenamefont
  {Mutuk}}]{Ovgun:2019yor}%
  \BibitemOpen
  \bibfield  {author} {\bibinfo {author} {\bibfnamefont {A.}~\bibnamefont
  {{\"O}vg{\"u}n}}, \bibinfo {author} {\bibfnamefont {I.}~\bibnamefont
  {Sakalli}}, \ and\ \bibinfo {author} {\bibfnamefont {H.}~\bibnamefont
  {Mutuk}},\ }\href@noop {} {\  (\bibinfo {year} {2019})},\ \Eprint
  {http://arxiv.org/abs/1904.09509} {arXiv:1904.09509 [gr-qc]} \BibitemShut
  {NoStop}%
\bibitem [{\citenamefont {Bambi}\ \emph {et~al.}(2019)\citenamefont {Bambi},
  \citenamefont {Freese}, \citenamefont {Vagnozzi},\ and\ \citenamefont
  {Visinelli}}]{Bambi:2019tjh}%
  \BibitemOpen
  \bibfield  {author} {\bibinfo {author} {\bibfnamefont {C.}~\bibnamefont
  {Bambi}}, \bibinfo {author} {\bibfnamefont {K.}~\bibnamefont {Freese}},
  \bibinfo {author} {\bibfnamefont {S.}~\bibnamefont {Vagnozzi}}, \ and\
  \bibinfo {author} {\bibfnamefont {L.}~\bibnamefont {Visinelli}},\ }\href
  {\doibase 10.1103/PhysRevD.100.044057} {\bibfield  {journal} {\bibinfo
  {journal} {Phys. Rev.}\ }\textbf {\bibinfo {volume} {D100}},\ \bibinfo
  {pages} {044057} (\bibinfo {year} {2019})},\ \Eprint
  {http://arxiv.org/abs/1904.12983} {arXiv:1904.12983 [gr-qc]} \BibitemShut
  {NoStop}%
\bibitem [{\citenamefont {Nemmen}(2019)}]{Nemmen:2019idv}%
  \BibitemOpen
  \bibfield  {author} {\bibinfo {author} {\bibfnamefont {R.}~\bibnamefont
  {Nemmen}},\ }\href {\doibase 10.3847/2041-8213/ab2fd3} {\bibfield  {journal}
  {\bibinfo  {journal} {Astrophys. J.}\ }\textbf {\bibinfo {volume} {880}},\
  \bibinfo {pages} {L26} (\bibinfo {year} {2019})},\ \Eprint
  {http://arxiv.org/abs/1905.02143} {arXiv:1905.02143 [astro-ph.HE]}
  \BibitemShut {NoStop}%
\bibitem [{\citenamefont {Churilova}(2019)}]{Churilova:2019jqx}%
  \BibitemOpen
  \bibfield  {author} {\bibinfo {author} {\bibfnamefont {M.~S.}\ \bibnamefont
  {Churilova}},\ }\href {\doibase 10.1140/epjc/s10052-019-7146-0} {\bibfield
  {journal} {\bibinfo  {journal} {Eur. Phys. J.}\ }\textbf {\bibinfo {volume}
  {C79}},\ \bibinfo {pages} {629} (\bibinfo {year} {2019})},\ \Eprint
  {http://arxiv.org/abs/1905.04536} {arXiv:1905.04536 [gr-qc]} \BibitemShut
  {NoStop}%
\bibitem [{\citenamefont {Safarzadeh}\ \emph {et~al.}(2019)\citenamefont
  {Safarzadeh}, \citenamefont {Loeb},\ and\ \citenamefont
  {Reid}}]{Safarzadeh:2019imq}%
  \BibitemOpen
  \bibfield  {author} {\bibinfo {author} {\bibfnamefont {M.}~\bibnamefont
  {Safarzadeh}}, \bibinfo {author} {\bibfnamefont {A.}~\bibnamefont {Loeb}}, \
  and\ \bibinfo {author} {\bibfnamefont {M.}~\bibnamefont {Reid}},\ }\href
  {\doibase 10.1093/mnrasl/slz108} {\bibfield  {journal} {\bibinfo  {journal}
  {Mon. Not. Roy. Astron. Soc.}\ }\textbf {\bibinfo {volume} {488}},\ \bibinfo
  {pages} {L90} (\bibinfo {year} {2019})},\ \Eprint
  {http://arxiv.org/abs/1905.06835} {arXiv:1905.06835 [astro-ph.GA]}
  \BibitemShut {NoStop}%
\bibitem [{\citenamefont {Firouzjaee}\ and\ \citenamefont
  {Allahyari}(2019)}]{Firouzjaee:2019aij}%
  \BibitemOpen
  \bibfield  {author} {\bibinfo {author} {\bibfnamefont {J.~T.}\ \bibnamefont
  {Firouzjaee}}\ and\ \bibinfo {author} {\bibfnamefont {A.}~\bibnamefont
  {Allahyari}},\ }\href {\doibase 10.1140/epjc/s10052-019-7464-2} {\bibfield
  {journal} {\bibinfo  {journal} {Eur. Phys. J.}\ }\textbf {\bibinfo {volume}
  {C79}},\ \bibinfo {pages} {930} (\bibinfo {year} {2019})},\ \Eprint
  {http://arxiv.org/abs/1905.07378} {arXiv:1905.07378 [astro-ph.CO]}
  \BibitemShut {NoStop}%
\bibitem [{\citenamefont {Konoplya}\ \emph {et~al.}(2019)\citenamefont
  {Konoplya}, \citenamefont {Posada}, \citenamefont {Stuchl{\'\i}k},\ and\
  \citenamefont {Zhidenko}}]{Konoplya:2019nzp}%
  \BibitemOpen
  \bibfield  {author} {\bibinfo {author} {\bibfnamefont {R.~A.}\ \bibnamefont
  {Konoplya}}, \bibinfo {author} {\bibfnamefont {C.}~\bibnamefont {Posada}},
  \bibinfo {author} {\bibfnamefont {Z.}~\bibnamefont {Stuchl{\'\i}k}}, \ and\
  \bibinfo {author} {\bibfnamefont {A.}~\bibnamefont {Zhidenko}},\ }\href
  {\doibase 10.1103/PhysRevD.100.044027} {\bibfield  {journal} {\bibinfo
  {journal} {Phys. Rev.}\ }\textbf {\bibinfo {volume} {D100}},\ \bibinfo
  {pages} {044027} (\bibinfo {year} {2019})},\ \Eprint
  {http://arxiv.org/abs/1905.08097} {arXiv:1905.08097 [gr-qc]} \BibitemShut
  {NoStop}%
\bibitem [{\citenamefont {Kawashima}\ \emph {et~al.}(2019)\citenamefont
  {Kawashima}, \citenamefont {Kino},\ and\ \citenamefont
  {Akiyama}}]{Kawashima:2019ljv}%
  \BibitemOpen
  \bibfield  {author} {\bibinfo {author} {\bibfnamefont {T.}~\bibnamefont
  {Kawashima}}, \bibinfo {author} {\bibfnamefont {M.}~\bibnamefont {Kino}}, \
  and\ \bibinfo {author} {\bibfnamefont {K.}~\bibnamefont {Akiyama}},\ }\href
  {\doibase 10.3847/1538-4357/ab19c0} {\bibfield  {journal} {\bibinfo
  {journal} {Astrophys. J.}\ }\textbf {\bibinfo {volume} {878}},\ \bibinfo
  {pages} {27} (\bibinfo {year} {2019})},\ \Eprint
  {http://arxiv.org/abs/1905.10717} {arXiv:1905.10717 [astro-ph.HE]}
  \BibitemShut {NoStop}%
\bibitem [{\citenamefont {Contreras}\ \emph
  {et~al.}(2019{\natexlab{a}})\citenamefont {Contreras}, \citenamefont
  {Ramirez-Velasquez}, \citenamefont {Rinc{\'o}n}, \citenamefont
  {Panotopoulos},\ and\ \citenamefont {Bargue{\~n}o}}]{Contreras:2019nih}%
  \BibitemOpen
  \bibfield  {author} {\bibinfo {author} {\bibfnamefont {E.}~\bibnamefont
  {Contreras}}, \bibinfo {author} {\bibfnamefont {J.~M.}\ \bibnamefont
  {Ramirez-Velasquez}}, \bibinfo {author} {\bibfnamefont {A.}~\bibnamefont
  {Rinc{\'o}n}}, \bibinfo {author} {\bibfnamefont {G.}~\bibnamefont
  {Panotopoulos}}, \ and\ \bibinfo {author} {\bibfnamefont {P.}~\bibnamefont
  {Bargue{\~n}o}},\ }\href {\doibase 10.1140/epjc/s10052-019-7309-z} {\bibfield
   {journal} {\bibinfo  {journal} {Eur. Phys. J.}\ }\textbf {\bibinfo {volume}
  {C79}},\ \bibinfo {pages} {802} (\bibinfo {year} {2019}{\natexlab{a}})},\
  \Eprint {http://arxiv.org/abs/1905.11443} {arXiv:1905.11443 [gr-qc]}
  \BibitemShut {NoStop}%
\bibitem [{\citenamefont {Bar}\ \emph {et~al.}(2019)\citenamefont {Bar},
  \citenamefont {Blum}, \citenamefont {Lacroix},\ and\ \citenamefont
  {Panci}}]{Bar:2019pnz}%
  \BibitemOpen
  \bibfield  {author} {\bibinfo {author} {\bibfnamefont {N.}~\bibnamefont
  {Bar}}, \bibinfo {author} {\bibfnamefont {K.}~\bibnamefont {Blum}}, \bibinfo
  {author} {\bibfnamefont {T.}~\bibnamefont {Lacroix}}, \ and\ \bibinfo
  {author} {\bibfnamefont {P.}~\bibnamefont {Panci}},\ }\href {\doibase
  10.1088/1475-7516/2019/07/045} {\bibfield  {journal} {\bibinfo  {journal}
  {JCAP}\ }\textbf {\bibinfo {volume} {1907}},\ \bibinfo {pages} {045}
  (\bibinfo {year} {2019})},\ \Eprint {http://arxiv.org/abs/1905.11745}
  {arXiv:1905.11745 [astro-ph.CO]} \BibitemShut {NoStop}%
\bibitem [{\citenamefont {Jusufi}\ \emph {et~al.}(2019)\citenamefont {Jusufi},
  \citenamefont {Jamil}, \citenamefont {Salucci}, \citenamefont {Zhu},\ and\
  \citenamefont {Haroon}}]{Jusufi:2019nrn}%
  \BibitemOpen
  \bibfield  {author} {\bibinfo {author} {\bibfnamefont {K.}~\bibnamefont
  {Jusufi}}, \bibinfo {author} {\bibfnamefont {M.}~\bibnamefont {Jamil}},
  \bibinfo {author} {\bibfnamefont {P.}~\bibnamefont {Salucci}}, \bibinfo
  {author} {\bibfnamefont {T.}~\bibnamefont {Zhu}}, \ and\ \bibinfo {author}
  {\bibfnamefont {S.}~\bibnamefont {Haroon}},\ }\href {\doibase
  10.1103/PhysRevD.100.044012} {\bibfield  {journal} {\bibinfo  {journal}
  {Phys. Rev.}\ }\textbf {\bibinfo {volume} {D100}},\ \bibinfo {pages} {044012}
  (\bibinfo {year} {2019})},\ \Eprint {http://arxiv.org/abs/1905.11803}
  {arXiv:1905.11803 [physics.gen-ph]} \BibitemShut {NoStop}%
\bibitem [{\citenamefont {Vagnozzi}\ and\ \citenamefont
  {Visinelli}(2019)}]{Vagnozzi:2019apd}%
  \BibitemOpen
  \bibfield  {author} {\bibinfo {author} {\bibfnamefont {S.}~\bibnamefont
  {Vagnozzi}}\ and\ \bibinfo {author} {\bibfnamefont {L.}~\bibnamefont
  {Visinelli}},\ }\href {\doibase 10.1103/PhysRevD.100.024020} {\bibfield
  {journal} {\bibinfo  {journal} {Phys. Rev.}\ }\textbf {\bibinfo {volume}
  {D100}},\ \bibinfo {pages} {024020} (\bibinfo {year} {2019})},\ \Eprint
  {http://arxiv.org/abs/1905.12421} {arXiv:1905.12421 [gr-qc]} \BibitemShut
  {NoStop}%
\bibitem [{\citenamefont {Banerjee}\ \emph
  {et~al.}(2019{\natexlab{a}})\citenamefont {Banerjee}, \citenamefont
  {Mandal},\ and\ \citenamefont {SenGupta}}]{Banerjee:2019cjk}%
  \BibitemOpen
  \bibfield  {author} {\bibinfo {author} {\bibfnamefont {I.}~\bibnamefont
  {Banerjee}}, \bibinfo {author} {\bibfnamefont {B.}~\bibnamefont {Mandal}}, \
  and\ \bibinfo {author} {\bibfnamefont {S.}~\bibnamefont {SenGupta}},\
  }\href@noop {} {\  (\bibinfo {year} {2019}{\natexlab{a}})},\ \Eprint
  {http://arxiv.org/abs/1905.12820} {arXiv:1905.12820 [gr-qc]} \BibitemShut
  {NoStop}%
\bibitem [{\citenamefont {Roy}\ and\ \citenamefont
  {Yajnik}(2019)}]{Roy:2019esk}%
  \BibitemOpen
  \bibfield  {author} {\bibinfo {author} {\bibfnamefont {R.}~\bibnamefont
  {Roy}}\ and\ \bibinfo {author} {\bibfnamefont {U.~A.}\ \bibnamefont
  {Yajnik}},\ }\href@noop {} {\  (\bibinfo {year} {2019})},\ \Eprint
  {http://arxiv.org/abs/1906.03190} {arXiv:1906.03190 [gr-qc]} \BibitemShut
  {NoStop}%
\bibitem [{\citenamefont {Ali}\ and\ \citenamefont {Amir}(2019)}]{Ali:2019khp}%
  \BibitemOpen
  \bibfield  {author} {\bibinfo {author} {\bibfnamefont {M.~S.}\ \bibnamefont
  {Ali}}\ and\ \bibinfo {author} {\bibfnamefont {M.}~\bibnamefont {Amir}},\
  }\href@noop {} {\  (\bibinfo {year} {2019})},\ \Eprint
  {http://arxiv.org/abs/1906.04146} {arXiv:1906.04146 [gr-qc]} \BibitemShut
  {NoStop}%
\bibitem [{\citenamefont {Long}\ \emph {et~al.}(2019)\citenamefont {Long},
  \citenamefont {Wang}, \citenamefont {Chen},\ and\ \citenamefont
  {Jing}}]{Long:2019nox}%
  \BibitemOpen
  \bibfield  {author} {\bibinfo {author} {\bibfnamefont {F.}~\bibnamefont
  {Long}}, \bibinfo {author} {\bibfnamefont {J.}~\bibnamefont {Wang}}, \bibinfo
  {author} {\bibfnamefont {S.}~\bibnamefont {Chen}}, \ and\ \bibinfo {author}
  {\bibfnamefont {J.}~\bibnamefont {Jing}},\ }\href {\doibase
  10.1007/JHEP10(2019)269} {\bibfield  {journal} {\bibinfo  {journal} {JHEP}\
  }\textbf {\bibinfo {volume} {10}},\ \bibinfo {pages} {269} (\bibinfo {year}
  {2019})},\ \Eprint {http://arxiv.org/abs/1906.04456} {arXiv:1906.04456
  [gr-qc]} \BibitemShut {NoStop}%
\bibitem [{\citenamefont {Zhu}\ \emph {et~al.}(2019)\citenamefont {Zhu},
  \citenamefont {Wu}, \citenamefont {Jamil},\ and\ \citenamefont
  {Jusufi}}]{Zhu:2019ura}%
  \BibitemOpen
  \bibfield  {author} {\bibinfo {author} {\bibfnamefont {T.}~\bibnamefont
  {Zhu}}, \bibinfo {author} {\bibfnamefont {Q.}~\bibnamefont {Wu}}, \bibinfo
  {author} {\bibfnamefont {M.}~\bibnamefont {Jamil}}, \ and\ \bibinfo {author}
  {\bibfnamefont {K.}~\bibnamefont {Jusufi}},\ }\href {\doibase
  10.1103/PhysRevD.100.044055} {\bibfield  {journal} {\bibinfo  {journal}
  {Phys. Rev.}\ }\textbf {\bibinfo {volume} {D100}},\ \bibinfo {pages} {044055}
  (\bibinfo {year} {2019})},\ \Eprint {http://arxiv.org/abs/1906.05673}
  {arXiv:1906.05673 [gr-qc]} \BibitemShut {NoStop}%
\bibitem [{\citenamefont {Contreras}\ \emph
  {et~al.}(2019{\natexlab{b}})\citenamefont {Contreras}, \citenamefont
  {Rinc{\'o}n}, \citenamefont {Panotopoulos}, \citenamefont {Bargue{\~n}o},\
  and\ \citenamefont {Koch}}]{Contreras:2019cmf}%
  \BibitemOpen
  \bibfield  {author} {\bibinfo {author} {\bibfnamefont {E.}~\bibnamefont
  {Contreras}}, \bibinfo {author} {\bibfnamefont {A.}~\bibnamefont
  {Rinc{\'o}n}}, \bibinfo {author} {\bibfnamefont {G.}~\bibnamefont
  {Panotopoulos}}, \bibinfo {author} {\bibfnamefont {P.}~\bibnamefont
  {Bargue{\~n}o}}, \ and\ \bibinfo {author} {\bibfnamefont {B.}~\bibnamefont
  {Koch}},\ }\href@noop {} {\  (\bibinfo {year} {2019}{\natexlab{b}})},\
  \Eprint {http://arxiv.org/abs/1906.06990} {arXiv:1906.06990 [gr-qc]}
  \BibitemShut {NoStop}%
\bibitem [{\citenamefont {Dokuchaev}\ and\ \citenamefont
  {Nazarova}(2019{\natexlab{b}})}]{Dokuchaev:2019pcx}%
  \BibitemOpen
  \bibfield  {author} {\bibinfo {author} {\bibfnamefont {V.~I.}\ \bibnamefont
  {Dokuchaev}}\ and\ \bibinfo {author} {\bibfnamefont {N.~O.}\ \bibnamefont
  {Nazarova}},\ }\href {\doibase 10.3390/universe5080183} {\bibfield  {journal}
  {\bibinfo  {journal} {Universe}\ }\textbf {\bibinfo {volume} {5}},\ \bibinfo
  {pages} {183} (\bibinfo {year} {2019}{\natexlab{b}})},\ \Eprint
  {http://arxiv.org/abs/1906.07171} {arXiv:1906.07171 [astro-ph.HE]}
  \BibitemShut {NoStop}%
\bibitem [{\citenamefont {Wang}\ \emph
  {et~al.}(2019{\natexlab{a}})\citenamefont {Wang}, \citenamefont {Zhao},
  \citenamefont {Zhang},\ and\ \citenamefont {Zhang}}]{Wang:2019tto}%
  \BibitemOpen
  \bibfield  {author} {\bibinfo {author} {\bibfnamefont {L.-F.}\ \bibnamefont
  {Wang}}, \bibinfo {author} {\bibfnamefont {Z.-W.}\ \bibnamefont {Zhao}},
  \bibinfo {author} {\bibfnamefont {J.-F.}\ \bibnamefont {Zhang}}, \ and\
  \bibinfo {author} {\bibfnamefont {X.}~\bibnamefont {Zhang}},\ }\href@noop {}
  {\  (\bibinfo {year} {2019}{\natexlab{a}})},\ \Eprint
  {http://arxiv.org/abs/1907.01838} {arXiv:1907.01838 [astro-ph.CO]}
  \BibitemShut {NoStop}%
\bibitem [{\citenamefont {Konoplya}\ and\ \citenamefont
  {Zhidenko}(2019)}]{Konoplya:2019goy}%
  \BibitemOpen
  \bibfield  {author} {\bibinfo {author} {\bibfnamefont {R.~A.}\ \bibnamefont
  {Konoplya}}\ and\ \bibinfo {author} {\bibfnamefont {A.}~\bibnamefont
  {Zhidenko}},\ }\href {\doibase 10.1103/PhysRevD.100.044015} {\bibfield
  {journal} {\bibinfo  {journal} {Phys. Rev.}\ }\textbf {\bibinfo {volume}
  {D100}},\ \bibinfo {pages} {044015} (\bibinfo {year} {2019})},\ \Eprint
  {http://arxiv.org/abs/1907.05551} {arXiv:1907.05551 [gr-qc]} \BibitemShut
  {NoStop}%
\bibitem [{\citenamefont {Roy}\ and\ \citenamefont
  {Biswas}(2019)}]{Roy:2019hqf}%
  \BibitemOpen
  \bibfield  {author} {\bibinfo {author} {\bibfnamefont {P.}~\bibnamefont
  {Roy}}\ and\ \bibinfo {author} {\bibfnamefont {R.}~\bibnamefont {Biswas}},\
  }\href@noop {} {\  (\bibinfo {year} {2019})},\ \Eprint
  {http://arxiv.org/abs/1907.06719} {arXiv:1907.06719 [gr-qc]} \BibitemShut
  {NoStop}%
\bibitem [{\citenamefont {Pavlovi{\'c}}\ \emph {et~al.}(2019)\citenamefont
  {Pavlovi{\'c}}, \citenamefont {Saveliev},\ and\ \citenamefont
  {Sossich}}]{Pavlovic:2019rim}%
  \BibitemOpen
  \bibfield  {author} {\bibinfo {author} {\bibfnamefont {P.}~\bibnamefont
  {Pavlovi{\'c}}}, \bibinfo {author} {\bibfnamefont {A.}~\bibnamefont
  {Saveliev}}, \ and\ \bibinfo {author} {\bibfnamefont {M.}~\bibnamefont
  {Sossich}},\ }\href {\doibase 10.1103/PhysRevD.100.084033} {\bibfield
  {journal} {\bibinfo  {journal} {Phys. Rev.}\ }\textbf {\bibinfo {volume}
  {D100}},\ \bibinfo {pages} {084033} (\bibinfo {year} {2019})},\ \Eprint
  {http://arxiv.org/abs/1908.01888} {arXiv:1908.01888 [gr-qc]} \BibitemShut
  {NoStop}%
\bibitem [{\citenamefont {Biswas}\ and\ \citenamefont
  {Dutta}(2019)}]{Biswas:2019gia}%
  \BibitemOpen
  \bibfield  {author} {\bibinfo {author} {\bibfnamefont {R.}~\bibnamefont
  {Biswas}}\ and\ \bibinfo {author} {\bibfnamefont {S.}~\bibnamefont {Dutta}},\
  }\href {\doibase 10.1140/epjc/s10052-019-7258-6} {\bibfield  {journal}
  {\bibinfo  {journal} {Eur. Phys. J.}\ }\textbf {\bibinfo {volume} {C79}},\
  \bibinfo {pages} {742} (\bibinfo {year} {2019})},\ \Eprint
  {http://arxiv.org/abs/1908.04268} {arXiv:1908.04268 [gr-qc]} \BibitemShut
  {NoStop}%
\bibitem [{\citenamefont {Wang}\ \emph
  {et~al.}(2019{\natexlab{b}})\citenamefont {Wang}, \citenamefont {Chen},\ and\
  \citenamefont {Jing}}]{Wang:2019skw}%
  \BibitemOpen
  \bibfield  {author} {\bibinfo {author} {\bibfnamefont {M.}~\bibnamefont
  {Wang}}, \bibinfo {author} {\bibfnamefont {S.}~\bibnamefont {Chen}}, \ and\
  \bibinfo {author} {\bibfnamefont {J.}~\bibnamefont {Jing}},\ }\href@noop {}
  {\  (\bibinfo {year} {2019}{\natexlab{b}})},\ \Eprint
  {http://arxiv.org/abs/1908.04527} {arXiv:1908.04527 [gr-qc]} \BibitemShut
  {NoStop}%
\bibitem [{\citenamefont {Nalewajko}\ \emph {et~al.}(2019)\citenamefont
  {Nalewajko}, \citenamefont {Sikora},\ and\ \citenamefont
  {R{\'o}{\.z}a{\'n}ska}}]{Nalewajko:2019mxh}%
  \BibitemOpen
  \bibfield  {author} {\bibinfo {author} {\bibfnamefont {K.}~\bibnamefont
  {Nalewajko}}, \bibinfo {author} {\bibfnamefont {M.}~\bibnamefont {Sikora}}, \
  and\ \bibinfo {author} {\bibfnamefont {A.}~\bibnamefont
  {R{\'o}{\.z}a{\'n}ska}},\ }\href@noop {} {\  (\bibinfo {year} {2019})},\
  \Eprint {http://arxiv.org/abs/1908.10376} {arXiv:1908.10376 [astro-ph.HE]}
  \BibitemShut {NoStop}%
\bibitem [{\citenamefont {Tian}\ and\ \citenamefont
  {Zhu}(2019)}]{Tian:2019yhn}%
  \BibitemOpen
  \bibfield  {author} {\bibinfo {author} {\bibfnamefont {S.~X.}\ \bibnamefont
  {Tian}}\ and\ \bibinfo {author} {\bibfnamefont {Z.-H.}\ \bibnamefont {Zhu}},\
  }\href {\doibase 10.1103/PhysRevD.100.064011} {\bibfield  {journal} {\bibinfo
   {journal} {Phys. Rev.}\ }\textbf {\bibinfo {volume} {D100}},\ \bibinfo
  {pages} {064011} (\bibinfo {year} {2019})},\ \Eprint
  {http://arxiv.org/abs/1908.11794} {arXiv:1908.11794 [gr-qc]} \BibitemShut
  {NoStop}%
\bibitem [{\citenamefont {Cunha}\ \emph {et~al.}(2019)\citenamefont {Cunha},
  \citenamefont {Herdeiro},\ and\ \citenamefont {Radu}}]{Cunha:2019ikd}%
  \BibitemOpen
  \bibfield  {author} {\bibinfo {author} {\bibfnamefont {P.~V.~P.}\
  \bibnamefont {Cunha}}, \bibinfo {author} {\bibfnamefont {C.~A.~R.}\
  \bibnamefont {Herdeiro}}, \ and\ \bibinfo {author} {\bibfnamefont
  {E.}~\bibnamefont {Radu}},\ }\href@noop {} {\  (\bibinfo {year} {2019})},\
  \Eprint {http://arxiv.org/abs/1909.08039} {arXiv:1909.08039 [gr-qc]}
  \BibitemShut {NoStop}%
\bibitem [{\citenamefont {Banerjee}\ \emph
  {et~al.}(2019{\natexlab{b}})\citenamefont {Banerjee}, \citenamefont
  {Chakraborty},\ and\ \citenamefont {SenGupta}}]{Banerjee:2019nnj}%
  \BibitemOpen
  \bibfield  {author} {\bibinfo {author} {\bibfnamefont {I.}~\bibnamefont
  {Banerjee}}, \bibinfo {author} {\bibfnamefont {S.}~\bibnamefont
  {Chakraborty}}, \ and\ \bibinfo {author} {\bibfnamefont {S.}~\bibnamefont
  {SenGupta}},\ }\href@noop {} {\  (\bibinfo {year} {2019}{\natexlab{b}})},\
  \Eprint {http://arxiv.org/abs/1909.09385} {arXiv:1909.09385 [gr-qc]}
  \BibitemShut {NoStop}%
\bibitem [{\citenamefont {Shaikh}\ and\ \citenamefont
  {Joshi}(2019)}]{Shaikh:2019hbm}%
  \BibitemOpen
  \bibfield  {author} {\bibinfo {author} {\bibfnamefont {R.}~\bibnamefont
  {Shaikh}}\ and\ \bibinfo {author} {\bibfnamefont {P.~S.}\ \bibnamefont
  {Joshi}},\ }\href {\doibase 10.1088/1475-7516/2019/10/064} {\bibfield
  {journal} {\bibinfo  {journal} {JCAP}\ }\textbf {\bibinfo {volume} {1910}},\
  \bibinfo {pages} {064} (\bibinfo {year} {2019})},\ \Eprint
  {http://arxiv.org/abs/1909.10322} {arXiv:1909.10322 [gr-qc]} \BibitemShut
  {NoStop}%
\bibitem [{\citenamefont {Vrba}\ \emph {et~al.}(2019)\citenamefont {Vrba},
  \citenamefont {Abdujabbarov}, \citenamefont {Tursunov}, \citenamefont
  {Ahmedov},\ and\ \citenamefont {Stuchl{\'\i}k}}]{Vrba:2019vqh}%
  \BibitemOpen
  \bibfield  {author} {\bibinfo {author} {\bibfnamefont {J.}~\bibnamefont
  {Vrba}}, \bibinfo {author} {\bibfnamefont {A.}~\bibnamefont {Abdujabbarov}},
  \bibinfo {author} {\bibfnamefont {A.}~\bibnamefont {Tursunov}}, \bibinfo
  {author} {\bibfnamefont {B.}~\bibnamefont {Ahmedov}}, \ and\ \bibinfo
  {author} {\bibfnamefont {Z.}~\bibnamefont {Stuchl{\'\i}k}},\ }\href {\doibase
  10.1140/epjc/s10052-019-7286-2} {\bibfield  {journal} {\bibinfo  {journal}
  {Eur. Phys. J.}\ }\textbf {\bibinfo {volume} {C79}},\ \bibinfo {pages} {778}
  (\bibinfo {year} {2019})},\ \Eprint {http://arxiv.org/abs/1909.12026}
  {arXiv:1909.12026 [gr-qc]} \BibitemShut {NoStop}%
\bibitem [{\citenamefont {Kumar}\ \emph
  {et~al.}(2019{\natexlab{a}})\citenamefont {Kumar}, \citenamefont {Ghosh},\
  and\ \citenamefont {Wang}}]{Kumar:2019pjp}%
  \BibitemOpen
  \bibfield  {author} {\bibinfo {author} {\bibfnamefont {R.}~\bibnamefont
  {Kumar}}, \bibinfo {author} {\bibfnamefont {S.~G.}\ \bibnamefont {Ghosh}}, \
  and\ \bibinfo {author} {\bibfnamefont {A.}~\bibnamefont {Wang}},\ }\href
  {\doibase 10.1103/PhysRevD.100.124024} {\bibfield  {journal} {\bibinfo
  {journal} {Phys. Rev.}\ }\textbf {\bibinfo {volume} {D100}},\ \bibinfo
  {pages} {124024} (\bibinfo {year} {2019}{\natexlab{a}})},\ \Eprint
  {http://arxiv.org/abs/1912.05154} {arXiv:1912.05154 [gr-qc]} \BibitemShut
  {NoStop}%
\bibitem [{\citenamefont {Allahyari}\ \emph {et~al.}(2020)\citenamefont
  {Allahyari}, \citenamefont {Khodadi}, \citenamefont {Vagnozzi},\ and\
  \citenamefont {Mota}}]{Allahyari:2019jqz}%
  \BibitemOpen
  \bibfield  {author} {\bibinfo {author} {\bibfnamefont {A.}~\bibnamefont
  {Allahyari}}, \bibinfo {author} {\bibfnamefont {M.}~\bibnamefont {Khodadi}},
  \bibinfo {author} {\bibfnamefont {S.}~\bibnamefont {Vagnozzi}}, \ and\
  \bibinfo {author} {\bibfnamefont {D.~F.}\ \bibnamefont {Mota}},\ }\href
  {\doibase 10.1088/1475-7516/2020/02/003} {\bibfield  {journal} {\bibinfo
  {journal} {JCAP}\ }\textbf {\bibinfo {volume} {2002}},\ \bibinfo {pages}
  {003} (\bibinfo {year} {2020})},\ \Eprint {http://arxiv.org/abs/1912.08231}
  {arXiv:1912.08231 [gr-qc]} \BibitemShut {NoStop}%
\bibitem [{\citenamefont {Li}\ \emph {et~al.}(2019)\citenamefont {Li},
  \citenamefont {Yan}, \citenamefont {Xue}, \citenamefont {Ren}, \citenamefont
  {Cai}, \citenamefont {Easson}, \citenamefont {Yuan},\ and\ \citenamefont
  {Zhao}}]{Li:2019lsm}%
  \BibitemOpen
  \bibfield  {author} {\bibinfo {author} {\bibfnamefont {C.}~\bibnamefont
  {Li}}, \bibinfo {author} {\bibfnamefont {S.-F.}\ \bibnamefont {Yan}},
  \bibinfo {author} {\bibfnamefont {L.}~\bibnamefont {Xue}}, \bibinfo {author}
  {\bibfnamefont {X.}~\bibnamefont {Ren}}, \bibinfo {author} {\bibfnamefont
  {Y.-F.}\ \bibnamefont {Cai}}, \bibinfo {author} {\bibfnamefont {D.~A.}\
  \bibnamefont {Easson}}, \bibinfo {author} {\bibfnamefont {Y.-F.}\
  \bibnamefont {Yuan}}, \ and\ \bibinfo {author} {\bibfnamefont
  {H.}~\bibnamefont {Zhao}},\ }\href@noop {} {\  (\bibinfo {year} {2019})},\
  \Eprint {http://arxiv.org/abs/1912.12629} {arXiv:1912.12629 [astro-ph.CO]}
  \BibitemShut {NoStop}%
\bibitem [{\citenamefont {Jusufi}(2019)}]{Jusufi:2019ltj}%
  \BibitemOpen
  \bibfield  {author} {\bibinfo {author} {\bibfnamefont {K.}~\bibnamefont
  {Jusufi}},\ }\href@noop {} {\  (\bibinfo {year} {2019})},\ \Eprint
  {http://arxiv.org/abs/1912.13320} {arXiv:1912.13320 [gr-qc]} \BibitemShut
  {NoStop}%
\bibitem [{\citenamefont {Rummel}\ and\ \citenamefont
  {Burgess}(2019)}]{Rummel:2019ads}%
  \BibitemOpen
  \bibfield  {author} {\bibinfo {author} {\bibfnamefont {M.}~\bibnamefont
  {Rummel}}\ and\ \bibinfo {author} {\bibfnamefont {C.~P.}\ \bibnamefont
  {Burgess}},\ }\href@noop {} {\  (\bibinfo {year} {2019})},\ \Eprint
  {http://arxiv.org/abs/2001.00041} {arXiv:2001.00041 [gr-qc]} \BibitemShut
  {NoStop}%
\bibitem [{\citenamefont {Kumar}\ \emph {et~al.}(2020)\citenamefont {Kumar},
  \citenamefont {Ghosh},\ and\ \citenamefont {Wang}}]{Kumar:2020hgm}%
  \BibitemOpen
  \bibfield  {author} {\bibinfo {author} {\bibfnamefont {R.}~\bibnamefont
  {Kumar}}, \bibinfo {author} {\bibfnamefont {S.~G.}\ \bibnamefont {Ghosh}}, \
  and\ \bibinfo {author} {\bibfnamefont {A.}~\bibnamefont {Wang}},\ }\href@noop
  {} {\  (\bibinfo {year} {2020})},\ \Eprint {http://arxiv.org/abs/2001.00460}
  {arXiv:2001.00460 [gr-qc]} \BibitemShut {NoStop}%
\bibitem [{\citenamefont {Einstein}\ and\ \citenamefont
  {Straus}(1945)}]{Einstein:1945id}%
  \BibitemOpen
  \bibfield  {author} {\bibinfo {author} {\bibfnamefont {A.}~\bibnamefont
  {Einstein}}\ and\ \bibinfo {author} {\bibfnamefont {E.~G.}\ \bibnamefont
  {Straus}},\ }\href {\doibase 10.1103/RevModPhys.17.120} {\bibfield  {journal}
  {\bibinfo  {journal} {Rev. Mod. Phys.}\ }\textbf {\bibinfo {volume} {17}},\
  \bibinfo {pages} {120} (\bibinfo {year} {1945})}\BibitemShut {NoStop}%
\bibitem [{\citenamefont {Einstein}\ and\ \citenamefont
  {Straus}(1946)}]{Einstein:1946zz}%
  \BibitemOpen
  \bibfield  {author} {\bibinfo {author} {\bibfnamefont {A.}~\bibnamefont
  {Einstein}}\ and\ \bibinfo {author} {\bibfnamefont {E.~G.}\ \bibnamefont
  {Straus}},\ }\href {\doibase 10.1103/RevModPhys.18.148} {\bibfield  {journal}
  {\bibinfo  {journal} {Rev. Mod. Phys.}\ }\textbf {\bibinfo {volume} {18}},\
  \bibinfo {pages} {148} (\bibinfo {year} {1946})}\BibitemShut {NoStop}%
\bibitem [{\citenamefont {McVittie}(1933)}]{McVittie:1933zz}%
  \BibitemOpen
  \bibfield  {author} {\bibinfo {author} {\bibfnamefont {G.~C.}\ \bibnamefont
  {McVittie}},\ }\href {\doibase 10.1093/mnras/93.5.325} {\bibfield  {journal}
  {\bibinfo  {journal} {Mon. Not. Roy. Astron. Soc.}\ }\textbf {\bibinfo
  {volume} {93}},\ \bibinfo {pages} {325} (\bibinfo {year} {1933})}\BibitemShut
  {NoStop}%
\bibitem [{\citenamefont {Nolan}(1998)}]{Nolan:1998xs}%
  \BibitemOpen
  \bibfield  {author} {\bibinfo {author} {\bibfnamefont {B.~C.}\ \bibnamefont
  {Nolan}},\ }\href {\doibase 10.1103/PhysRevD.58.064006} {\bibfield  {journal}
  {\bibinfo  {journal} {Phys. Rev.}\ }\textbf {\bibinfo {volume} {D58}},\
  \bibinfo {pages} {064006} (\bibinfo {year} {1998})},\ \Eprint
  {http://arxiv.org/abs/gr-qc/9805041} {arXiv:gr-qc/9805041 [gr-qc]}
  \BibitemShut {NoStop}%
\bibitem [{\citenamefont {Nolan}(1999{\natexlab{a}})}]{Nolan:1999kk}%
  \BibitemOpen
  \bibfield  {author} {\bibinfo {author} {\bibfnamefont {B.~C.}\ \bibnamefont
  {Nolan}},\ }\href {\doibase 10.1088/0264-9381/16/4/012} {\bibfield  {journal}
  {\bibinfo  {journal} {Class. Quant. Grav.}\ }\textbf {\bibinfo {volume}
  {16}},\ \bibinfo {pages} {1227} (\bibinfo {year}
  {1999}{\natexlab{a}})}\BibitemShut {NoStop}%
\bibitem [{\citenamefont {Nolan}(1999{\natexlab{b}})}]{Nolan:1999wf}%
  \BibitemOpen
  \bibfield  {author} {\bibinfo {author} {\bibfnamefont {B.~C.}\ \bibnamefont
  {Nolan}},\ }\href {\doibase 10.1088/0264-9381/16/10/310} {\bibfield
  {journal} {\bibinfo  {journal} {Class. Quant. Grav.}\ }\textbf {\bibinfo
  {volume} {16}},\ \bibinfo {pages} {3183} (\bibinfo {year}
  {1999}{\natexlab{b}})},\ \Eprint {http://arxiv.org/abs/gr-qc/9907018}
  {arXiv:gr-qc/9907018 [gr-qc]} \BibitemShut {NoStop}%
\bibitem [{\citenamefont {Carrera}\ and\ \citenamefont
  {Giulini}(2008)}]{Carrera:2008pi}%
  \BibitemOpen
  \bibfield  {author} {\bibinfo {author} {\bibfnamefont {M.}~\bibnamefont
  {Carrera}}\ and\ \bibinfo {author} {\bibfnamefont {D.}~\bibnamefont
  {Giulini}},\ }\href {\doibase 10.1103/RevModPhys.82.169} {\  (\bibinfo {year}
  {2008}),\ 10.1103/RevModPhys.82.169},\ \bibinfo {note} {[Rev. Mod.
  Phys.82,169(2010)]},\ \Eprint {http://arxiv.org/abs/0810.2712}
  {arXiv:0810.2712 [gr-qc]} \BibitemShut {NoStop}%
\bibitem [{\citenamefont {Gibbons}\ and\ \citenamefont
  {Maeda}(2010)}]{Gibbons:2009dr}%
  \BibitemOpen
  \bibfield  {author} {\bibinfo {author} {\bibfnamefont {G.~W.}\ \bibnamefont
  {Gibbons}}\ and\ \bibinfo {author} {\bibfnamefont {K.-i.}\ \bibnamefont
  {Maeda}},\ }\href {\doibase 10.1103/PhysRevLett.104.131101} {\bibfield
  {journal} {\bibinfo  {journal} {Phys. Rev. Lett.}\ }\textbf {\bibinfo
  {volume} {104}},\ \bibinfo {pages} {131101} (\bibinfo {year} {2010})},\
  \Eprint {http://arxiv.org/abs/0912.2809} {arXiv:0912.2809 [gr-qc]}
  \BibitemShut {NoStop}%
\bibitem [{\citenamefont {Nandra}\ \emph
  {et~al.}(2012{\natexlab{a}})\citenamefont {Nandra}, \citenamefont {Lasenby},\
  and\ \citenamefont {Hobson}}]{Nandra:2011ug}%
  \BibitemOpen
  \bibfield  {author} {\bibinfo {author} {\bibfnamefont {R.}~\bibnamefont
  {Nandra}}, \bibinfo {author} {\bibfnamefont {A.~N.}\ \bibnamefont {Lasenby}},
  \ and\ \bibinfo {author} {\bibfnamefont {M.~P.}\ \bibnamefont {Hobson}},\
  }\href {\doibase 10.1111/j.1365-2966.2012.20618.x} {\bibfield  {journal}
  {\bibinfo  {journal} {Mon. Not. Roy. Astron. Soc.}\ }\textbf {\bibinfo
  {volume} {422}},\ \bibinfo {pages} {2931} (\bibinfo {year}
  {2012}{\natexlab{a}})},\ \Eprint {http://arxiv.org/abs/1104.4447}
  {arXiv:1104.4447 [gr-qc]} \BibitemShut {NoStop}%
\bibitem [{\citenamefont {Nandra}\ \emph
  {et~al.}(2012{\natexlab{b}})\citenamefont {Nandra}, \citenamefont {Lasenby},\
  and\ \citenamefont {Hobson}}]{Nandra:2011ui}%
  \BibitemOpen
  \bibfield  {author} {\bibinfo {author} {\bibfnamefont {R.}~\bibnamefont
  {Nandra}}, \bibinfo {author} {\bibfnamefont {A.~N.}\ \bibnamefont {Lasenby}},
  \ and\ \bibinfo {author} {\bibfnamefont {M.~P.}\ \bibnamefont {Hobson}},\
  }\href {\doibase 10.1111/j.1365-2966.2012.20617.x} {\bibfield  {journal}
  {\bibinfo  {journal} {Mon. Not. Roy. Astron. Soc.}\ }\textbf {\bibinfo
  {volume} {422}},\ \bibinfo {pages} {2945} (\bibinfo {year}
  {2012}{\natexlab{b}})},\ \Eprint {http://arxiv.org/abs/1104.4458}
  {arXiv:1104.4458 [gr-qc]} \BibitemShut {NoStop}%
\bibitem [{\citenamefont {Stuchlik}\ and\ \citenamefont
  {Hledik}(1999)}]{Stuchlik:1999qk}%
  \BibitemOpen
  \bibfield  {author} {\bibinfo {author} {\bibfnamefont {Z.}~\bibnamefont
  {Stuchlik}}\ and\ \bibinfo {author} {\bibfnamefont {S.}~\bibnamefont
  {Hledik}},\ }\href {\doibase 10.1103/PhysRevD.60.044006} {\bibfield
  {journal} {\bibinfo  {journal} {Phys. Rev.}\ }\textbf {\bibinfo {volume}
  {D60}},\ \bibinfo {pages} {044006} (\bibinfo {year} {1999})}\BibitemShut
  {NoStop}%
\bibitem [{\citenamefont {Bakala}\ \emph {et~al.}(2007)\citenamefont {Bakala},
  \citenamefont {Cermak}, \citenamefont {Hledik}, \citenamefont {Stuchlik},\
  and\ \citenamefont {Truparova}}]{Bakala:2007pw}%
  \BibitemOpen
  \bibfield  {author} {\bibinfo {author} {\bibfnamefont {P.}~\bibnamefont
  {Bakala}}, \bibinfo {author} {\bibfnamefont {P.}~\bibnamefont {Cermak}},
  \bibinfo {author} {\bibfnamefont {S.}~\bibnamefont {Hledik}}, \bibinfo
  {author} {\bibfnamefont {Z.}~\bibnamefont {Stuchlik}}, \ and\ \bibinfo
  {author} {\bibfnamefont {K.}~\bibnamefont {Truparova}},\ }\href {\doibase
  10.2478/s11534-007-0033-6} {\bibfield  {journal} {\bibinfo  {journal}
  {Central Eur. J. Phys.}\ }\textbf {\bibinfo {volume} {5}},\ \bibinfo {pages}
  {599} (\bibinfo {year} {2007})},\ \Eprint {http://arxiv.org/abs/0709.4274}
  {arXiv:0709.4274 [astro-ph]} \BibitemShut {NoStop}%
\bibitem [{\citenamefont {Stuchl{\'\i}k}\ \emph {et~al.}(2018)\citenamefont
  {Stuchl{\'\i}k}, \citenamefont {Charbul{\'a}k},\ and\ \citenamefont
  {Schee}}]{Stuchlik:2018qyz}%
  \BibitemOpen
  \bibfield  {author} {\bibinfo {author} {\bibfnamefont {Z.}~\bibnamefont
  {Stuchl{\'\i}k}}, \bibinfo {author} {\bibfnamefont {D.}~\bibnamefont
  {Charbul{\'a}k}}, \ and\ \bibinfo {author} {\bibfnamefont {J.}~\bibnamefont
  {Schee}},\ }\href {\doibase 10.1140/epjc/s10052-018-5578-6} {\bibfield
  {journal} {\bibinfo  {journal} {Eur. Phys. J.}\ }\textbf {\bibinfo {volume}
  {C78}},\ \bibinfo {pages} {180} (\bibinfo {year} {2018})},\ \Eprint
  {http://arxiv.org/abs/1811.00072} {arXiv:1811.00072 [gr-qc]} \BibitemShut
  {NoStop}%
\bibitem [{\citenamefont {Perlick}\ \emph {et~al.}(2018)\citenamefont
  {Perlick}, \citenamefont {Tsupko},\ and\ \citenamefont
  {Bisnovatyi-Kogan}}]{Perlick:2018iye}%
  \BibitemOpen
  \bibfield  {author} {\bibinfo {author} {\bibfnamefont {V.}~\bibnamefont
  {Perlick}}, \bibinfo {author} {\bibfnamefont {O.~{\relax Yu}.}\ \bibnamefont
  {Tsupko}}, \ and\ \bibinfo {author} {\bibfnamefont {G.~S.}\ \bibnamefont
  {Bisnovatyi-Kogan}},\ }\href {\doibase 10.1103/PhysRevD.97.104062} {\bibfield
   {journal} {\bibinfo  {journal} {Phys. Rev.}\ }\textbf {\bibinfo {volume}
  {D97}},\ \bibinfo {pages} {104062} (\bibinfo {year} {2018})},\ \Eprint
  {http://arxiv.org/abs/1804.04898} {arXiv:1804.04898 [gr-qc]} \BibitemShut
  {NoStop}%
\bibitem [{\citenamefont {Bisnovatyi-Kogan}\ and\ \citenamefont
  {Tsupko}(2018)}]{Bisnovatyi-Kogan:2018vxl}%
  \BibitemOpen
  \bibfield  {author} {\bibinfo {author} {\bibfnamefont {G.~S.}\ \bibnamefont
  {Bisnovatyi-Kogan}}\ and\ \bibinfo {author} {\bibfnamefont {O.~{\relax Yu}.}\
  \bibnamefont {Tsupko}},\ }\href {\doibase 10.1103/PhysRevD.98.084020}
  {\bibfield  {journal} {\bibinfo  {journal} {Phys. Rev.}\ }\textbf {\bibinfo
  {volume} {D98}},\ \bibinfo {pages} {084020} (\bibinfo {year} {2018})},\
  \Eprint {http://arxiv.org/abs/1805.03311} {arXiv:1805.03311 [gr-qc]}
  \BibitemShut {NoStop}%
\bibitem [{\citenamefont {Bisnovatyi-Kogan}\ \emph {et~al.}(2019)\citenamefont
  {Bisnovatyi-Kogan}, \citenamefont {Tsupko},\ and\ \citenamefont
  {Perlick}}]{Bisnovatyi-Kogan:2019wdd}%
  \BibitemOpen
  \bibfield  {author} {\bibinfo {author} {\bibfnamefont {G.~S.}\ \bibnamefont
  {Bisnovatyi-Kogan}}, \bibinfo {author} {\bibfnamefont {O.~{\relax Yu}.}\
  \bibnamefont {Tsupko}}, \ and\ \bibinfo {author} {\bibfnamefont
  {V.}~\bibnamefont {Perlick}},\ }in\ \href@noop {} {\emph {\bibinfo
  {booktitle} {{13th Frascati Workshop on Multifrequency Behaviour of High
  Energy Cosmic Sources (MULTIF2019) Palermo, Italy, June 3-8, 2019}}}}\
  (\bibinfo {year} {2019})\ \Eprint {http://arxiv.org/abs/1910.10514}
  {arXiv:1910.10514 [gr-qc]} \BibitemShut {NoStop}%
\bibitem [{\citenamefont {Shemmer}\ \emph {et~al.}(2004)\citenamefont
  {Shemmer}, \citenamefont {Netzer}, \citenamefont {Maiolino}, \citenamefont
  {Oliva}, \citenamefont {Croom}, \citenamefont {Corbett},\ and\ \citenamefont
  {di~Fabrizio}}]{Shemmer:2004ph}%
  \BibitemOpen
  \bibfield  {author} {\bibinfo {author} {\bibfnamefont {O.}~\bibnamefont
  {Shemmer}}, \bibinfo {author} {\bibfnamefont {H.}~\bibnamefont {Netzer}},
  \bibinfo {author} {\bibfnamefont {R.}~\bibnamefont {Maiolino}}, \bibinfo
  {author} {\bibfnamefont {E.}~\bibnamefont {Oliva}}, \bibinfo {author}
  {\bibfnamefont {S.}~\bibnamefont {Croom}}, \bibinfo {author} {\bibfnamefont
  {E.}~\bibnamefont {Corbett}}, \ and\ \bibinfo {author} {\bibfnamefont
  {L.}~\bibnamefont {di~Fabrizio}},\ }\href {\doibase 10.1086/423607}
  {\bibfield  {journal} {\bibinfo  {journal} {Astrophys. J.}\ }\textbf
  {\bibinfo {volume} {614}},\ \bibinfo {pages} {547} (\bibinfo {year}
  {2004})},\ \Eprint {http://arxiv.org/abs/astro-ph/0406559}
  {arXiv:astro-ph/0406559 [astro-ph]} \BibitemShut {NoStop}%
\bibitem [{\citenamefont {{Mehrgan}}\ \emph {et~al.}(2019)\citenamefont
  {{Mehrgan}}, \citenamefont {{Thomas}}, \citenamefont {{Saglia}},
  \citenamefont {{Mazzalay}}, \citenamefont {{Erwin}}, \citenamefont
  {{Bender}}, \citenamefont {{Kluge}},\ and\ \citenamefont
  {{Fabricius}}}]{2019ApJ...887..195M}%
  \BibitemOpen
  \bibfield  {author} {\bibinfo {author} {\bibfnamefont {K.}~\bibnamefont
  {{Mehrgan}}}, \bibinfo {author} {\bibfnamefont {J.}~\bibnamefont {{Thomas}}},
  \bibinfo {author} {\bibfnamefont {R.}~\bibnamefont {{Saglia}}}, \bibinfo
  {author} {\bibfnamefont {X.}~\bibnamefont {{Mazzalay}}}, \bibinfo {author}
  {\bibfnamefont {P.}~\bibnamefont {{Erwin}}}, \bibinfo {author} {\bibfnamefont
  {R.}~\bibnamefont {{Bender}}}, \bibinfo {author} {\bibfnamefont
  {M.}~\bibnamefont {{Kluge}}}, \ and\ \bibinfo {author} {\bibfnamefont
  {M.}~\bibnamefont {{Fabricius}}},\ }\href {\doibase 10.3847/1538-4357/ab5856}
  {\bibfield  {journal} {\bibinfo  {journal} {Astrophys. J.}\ }\textbf
  {\bibinfo {volume} {887}},\ \bibinfo {eid} {195} (\bibinfo {year} {2019})},\
  \Eprint {http://arxiv.org/abs/1907.10608} {arXiv:1907.10608 [astro-ph.GA]}
  \BibitemShut {NoStop}%
\bibitem [{\citenamefont {{Dullo}}\ \emph {et~al.}(2017)\citenamefont
  {{Dullo}}, \citenamefont {{Graham}},\ and\ \citenamefont
  {{Knapen}}}]{2017MNRAS.471.2321D}%
  \BibitemOpen
  \bibfield  {author} {\bibinfo {author} {\bibfnamefont {B.~T.}\ \bibnamefont
  {{Dullo}}}, \bibinfo {author} {\bibfnamefont {A.~W.}\ \bibnamefont
  {{Graham}}}, \ and\ \bibinfo {author} {\bibfnamefont {J.~H.}\ \bibnamefont
  {{Knapen}}},\ }\href {\doibase 10.1093/mnras/stx1635} {\bibfield  {journal}
  {\bibinfo  {journal} {Mon. Not. Roy. Astron. Soc.}\ }\textbf {\bibinfo
  {volume} {471}},\ \bibinfo {pages} {2321} (\bibinfo {year} {2017})},\ \Eprint
  {http://arxiv.org/abs/1707.02277} {arXiv:1707.02277 [astro-ph.GA]}
  \BibitemShut {NoStop}%
\bibitem [{\citenamefont {{Ghisellini}}\ \emph {et~al.}(2009)\citenamefont
  {{Ghisellini}}, \citenamefont {{Foschini}}, \citenamefont {{Volonteri}},
  \citenamefont {{Ghirland a}}, \citenamefont {{Haardt}}, \citenamefont
  {{Burlon}},\ and\ \citenamefont {{Tavecchio}}}]{2009MNRAS.399L..24G}%
  \BibitemOpen
  \bibfield  {author} {\bibinfo {author} {\bibfnamefont {G.}~\bibnamefont
  {{Ghisellini}}}, \bibinfo {author} {\bibfnamefont {L.}~\bibnamefont
  {{Foschini}}}, \bibinfo {author} {\bibfnamefont {M.}~\bibnamefont
  {{Volonteri}}}, \bibinfo {author} {\bibfnamefont {G.}~\bibnamefont {{Ghirland
  a}}}, \bibinfo {author} {\bibfnamefont {F.}~\bibnamefont {{Haardt}}},
  \bibinfo {author} {\bibfnamefont {D.}~\bibnamefont {{Burlon}}}, \ and\
  \bibinfo {author} {\bibfnamefont {F.}~\bibnamefont {{Tavecchio}}},\ }\href
  {\doibase 10.1111/j.1745-3933.2009.00716.x} {\bibfield  {journal} {\bibinfo
  {journal} {Mon. Not. Roy. Astron. Soc.}\ }\textbf {\bibinfo {volume} {399}},\
  \bibinfo {pages} {L24} (\bibinfo {year} {2009})},\ \Eprint
  {http://arxiv.org/abs/0906.0575} {arXiv:0906.0575 [astro-ph.CO]} \BibitemShut
  {NoStop}%
\bibitem [{\citenamefont {{Ghisellini}}\ \emph {et~al.}(2010)\citenamefont
  {{Ghisellini}}, \citenamefont {{Della Ceca}}, \citenamefont {{Volonteri}},
  \citenamefont {{Ghirland a}}, \citenamefont {{Tavecchio}}, \citenamefont
  {{Foschini}}, \citenamefont {{Tagliaferri}}, \citenamefont {{Haardt}},
  \citenamefont {{Pareschi}},\ and\ \citenamefont
  {{Grindlay}}}]{2010MNRAS.405..387G}%
  \BibitemOpen
  \bibfield  {author} {\bibinfo {author} {\bibfnamefont {G.}~\bibnamefont
  {{Ghisellini}}}, \bibinfo {author} {\bibfnamefont {R.}~\bibnamefont {{Della
  Ceca}}}, \bibinfo {author} {\bibfnamefont {M.}~\bibnamefont {{Volonteri}}},
  \bibinfo {author} {\bibfnamefont {G.}~\bibnamefont {{Ghirland a}}}, \bibinfo
  {author} {\bibfnamefont {F.}~\bibnamefont {{Tavecchio}}}, \bibinfo {author}
  {\bibfnamefont {L.}~\bibnamefont {{Foschini}}}, \bibinfo {author}
  {\bibfnamefont {G.}~\bibnamefont {{Tagliaferri}}}, \bibinfo {author}
  {\bibfnamefont {F.}~\bibnamefont {{Haardt}}}, \bibinfo {author}
  {\bibfnamefont {G.}~\bibnamefont {{Pareschi}}}, \ and\ \bibinfo {author}
  {\bibfnamefont {J.}~\bibnamefont {{Grindlay}}},\ }\href {\doibase
  10.1111/j.1365-2966.2010.16449.x} {\bibfield  {journal} {\bibinfo  {journal}
  {Mon. Not. Roy. Astron. Soc.}\ }\textbf {\bibinfo {volume} {405}},\ \bibinfo
  {pages} {387} (\bibinfo {year} {2010})},\ \Eprint
  {http://arxiv.org/abs/0912.0001} {arXiv:0912.0001 [astro-ph.HE]} \BibitemShut
  {NoStop}%
\bibitem [{\citenamefont {Sprenger}\ \emph {et~al.}(2019)\citenamefont
  {Sprenger}, \citenamefont {Archidiacono}, \citenamefont {Brinckmann},
  \citenamefont {Clesse},\ and\ \citenamefont
  {Lesgourgues}}]{Sprenger:2018tdb}%
  \BibitemOpen
  \bibfield  {author} {\bibinfo {author} {\bibfnamefont {T.}~\bibnamefont
  {Sprenger}}, \bibinfo {author} {\bibfnamefont {M.}~\bibnamefont
  {Archidiacono}}, \bibinfo {author} {\bibfnamefont {T.}~\bibnamefont
  {Brinckmann}}, \bibinfo {author} {\bibfnamefont {S.}~\bibnamefont {Clesse}},
  \ and\ \bibinfo {author} {\bibfnamefont {J.}~\bibnamefont {Lesgourgues}},\
  }\href {\doibase 10.1088/1475-7516/2019/02/047} {\bibfield  {journal}
  {\bibinfo  {journal} {JCAP}\ }\textbf {\bibinfo {volume} {1902}},\ \bibinfo
  {pages} {047} (\bibinfo {year} {2019})},\ \Eprint
  {http://arxiv.org/abs/1801.08331} {arXiv:1801.08331 [astro-ph.CO]}
  \BibitemShut {NoStop}%
\bibitem [{\citenamefont {Brinckmann}\ \emph {et~al.}(2019)\citenamefont
  {Brinckmann}, \citenamefont {Hooper}, \citenamefont {Archidiacono},
  \citenamefont {Lesgourgues},\ and\ \citenamefont
  {Sprenger}}]{Brinckmann:2018owf}%
  \BibitemOpen
  \bibfield  {author} {\bibinfo {author} {\bibfnamefont {T.}~\bibnamefont
  {Brinckmann}}, \bibinfo {author} {\bibfnamefont {D.~C.}\ \bibnamefont
  {Hooper}}, \bibinfo {author} {\bibfnamefont {M.}~\bibnamefont
  {Archidiacono}}, \bibinfo {author} {\bibfnamefont {J.}~\bibnamefont
  {Lesgourgues}}, \ and\ \bibinfo {author} {\bibfnamefont {T.}~\bibnamefont
  {Sprenger}},\ }\href {\doibase 10.1088/1475-7516/2019/01/059} {\bibfield
  {journal} {\bibinfo  {journal} {JCAP}\ }\textbf {\bibinfo {volume} {1901}},\
  \bibinfo {pages} {059} (\bibinfo {year} {2019})},\ \Eprint
  {http://arxiv.org/abs/1808.05955} {arXiv:1808.05955 [astro-ph.CO]}
  \BibitemShut {NoStop}%
\bibitem [{\citenamefont {Muñoz}\ \emph {et~al.}(2019)\citenamefont {Muñoz},
  \citenamefont {Dvorkin},\ and\ \citenamefont {Cyr-Racine}}]{Munoz:2019hjh}%
  \BibitemOpen
  \bibfield  {author} {\bibinfo {author} {\bibfnamefont {J.~B.}\ \bibnamefont
  {Muñoz}}, \bibinfo {author} {\bibfnamefont {C.}~\bibnamefont {Dvorkin}}, \
  and\ \bibinfo {author} {\bibfnamefont {F.-Y.}\ \bibnamefont {Cyr-Racine}},\
  }\href@noop {} {\  (\bibinfo {year} {2019})},\ \Eprint
  {http://arxiv.org/abs/1911.11144} {arXiv:1911.11144 [astro-ph.CO]}
  \BibitemShut {NoStop}%
\bibitem [{\citenamefont {Gebhardt}\ \emph {et~al.}(2011)\citenamefont
  {Gebhardt}, \citenamefont {Adams}, \citenamefont {Richstone}, \citenamefont
  {Lauer}, \citenamefont {Faber}, \citenamefont {Gultekin}, \citenamefont
  {Murphy},\ and\ \citenamefont {Tremaine}}]{Gebhardt:2011yw}%
  \BibitemOpen
  \bibfield  {author} {\bibinfo {author} {\bibfnamefont {K.}~\bibnamefont
  {Gebhardt}}, \bibinfo {author} {\bibfnamefont {J.}~\bibnamefont {Adams}},
  \bibinfo {author} {\bibfnamefont {D.}~\bibnamefont {Richstone}}, \bibinfo
  {author} {\bibfnamefont {T.~R.}\ \bibnamefont {Lauer}}, \bibinfo {author}
  {\bibfnamefont {S.~M.}\ \bibnamefont {Faber}}, \bibinfo {author}
  {\bibfnamefont {K.}~\bibnamefont {Gultekin}}, \bibinfo {author}
  {\bibfnamefont {J.}~\bibnamefont {Murphy}}, \ and\ \bibinfo {author}
  {\bibfnamefont {S.}~\bibnamefont {Tremaine}},\ }\href {\doibase
  10.1088/0004-637X/729/2/119} {\bibfield  {journal} {\bibinfo  {journal}
  {Astrophys. J.}\ }\textbf {\bibinfo {volume} {729}},\ \bibinfo {pages} {119}
  (\bibinfo {year} {2011})},\ \Eprint {http://arxiv.org/abs/1101.1954}
  {arXiv:1101.1954 [astro-ph.CO]} \BibitemShut {NoStop}%
\bibitem [{\citenamefont {Walsh}\ \emph {et~al.}(2013)\citenamefont {Walsh},
  \citenamefont {Barth}, \citenamefont {Ho},\ and\ \citenamefont
  {Sarzi}}]{Walsh:2013uua}%
  \BibitemOpen
  \bibfield  {author} {\bibinfo {author} {\bibfnamefont {J.~L.}\ \bibnamefont
  {Walsh}}, \bibinfo {author} {\bibfnamefont {A.~J.}\ \bibnamefont {Barth}},
  \bibinfo {author} {\bibfnamefont {L.~C.}\ \bibnamefont {Ho}}, \ and\ \bibinfo
  {author} {\bibfnamefont {M.}~\bibnamefont {Sarzi}},\ }\href {\doibase
  10.1088/0004-637X/770/2/86} {\bibfield  {journal} {\bibinfo  {journal}
  {Astrophys. J.}\ }\textbf {\bibinfo {volume} {770}},\ \bibinfo {pages} {86}
  (\bibinfo {year} {2013})},\ \Eprint {http://arxiv.org/abs/1304.7273}
  {arXiv:1304.7273 [astro-ph.CO]} \BibitemShut {NoStop}%
\bibitem [{\citenamefont {Jeter}\ \emph {et~al.}(2019)\citenamefont {Jeter},
  \citenamefont {Broderick},\ and\ \citenamefont {McNamara}}]{Jeter:2018eoh}%
  \BibitemOpen
  \bibfield  {author} {\bibinfo {author} {\bibfnamefont {B.}~\bibnamefont
  {Jeter}}, \bibinfo {author} {\bibfnamefont {A.~E.}\ \bibnamefont
  {Broderick}}, \ and\ \bibinfo {author} {\bibfnamefont {B.~R.}\ \bibnamefont
  {McNamara}},\ }\href {\doibase 10.3847/1538-4357/ab3221} {\bibfield
  {journal} {\bibinfo  {journal} {Astrophys. J.}\ }\textbf {\bibinfo {volume}
  {882}},\ \bibinfo {pages} {82} (\bibinfo {year} {2019})},\ \Eprint
  {http://arxiv.org/abs/1810.05238} {arXiv:1810.05238 [astro-ph.HE]}
  \BibitemShut {NoStop}%
\bibitem [{\citenamefont {{Blandford}}\ and\ \citenamefont
  {{McKee}}(1982)}]{1982ApJ...255..419B}%
  \BibitemOpen
  \bibfield  {author} {\bibinfo {author} {\bibfnamefont {R.~D.}\ \bibnamefont
  {{Blandford}}}\ and\ \bibinfo {author} {\bibfnamefont {C.~F.}\ \bibnamefont
  {{McKee}}},\ }\href {\doibase 10.1086/159843} {\bibfield  {journal} {\bibinfo
   {journal} {\apj}\ }\textbf {\bibinfo {volume} {255}},\ \bibinfo {pages}
  {419} (\bibinfo {year} {1982})}\BibitemShut {NoStop}%
\bibitem [{\citenamefont {{Peterson}}(2001)}]{2001sac..conf....3P}%
  \BibitemOpen
  \bibfield  {author} {\bibinfo {author} {\bibfnamefont {B.~M.}\ \bibnamefont
  {{Peterson}}},\ }in\ \href {\doibase 10.1142/9789812811318_0002} {\emph
  {\bibinfo {booktitle} {Advanced Lectures on the Starburst-AGN}}},\ \bibinfo
  {editor} {edited by\ \bibinfo {editor} {\bibfnamefont {I.}~\bibnamefont
  {{Aretxaga}}}, \bibinfo {editor} {\bibfnamefont {D.}~\bibnamefont {{Kunth}}},
  \ and\ \bibinfo {editor} {\bibfnamefont {R.}~\bibnamefont {{M{\'u}jica}}}}\
  (\bibinfo {year} {2001})\ p.~\bibinfo {pages} {3},\ \Eprint
  {http://arxiv.org/abs/astro-ph/0109495} {arXiv:astro-ph/0109495 [astro-ph]}
  \BibitemShut {NoStop}%
\bibitem [{\citenamefont {Denney}\ \emph {et~al.}(2009)\citenamefont {Denney},
  \citenamefont {Peterson}, \citenamefont {Dietrich}, \citenamefont
  {Vestergaard},\ and\ \citenamefont {Bentz}}]{Denney:2008gk}%
  \BibitemOpen
  \bibfield  {author} {\bibinfo {author} {\bibfnamefont {K.~D.}\ \bibnamefont
  {Denney}}, \bibinfo {author} {\bibfnamefont {B.~M.}\ \bibnamefont
  {Peterson}}, \bibinfo {author} {\bibfnamefont {M.}~\bibnamefont {Dietrich}},
  \bibinfo {author} {\bibfnamefont {M.}~\bibnamefont {Vestergaard}}, \ and\
  \bibinfo {author} {\bibfnamefont {M.~C.}\ \bibnamefont {Bentz}},\ }\href
  {\doibase 10.1088/0004-637X/692/1/246} {\bibfield  {journal} {\bibinfo
  {journal} {Astrophys. J.}\ }\textbf {\bibinfo {volume} {692}},\ \bibinfo
  {pages} {246} (\bibinfo {year} {2009})},\ \Eprint
  {http://arxiv.org/abs/0810.3234} {arXiv:0810.3234 [astro-ph]} \BibitemShut
  {NoStop}%
\bibitem [{\citenamefont {Shen}(2013)}]{Shen:2013pea}%
  \BibitemOpen
  \bibfield  {author} {\bibinfo {author} {\bibfnamefont {Y.}~\bibnamefont
  {Shen}},\ }\href@noop {} {\bibfield  {journal} {\bibinfo  {journal} {Bull.
  Astron. Soc. India}\ }\textbf {\bibinfo {volume} {41}},\ \bibinfo {pages}
  {61} (\bibinfo {year} {2013})},\ \Eprint {http://arxiv.org/abs/1302.2643}
  {arXiv:1302.2643 [astro-ph.CO]} \BibitemShut {NoStop}%
\bibitem [{\citenamefont {Campitiello}\ \emph {et~al.}(2019)\citenamefont
  {Campitiello}, \citenamefont {Celotti}, \citenamefont {Ghisellini},\ and\
  \citenamefont {Sbarrato}}]{Campitiello:2019otf}%
  \BibitemOpen
  \bibfield  {author} {\bibinfo {author} {\bibfnamefont {S.}~\bibnamefont
  {Campitiello}}, \bibinfo {author} {\bibfnamefont {A.}~\bibnamefont
  {Celotti}}, \bibinfo {author} {\bibfnamefont {G.}~\bibnamefont {Ghisellini}},
  \ and\ \bibinfo {author} {\bibfnamefont {T.}~\bibnamefont {Sbarrato}},\
  }\href@noop {} {\  (\bibinfo {year} {2019})},\ \Eprint
  {http://arxiv.org/abs/1907.00986} {arXiv:1907.00986 [astro-ph.HE]}
  \BibitemShut {NoStop}%
\bibitem [{\citenamefont {Kardashev}\ \emph {et~al.}(2014)\citenamefont
  {Kardashev} \emph {et~al.}}]{Kardashev:2015xua}%
  \BibitemOpen
  \bibfield  {author} {\bibinfo {author} {\bibfnamefont {N.~S.}\ \bibnamefont
  {Kardashev}} \emph {et~al.},\ }\href {\doibase
  10.3367/UFNe.0184.201412c.1319, 10.3367/UFNr.0184.201412c.1319} {\bibfield
  {journal} {\bibinfo  {journal} {Phys. Usp.}\ }\textbf {\bibinfo {volume}
  {57}},\ \bibinfo {pages} {1199} (\bibinfo {year} {2014})},\ \bibinfo {note}
  {[Usp. Fiz. Nauk184,no.12,1319(2014)]},\ \Eprint
  {http://arxiv.org/abs/1502.06071} {arXiv:1502.06071 [astro-ph.IM]}
  \BibitemShut {NoStop}%
\bibitem [{\citenamefont {{Fish}}\ \emph {et~al.}(2019)\citenamefont {{Fish}},
  \citenamefont {{Shea}},\ and\ \citenamefont
  {{Akiyama}}}]{2019arXiv190309539F}%
  \BibitemOpen
  \bibfield  {author} {\bibinfo {author} {\bibfnamefont {V.~L.}\ \bibnamefont
  {{Fish}}}, \bibinfo {author} {\bibfnamefont {M.}~\bibnamefont {{Shea}}}, \
  and\ \bibinfo {author} {\bibfnamefont {K.}~\bibnamefont {{Akiyama}}},\
  }\href@noop {} {\bibfield  {journal} {\bibinfo  {journal} {arXiv e-prints}\
  ,\ \bibinfo {eid} {arXiv:1903.09539}} (\bibinfo {year} {2019})},\ \Eprint
  {http://arxiv.org/abs/1903.09539} {arXiv:1903.09539 [astro-ph.IM]}
  \BibitemShut {NoStop}%
\bibitem [{\citenamefont {{Palumbo}}\ \emph {et~al.}(2019)\citenamefont
  {{Palumbo}}, \citenamefont {{Doeleman}}, \citenamefont {{Johnson}},
  \citenamefont {{Bouman}},\ and\ \citenamefont
  {{Chael}}}]{2019ApJ...881...62P}%
  \BibitemOpen
  \bibfield  {author} {\bibinfo {author} {\bibfnamefont {D.~C.~M.}\
  \bibnamefont {{Palumbo}}}, \bibinfo {author} {\bibfnamefont {S.~S.}\
  \bibnamefont {{Doeleman}}}, \bibinfo {author} {\bibfnamefont {M.~D.}\
  \bibnamefont {{Johnson}}}, \bibinfo {author} {\bibfnamefont {K.~L.}\
  \bibnamefont {{Bouman}}}, \ and\ \bibinfo {author} {\bibfnamefont {A.~A.}\
  \bibnamefont {{Chael}}},\ }\href {\doibase 10.3847/1538-4357/ab2bed}
  {\bibfield  {journal} {\bibinfo  {journal} {\apj}\ }\textbf {\bibinfo
  {volume} {881}},\ \bibinfo {eid} {62} (\bibinfo {year} {2019})},\ \Eprint
  {http://arxiv.org/abs/1906.08828} {arXiv:1906.08828 [astro-ph.IM]}
  \BibitemShut {NoStop}%
\bibitem [{\citenamefont {Uttley}\ \emph {et~al.}(2019)\citenamefont {Uttley}
  \emph {et~al.}}]{Uttley:2019ngm}%
  \BibitemOpen
  \bibfield  {author} {\bibinfo {author} {\bibfnamefont {P.}~\bibnamefont
  {Uttley}} \emph {et~al.},\ }\href@noop {} {\  (\bibinfo {year} {2019})},\
  \Eprint {http://arxiv.org/abs/1908.03144} {arXiv:1908.03144 [astro-ph.HE]}
  \BibitemShut {NoStop}%
\bibitem [{\citenamefont {Narayan}\ and\ \citenamefont
  {Yi}(1995{\natexlab{a}})}]{Narayan:1994et}%
  \BibitemOpen
  \bibfield  {author} {\bibinfo {author} {\bibfnamefont {R.}~\bibnamefont
  {Narayan}}\ and\ \bibinfo {author} {\bibfnamefont {I.-s.}\ \bibnamefont
  {Yi}},\ }\href {\doibase 10.1086/175599} {\bibfield  {journal} {\bibinfo
  {journal} {Astrophys. J.}\ }\textbf {\bibinfo {volume} {444}},\ \bibinfo
  {pages} {231} (\bibinfo {year} {1995}{\natexlab{a}})},\ \Eprint
  {http://arxiv.org/abs/astro-ph/9411058} {arXiv:astro-ph/9411058 [astro-ph]}
  \BibitemShut {NoStop}%
\bibitem [{\citenamefont {Narayan}\ and\ \citenamefont
  {Yi}(1995{\natexlab{b}})}]{Narayan:1994is}%
  \BibitemOpen
  \bibfield  {author} {\bibinfo {author} {\bibfnamefont {R.}~\bibnamefont
  {Narayan}}\ and\ \bibinfo {author} {\bibfnamefont {I.}~\bibnamefont {Yi}},\
  }\href {\doibase 10.1086/176343} {\bibfield  {journal} {\bibinfo  {journal}
  {Astrophys. J.}\ }\textbf {\bibinfo {volume} {452}},\ \bibinfo {pages} {710}
  (\bibinfo {year} {1995}{\natexlab{b}})},\ \Eprint
  {http://arxiv.org/abs/astro-ph/9411059} {arXiv:astro-ph/9411059 [astro-ph]}
  \BibitemShut {NoStop}%
\bibitem [{\citenamefont {Yuan}\ and\ \citenamefont
  {Narayan}(2014)}]{Yuan:2014gma}%
  \BibitemOpen
  \bibfield  {author} {\bibinfo {author} {\bibfnamefont {F.}~\bibnamefont
  {Yuan}}\ and\ \bibinfo {author} {\bibfnamefont {R.}~\bibnamefont {Narayan}},\
  }\href {\doibase 10.1146/annurev-astro-082812-141003} {\bibfield  {journal}
  {\bibinfo  {journal} {Ann. Rev. Astron. Astrophys.}\ }\textbf {\bibinfo
  {volume} {52}},\ \bibinfo {pages} {529} (\bibinfo {year} {2014})},\ \Eprint
  {http://arxiv.org/abs/1401.0586} {arXiv:1401.0586 [astro-ph.HE]} \BibitemShut
  {NoStop}%
\bibitem [{\citenamefont {Mortlock}\ \emph {et~al.}(2011)\citenamefont
  {Mortlock} \emph {et~al.}}]{Mortlock:2011va}%
  \BibitemOpen
  \bibfield  {author} {\bibinfo {author} {\bibfnamefont {D.~J.}\ \bibnamefont
  {Mortlock}} \emph {et~al.},\ }\href {\doibase 10.1038/nature10159} {\bibfield
   {journal} {\bibinfo  {journal} {Nature}\ }\textbf {\bibinfo {volume}
  {474}},\ \bibinfo {pages} {616} (\bibinfo {year} {2011})},\ \Eprint
  {http://arxiv.org/abs/1106.6088} {arXiv:1106.6088 [astro-ph.CO]} \BibitemShut
  {NoStop}%
\bibitem [{\citenamefont {De~Rosa}\ \emph {et~al.}(2014)\citenamefont {De~Rosa}
  \emph {et~al.}}]{DeRosa:2013iia}%
  \BibitemOpen
  \bibfield  {author} {\bibinfo {author} {\bibfnamefont {G.}~\bibnamefont
  {De~Rosa}} \emph {et~al.},\ }\href {\doibase 10.1088/0004-637X/790/2/145}
  {\bibfield  {journal} {\bibinfo  {journal} {Astrophys. J.}\ }\textbf
  {\bibinfo {volume} {790}},\ \bibinfo {pages} {145} (\bibinfo {year}
  {2014})},\ \Eprint {http://arxiv.org/abs/1311.3260} {arXiv:1311.3260
  [astro-ph.CO]} \BibitemShut {NoStop}%
\bibitem [{\citenamefont {{Wu}}\ \emph {et~al.}(2015)\citenamefont {{Wu}},
  \citenamefont {{Wang}}, \citenamefont {{Fan}}, \citenamefont {{Yi}},
  \citenamefont {{Zuo}}, \citenamefont {{Bian}}, \citenamefont {{Jiang}},
  \citenamefont {{McGreer}}, \citenamefont {{Wang}}, \citenamefont {{Yang}},
  \citenamefont {{Yang}}, \citenamefont {{Thompson}},\ and\ \citenamefont
  {{Beletsky}}}]{2015Natur.518..512W}%
  \BibitemOpen
  \bibfield  {author} {\bibinfo {author} {\bibfnamefont {X.-B.}\ \bibnamefont
  {{Wu}}}, \bibinfo {author} {\bibfnamefont {F.}~\bibnamefont {{Wang}}},
  \bibinfo {author} {\bibfnamefont {X.}~\bibnamefont {{Fan}}}, \bibinfo
  {author} {\bibfnamefont {W.}~\bibnamefont {{Yi}}}, \bibinfo {author}
  {\bibfnamefont {W.}~\bibnamefont {{Zuo}}}, \bibinfo {author} {\bibfnamefont
  {F.}~\bibnamefont {{Bian}}}, \bibinfo {author} {\bibfnamefont
  {L.}~\bibnamefont {{Jiang}}}, \bibinfo {author} {\bibfnamefont {I.~D.}\
  \bibnamefont {{McGreer}}}, \bibinfo {author} {\bibfnamefont {R.}~\bibnamefont
  {{Wang}}}, \bibinfo {author} {\bibfnamefont {J.}~\bibnamefont {{Yang}}},
  \bibinfo {author} {\bibfnamefont {Q.}~\bibnamefont {{Yang}}}, \bibinfo
  {author} {\bibfnamefont {D.}~\bibnamefont {{Thompson}}}, \ and\ \bibinfo
  {author} {\bibfnamefont {Y.}~\bibnamefont {{Beletsky}}},\ }\href {\doibase
  10.1038/nature14241} {\bibfield  {journal} {\bibinfo  {journal} {\nat}\
  }\textbf {\bibinfo {volume} {518}},\ \bibinfo {pages} {512} (\bibinfo {year}
  {2015})},\ \Eprint {http://arxiv.org/abs/1502.07418} {arXiv:1502.07418
  [astro-ph.GA]} \BibitemShut {NoStop}%
\bibitem [{\citenamefont {Fan}\ \emph {et~al.}(2006)\citenamefont {Fan} \emph
  {et~al.}}]{Fan:2005eq}%
  \BibitemOpen
  \bibfield  {author} {\bibinfo {author} {\bibfnamefont {X.-H.}\ \bibnamefont
  {Fan}} \emph {et~al.} (\bibinfo {collaboration} {SDSS}),\ }\href {\doibase
  10.1086/500296} {\bibfield  {journal} {\bibinfo  {journal} {Astron. J.}\
  }\textbf {\bibinfo {volume} {131}},\ \bibinfo {pages} {1203} (\bibinfo {year}
  {2006})},\ \Eprint {http://arxiv.org/abs/astro-ph/0512080}
  {arXiv:astro-ph/0512080 [astro-ph]} \BibitemShut {NoStop}%
\bibitem [{\citenamefont {Volonteri}(2010)}]{Volonteri:2010wz}%
  \BibitemOpen
  \bibfield  {author} {\bibinfo {author} {\bibfnamefont {M.}~\bibnamefont
  {Volonteri}},\ }\href {\doibase 10.1007/s00159-010-0029-x} {\bibfield
  {journal} {\bibinfo  {journal} {Astron. Astrophys. Rev.}\ }\textbf {\bibinfo
  {volume} {18}},\ \bibinfo {pages} {279} (\bibinfo {year} {2010})},\ \Eprint
  {http://arxiv.org/abs/1003.4404} {arXiv:1003.4404 [astro-ph.CO]} \BibitemShut
  {NoStop}%
\bibitem [{\citenamefont {Volonteri}\ and\ \citenamefont
  {Rees}(2005)}]{Volonteri:2005fj}%
  \BibitemOpen
  \bibfield  {author} {\bibinfo {author} {\bibfnamefont {M.}~\bibnamefont
  {Volonteri}}\ and\ \bibinfo {author} {\bibfnamefont {M.~J.}\ \bibnamefont
  {Rees}},\ }\href {\doibase 10.1086/466521} {\bibfield  {journal} {\bibinfo
  {journal} {Astrophys. J.}\ }\textbf {\bibinfo {volume} {633}},\ \bibinfo
  {pages} {624} (\bibinfo {year} {2005})},\ \Eprint
  {http://arxiv.org/abs/astro-ph/0506040} {arXiv:astro-ph/0506040 [astro-ph]}
  \BibitemShut {NoStop}%
\bibitem [{\citenamefont {Madau}\ \emph {et~al.}(2014)\citenamefont {Madau},
  \citenamefont {Haardt},\ and\ \citenamefont {Dotti}}]{Madau:2014pta}%
  \BibitemOpen
  \bibfield  {author} {\bibinfo {author} {\bibfnamefont {P.}~\bibnamefont
  {Madau}}, \bibinfo {author} {\bibfnamefont {F.}~\bibnamefont {Haardt}}, \
  and\ \bibinfo {author} {\bibfnamefont {M.}~\bibnamefont {Dotti}},\ }\href
  {\doibase 10.1088/2041-8205/784/2/L38} {\bibfield  {journal} {\bibinfo
  {journal} {Astrophys. J.}\ }\textbf {\bibinfo {volume} {784}},\ \bibinfo
  {pages} {L38} (\bibinfo {year} {2014})},\ \Eprint
  {http://arxiv.org/abs/1402.6995} {arXiv:1402.6995 [astro-ph.CO]} \BibitemShut
  {NoStop}%
\bibitem [{\citenamefont {Alexander}\ and\ \citenamefont
  {Natarajan}(2014)}]{Alexander:2014noa}%
  \BibitemOpen
  \bibfield  {author} {\bibinfo {author} {\bibfnamefont {T.}~\bibnamefont
  {Alexander}}\ and\ \bibinfo {author} {\bibfnamefont {P.}~\bibnamefont
  {Natarajan}},\ }\href {\doibase 10.1126/science.1251053} {\bibfield
  {journal} {\bibinfo  {journal} {Science}\ }\textbf {\bibinfo {volume}
  {345}},\ \bibinfo {pages} {1330} (\bibinfo {year} {2014})},\ \Eprint
  {http://arxiv.org/abs/1408.1718} {arXiv:1408.1718 [astro-ph.GA]} \BibitemShut
  {NoStop}%
\bibitem [{\citenamefont {Haehnelt}\ \emph {et~al.}(1998)\citenamefont
  {Haehnelt}, \citenamefont {Natarajan},\ and\ \citenamefont
  {Rees}}]{Haehnelt:1997js}%
  \BibitemOpen
  \bibfield  {author} {\bibinfo {author} {\bibfnamefont {M.~G.}\ \bibnamefont
  {Haehnelt}}, \bibinfo {author} {\bibfnamefont {P.}~\bibnamefont {Natarajan}},
  \ and\ \bibinfo {author} {\bibfnamefont {M.~J.}\ \bibnamefont {Rees}},\
  }\href {\doibase 10.1046/j.1365-8711.1998.01951.x} {\bibfield  {journal}
  {\bibinfo  {journal} {Mon. Not. Roy. Astron. Soc.}\ }\textbf {\bibinfo
  {volume} {300}},\ \bibinfo {pages} {817} (\bibinfo {year} {1998})},\ \Eprint
  {http://arxiv.org/abs/astro-ph/9712259} {arXiv:astro-ph/9712259 [astro-ph]}
  \BibitemShut {NoStop}%
\bibitem [{\citenamefont {Nulsen}\ and\ \citenamefont
  {Fabian}(2000)}]{Nulsen:1999mt}%
  \BibitemOpen
  \bibfield  {author} {\bibinfo {author} {\bibfnamefont {P.~E.~J.}\
  \bibnamefont {Nulsen}}\ and\ \bibinfo {author} {\bibfnamefont {A.~C.}\
  \bibnamefont {Fabian}},\ }\href {\doibase 10.1046/j.1365-8711.2000.03038.x}
  {\bibfield  {journal} {\bibinfo  {journal} {Mon. Not. Roy. Astron. Soc.}\
  }\textbf {\bibinfo {volume} {311}},\ \bibinfo {pages} {346} (\bibinfo {year}
  {2000})},\ \Eprint {http://arxiv.org/abs/astro-ph/9908282}
  {arXiv:astro-ph/9908282 [astro-ph]} \BibitemShut {NoStop}%
\bibitem [{\citenamefont {Barausse}(2012)}]{Barausse:2012fy}%
  \BibitemOpen
  \bibfield  {author} {\bibinfo {author} {\bibfnamefont {E.}~\bibnamefont
  {Barausse}},\ }\href {\doibase 10.1111/j.1365-2966.2012.21057.x} {\bibfield
  {journal} {\bibinfo  {journal} {Mon. Not. Roy. Astron. Soc.}\ }\textbf
  {\bibinfo {volume} {423}},\ \bibinfo {pages} {2533} (\bibinfo {year}
  {2012})},\ \Eprint {http://arxiv.org/abs/1201.5888} {arXiv:1201.5888
  [astro-ph.CO]} \BibitemShut {NoStop}%
\bibitem [{\citenamefont {Volonteri}\ \emph {et~al.}(2015)\citenamefont
  {Volonteri}, \citenamefont {Silk},\ and\ \citenamefont
  {Dubus}}]{Volonteri:2014lja}%
  \BibitemOpen
  \bibfield  {author} {\bibinfo {author} {\bibfnamefont {M.}~\bibnamefont
  {Volonteri}}, \bibinfo {author} {\bibfnamefont {J.}~\bibnamefont {Silk}}, \
  and\ \bibinfo {author} {\bibfnamefont {G.}~\bibnamefont {Dubus}},\ }\href
  {\doibase 10.1088/0004-637X/804/2/148} {\bibfield  {journal} {\bibinfo
  {journal} {Astrophys. J.}\ }\textbf {\bibinfo {volume} {804}},\ \bibinfo
  {pages} {148} (\bibinfo {year} {2015})},\ \Eprint
  {http://arxiv.org/abs/1401.3513} {arXiv:1401.3513 [astro-ph.GA]} \BibitemShut
  {NoStop}%
\bibitem [{\citenamefont {Pacucci}\ and\ \citenamefont
  {Ferrara}(2015)}]{Pacucci:2015efa}%
  \BibitemOpen
  \bibfield  {author} {\bibinfo {author} {\bibfnamefont {F.}~\bibnamefont
  {Pacucci}}\ and\ \bibinfo {author} {\bibfnamefont {A.}~\bibnamefont
  {Ferrara}},\ }\href {\doibase 10.1093/mnras/stv018} {\bibfield  {journal}
  {\bibinfo  {journal} {Mon. Not. Roy. Astron. Soc.}\ }\textbf {\bibinfo
  {volume} {448}},\ \bibinfo {pages} {104} (\bibinfo {year} {2015})},\ \Eprint
  {http://arxiv.org/abs/1501.00989} {arXiv:1501.00989 [astro-ph.HE]}
  \BibitemShut {NoStop}%
\bibitem [{\citenamefont {Pacucci}\ \emph
  {et~al.}(2015{\natexlab{a}})\citenamefont {Pacucci}, \citenamefont
  {Volonteri},\ and\ \citenamefont {Ferrara}}]{Pacucci:2015rwa}%
  \BibitemOpen
  \bibfield  {author} {\bibinfo {author} {\bibfnamefont {F.}~\bibnamefont
  {Pacucci}}, \bibinfo {author} {\bibfnamefont {M.}~\bibnamefont {Volonteri}},
  \ and\ \bibinfo {author} {\bibfnamefont {A.}~\bibnamefont {Ferrara}},\ }\href
  {\doibase 10.1093/mnras/stv1465} {\bibfield  {journal} {\bibinfo  {journal}
  {Mon. Not. Roy. Astron. Soc.}\ }\textbf {\bibinfo {volume} {452}},\ \bibinfo
  {pages} {1922} (\bibinfo {year} {2015}{\natexlab{a}})},\ \Eprint
  {http://arxiv.org/abs/1506.04750} {arXiv:1506.04750 [astro-ph.GA]}
  \BibitemShut {NoStop}%
\bibitem [{\citenamefont {Pacucci}\ \emph
  {et~al.}(2015{\natexlab{b}})\citenamefont {Pacucci}, \citenamefont {Ferrara},
  \citenamefont {Volonteri},\ and\ \citenamefont {Dubus}}]{Pacucci:2015wea}%
  \BibitemOpen
  \bibfield  {author} {\bibinfo {author} {\bibfnamefont {F.}~\bibnamefont
  {Pacucci}}, \bibinfo {author} {\bibfnamefont {A.}~\bibnamefont {Ferrara}},
  \bibinfo {author} {\bibfnamefont {M.}~\bibnamefont {Volonteri}}, \ and\
  \bibinfo {author} {\bibfnamefont {G.}~\bibnamefont {Dubus}},\ }\href
  {\doibase 10.1093/mnras/stv2196} {\bibfield  {journal} {\bibinfo  {journal}
  {Mon. Not. Roy. Astron. Soc.}\ }\textbf {\bibinfo {volume} {454}},\ \bibinfo
  {pages} {3771} (\bibinfo {year} {2015}{\natexlab{b}})},\ \Eprint
  {http://arxiv.org/abs/1506.05299} {arXiv:1506.05299 [astro-ph.HE]}
  \BibitemShut {NoStop}%
\bibitem [{\citenamefont {Paliya}\ \emph {et~al.}(2019)\citenamefont {Paliya}
  \emph {et~al.}}]{Paliya:2019oyn}%
  \BibitemOpen
  \bibfield  {author} {\bibinfo {author} {\bibfnamefont {V.~S.}\ \bibnamefont
  {Paliya}} \emph {et~al.},\ }\href@noop {} {\  (\bibinfo {year} {2019})},\
  \Eprint {http://arxiv.org/abs/1903.06106} {arXiv:1903.06106 [astro-ph.HE]}
  \BibitemShut {NoStop}%
\bibitem [{\citenamefont {Bartelmann}\ and\ \citenamefont
  {Schneider}(2001)}]{Bartelmann:1999yn}%
  \BibitemOpen
  \bibfield  {author} {\bibinfo {author} {\bibfnamefont {M.}~\bibnamefont
  {Bartelmann}}\ and\ \bibinfo {author} {\bibfnamefont {P.}~\bibnamefont
  {Schneider}},\ }\href {\doibase 10.1016/S0370-1573(00)00082-X} {\bibfield
  {journal} {\bibinfo  {journal} {Phys. Rept.}\ }\textbf {\bibinfo {volume}
  {340}},\ \bibinfo {pages} {291} (\bibinfo {year} {2001})},\ \Eprint
  {http://arxiv.org/abs/astro-ph/9912508} {arXiv:astro-ph/9912508 [astro-ph]}
  \BibitemShut {NoStop}%
\bibitem [{\citenamefont {Refregier}(2003)}]{Refregier:2003ct}%
  \BibitemOpen
  \bibfield  {author} {\bibinfo {author} {\bibfnamefont {A.}~\bibnamefont
  {Refregier}},\ }\href {\doibase 10.1146/annurev.astro.41.111302.102207}
  {\bibfield  {journal} {\bibinfo  {journal} {Ann. Rev. Astron. Astrophys.}\
  }\textbf {\bibinfo {volume} {41}},\ \bibinfo {pages} {645} (\bibinfo {year}
  {2003})},\ \Eprint {http://arxiv.org/abs/astro-ph/0307212}
  {arXiv:astro-ph/0307212 [astro-ph]} \BibitemShut {NoStop}%
\bibitem [{\citenamefont {Mandelbaum}(2018)}]{Mandelbaum:2017jpr}%
  \BibitemOpen
  \bibfield  {author} {\bibinfo {author} {\bibfnamefont {R.}~\bibnamefont
  {Mandelbaum}},\ }\href {\doibase 10.1146/annurev-astro-081817-051928}
  {\bibfield  {journal} {\bibinfo  {journal} {Ann. Rev. Astron. Astrophys.}\
  }\textbf {\bibinfo {volume} {56}},\ \bibinfo {pages} {393} (\bibinfo {year}
  {2018})},\ \Eprint {http://arxiv.org/abs/1710.03235} {arXiv:1710.03235
  [astro-ph.CO]} \BibitemShut {NoStop}%
\bibitem [{\citenamefont {Hu}\ and\ \citenamefont {Sawicki}(2007)}]{Hu:2007nk}%
  \BibitemOpen
  \bibfield  {author} {\bibinfo {author} {\bibfnamefont {W.}~\bibnamefont
  {Hu}}\ and\ \bibinfo {author} {\bibfnamefont {I.}~\bibnamefont {Sawicki}},\
  }\href {\doibase 10.1103/PhysRevD.76.064004} {\bibfield  {journal} {\bibinfo
  {journal} {Phys. Rev.}\ }\textbf {\bibinfo {volume} {D76}},\ \bibinfo {pages}
  {064004} (\bibinfo {year} {2007})},\ \Eprint {http://arxiv.org/abs/0705.1158}
  {arXiv:0705.1158 [astro-ph]} \BibitemShut {NoStop}%
\bibitem [{\citenamefont {Boehmer}\ \emph {et~al.}(2008)\citenamefont
  {Boehmer}, \citenamefont {Harko},\ and\ \citenamefont
  {Lobo}}]{Boehmer:2007kx}%
  \BibitemOpen
  \bibfield  {author} {\bibinfo {author} {\bibfnamefont {C.~G.}\ \bibnamefont
  {Boehmer}}, \bibinfo {author} {\bibfnamefont {T.}~\bibnamefont {Harko}}, \
  and\ \bibinfo {author} {\bibfnamefont {F.~S.~N.}\ \bibnamefont {Lobo}},\
  }\href {\doibase 10.1016/j.astropartphys.2008.04.003} {\bibfield  {journal}
  {\bibinfo  {journal} {Astropart. Phys.}\ }\textbf {\bibinfo {volume} {29}},\
  \bibinfo {pages} {386} (\bibinfo {year} {2008})},\ \Eprint
  {http://arxiv.org/abs/0709.0046} {arXiv:0709.0046 [gr-qc]} \BibitemShut
  {NoStop}%
\bibitem [{\citenamefont {Saridakis}(2010)}]{Saridakis:2009bv}%
  \BibitemOpen
  \bibfield  {author} {\bibinfo {author} {\bibfnamefont {E.~N.}\ \bibnamefont
  {Saridakis}},\ }\href {\doibase 10.1140/epjc/s10052-010-1294-6} {\bibfield
  {journal} {\bibinfo  {journal} {Eur. Phys. J.}\ }\textbf {\bibinfo {volume}
  {C67}},\ \bibinfo {pages} {229} (\bibinfo {year} {2010})},\ \Eprint
  {http://arxiv.org/abs/0905.3532} {arXiv:0905.3532 [hep-th]} \BibitemShut
  {NoStop}%
\bibitem [{\citenamefont {Clifton}\ \emph {et~al.}(2012)\citenamefont
  {Clifton}, \citenamefont {Ferreira}, \citenamefont {Padilla},\ and\
  \citenamefont {Skordis}}]{Clifton:2011jh}%
  \BibitemOpen
  \bibfield  {author} {\bibinfo {author} {\bibfnamefont {T.}~\bibnamefont
  {Clifton}}, \bibinfo {author} {\bibfnamefont {P.~G.}\ \bibnamefont
  {Ferreira}}, \bibinfo {author} {\bibfnamefont {A.}~\bibnamefont {Padilla}}, \
  and\ \bibinfo {author} {\bibfnamefont {C.}~\bibnamefont {Skordis}},\ }\href
  {\doibase 10.1016/j.physrep.2012.01.001} {\bibfield  {journal} {\bibinfo
  {journal} {Phys. Rept.}\ }\textbf {\bibinfo {volume} {513}},\ \bibinfo
  {pages} {1} (\bibinfo {year} {2012})},\ \Eprint
  {http://arxiv.org/abs/1106.2476} {arXiv:1106.2476 [astro-ph.CO]} \BibitemShut
  {NoStop}%
\bibitem [{\citenamefont {Capozziello}\ and\ \citenamefont
  {De~Laurentis}(2011)}]{Capozziello:2011et}%
  \BibitemOpen
  \bibfield  {author} {\bibinfo {author} {\bibfnamefont {S.}~\bibnamefont
  {Capozziello}}\ and\ \bibinfo {author} {\bibfnamefont {M.}~\bibnamefont
  {De~Laurentis}},\ }\href {\doibase 10.1016/j.physrep.2011.09.003} {\bibfield
  {journal} {\bibinfo  {journal} {Phys. Rept.}\ }\textbf {\bibinfo {volume}
  {509}},\ \bibinfo {pages} {167} (\bibinfo {year} {2011})},\ \Eprint
  {http://arxiv.org/abs/1108.6266} {arXiv:1108.6266 [gr-qc]} \BibitemShut
  {NoStop}%
\bibitem [{\citenamefont {Chamseddine}\ and\ \citenamefont
  {Mukhanov}(2013)}]{Chamseddine:2013kea}%
  \BibitemOpen
  \bibfield  {author} {\bibinfo {author} {\bibfnamefont {A.~H.}\ \bibnamefont
  {Chamseddine}}\ and\ \bibinfo {author} {\bibfnamefont {V.}~\bibnamefont
  {Mukhanov}},\ }\href {\doibase 10.1007/JHEP11(2013)135} {\bibfield  {journal}
  {\bibinfo  {journal} {JHEP}\ }\textbf {\bibinfo {volume} {11}},\ \bibinfo
  {pages} {135} (\bibinfo {year} {2013})},\ \Eprint
  {http://arxiv.org/abs/1308.5410} {arXiv:1308.5410 [astro-ph.CO]} \BibitemShut
  {NoStop}%
\bibitem [{\citenamefont {Myrzakulov}\ \emph {et~al.}(2015)\citenamefont
  {Myrzakulov}, \citenamefont {Sebastiani},\ and\ \citenamefont
  {Vagnozzi}}]{Myrzakulov:2015qaa}%
  \BibitemOpen
  \bibfield  {author} {\bibinfo {author} {\bibfnamefont {R.}~\bibnamefont
  {Myrzakulov}}, \bibinfo {author} {\bibfnamefont {L.}~\bibnamefont
  {Sebastiani}}, \ and\ \bibinfo {author} {\bibfnamefont {S.}~\bibnamefont
  {Vagnozzi}},\ }\href {\doibase 10.1140/epjc/s10052-015-3672-6} {\bibfield
  {journal} {\bibinfo  {journal} {Eur. Phys. J.}\ }\textbf {\bibinfo {volume}
  {C75}},\ \bibinfo {pages} {444} (\bibinfo {year} {2015})},\ \Eprint
  {http://arxiv.org/abs/1504.07984} {arXiv:1504.07984 [gr-qc]} \BibitemShut
  {NoStop}%
\bibitem [{\citenamefont {Myrzakulov}\ \emph {et~al.}(2016)\citenamefont
  {Myrzakulov}, \citenamefont {Sebastiani}, \citenamefont {Vagnozzi},\ and\
  \citenamefont {Zerbini}}]{Myrzakulov:2015kda}%
  \BibitemOpen
  \bibfield  {author} {\bibinfo {author} {\bibfnamefont {R.}~\bibnamefont
  {Myrzakulov}}, \bibinfo {author} {\bibfnamefont {L.}~\bibnamefont
  {Sebastiani}}, \bibinfo {author} {\bibfnamefont {S.}~\bibnamefont
  {Vagnozzi}}, \ and\ \bibinfo {author} {\bibfnamefont {S.}~\bibnamefont
  {Zerbini}},\ }\href {\doibase 10.1088/0264-9381/33/12/125005} {\bibfield
  {journal} {\bibinfo  {journal} {Class. Quant. Grav.}\ }\textbf {\bibinfo
  {volume} {33}},\ \bibinfo {pages} {125005} (\bibinfo {year} {2016})},\
  \Eprint {http://arxiv.org/abs/1510.02284} {arXiv:1510.02284 [gr-qc]}
  \BibitemShut {NoStop}%
\bibitem [{\citenamefont {Cai}\ \emph {et~al.}(2016)\citenamefont {Cai},
  \citenamefont {Capozziello}, \citenamefont {De~Laurentis},\ and\
  \citenamefont {Saridakis}}]{Cai:2015emx}%
  \BibitemOpen
  \bibfield  {author} {\bibinfo {author} {\bibfnamefont {Y.-F.}\ \bibnamefont
  {Cai}}, \bibinfo {author} {\bibfnamefont {S.}~\bibnamefont {Capozziello}},
  \bibinfo {author} {\bibfnamefont {M.}~\bibnamefont {De~Laurentis}}, \ and\
  \bibinfo {author} {\bibfnamefont {E.~N.}\ \bibnamefont {Saridakis}},\ }\href
  {\doibase 10.1088/0034-4885/79/10/106901} {\bibfield  {journal} {\bibinfo
  {journal} {Rept. Prog. Phys.}\ }\textbf {\bibinfo {volume} {79}},\ \bibinfo
  {pages} {106901} (\bibinfo {year} {2016})},\ \Eprint
  {http://arxiv.org/abs/1511.07586} {arXiv:1511.07586 [gr-qc]} \BibitemShut
  {NoStop}%
\bibitem [{\citenamefont {Rinaldi}(2017)}]{Rinaldi:2016oqp}%
  \BibitemOpen
  \bibfield  {author} {\bibinfo {author} {\bibfnamefont {M.}~\bibnamefont
  {Rinaldi}},\ }\href {\doibase 10.1016/j.dark.2017.02.003} {\bibfield
  {journal} {\bibinfo  {journal} {Phys. Dark Univ.}\ }\textbf {\bibinfo
  {volume} {16}},\ \bibinfo {pages} {14} (\bibinfo {year} {2017})},\ \Eprint
  {http://arxiv.org/abs/1608.03839} {arXiv:1608.03839 [gr-qc]} \BibitemShut
  {NoStop}%
\bibitem [{\citenamefont {Sebastiani}\ \emph {et~al.}(2017)\citenamefont
  {Sebastiani}, \citenamefont {Vagnozzi},\ and\ \citenamefont
  {Myrzakulov}}]{Sebastiani:2016ras}%
  \BibitemOpen
  \bibfield  {author} {\bibinfo {author} {\bibfnamefont {L.}~\bibnamefont
  {Sebastiani}}, \bibinfo {author} {\bibfnamefont {S.}~\bibnamefont
  {Vagnozzi}}, \ and\ \bibinfo {author} {\bibfnamefont {R.}~\bibnamefont
  {Myrzakulov}},\ }\href {\doibase 10.1155/2017/3156915} {\bibfield  {journal}
  {\bibinfo  {journal} {Adv. High Energy Phys.}\ }\textbf {\bibinfo {volume}
  {2017}},\ \bibinfo {pages} {3156915} (\bibinfo {year} {2017})},\ \Eprint
  {http://arxiv.org/abs/1612.08661} {arXiv:1612.08661 [gr-qc]} \BibitemShut
  {NoStop}%
\bibitem [{\citenamefont {Capozziello}\ \emph {et~al.}(2017)\citenamefont
  {Capozziello}, \citenamefont {Jovanović}, \citenamefont {Jovanović},\ and\
  \citenamefont {Borka}}]{Capozziello:2017rvz}%
  \BibitemOpen
  \bibfield  {author} {\bibinfo {author} {\bibfnamefont {S.}~\bibnamefont
  {Capozziello}}, \bibinfo {author} {\bibfnamefont {P.}~\bibnamefont
  {Jovanović}}, \bibinfo {author} {\bibfnamefont {V.~B.}\ \bibnamefont
  {Jovanović}}, \ and\ \bibinfo {author} {\bibfnamefont {D.}~\bibnamefont
  {Borka}},\ }\href {\doibase 10.1088/1475-7516/2017/06/044} {\bibfield
  {journal} {\bibinfo  {journal} {JCAP}\ }\textbf {\bibinfo {volume} {1706}},\
  \bibinfo {pages} {044} (\bibinfo {year} {2017})},\ \Eprint
  {http://arxiv.org/abs/1702.03430} {arXiv:1702.03430 [gr-qc]} \BibitemShut
  {NoStop}%
\bibitem [{\citenamefont {Nojiri}\ \emph {et~al.}(2017)\citenamefont {Nojiri},
  \citenamefont {Odintsov},\ and\ \citenamefont {Oikonomou}}]{Nojiri:2017ncd}%
  \BibitemOpen
  \bibfield  {author} {\bibinfo {author} {\bibfnamefont {S.}~\bibnamefont
  {Nojiri}}, \bibinfo {author} {\bibfnamefont {S.~D.}\ \bibnamefont
  {Odintsov}}, \ and\ \bibinfo {author} {\bibfnamefont {V.~K.}\ \bibnamefont
  {Oikonomou}},\ }\href {\doibase 10.1016/j.physrep.2017.06.001} {\bibfield
  {journal} {\bibinfo  {journal} {Phys. Rept.}\ }\textbf {\bibinfo {volume}
  {692}},\ \bibinfo {pages} {1} (\bibinfo {year} {2017})},\ \Eprint
  {http://arxiv.org/abs/1705.11098} {arXiv:1705.11098 [gr-qc]} \BibitemShut
  {NoStop}%
\bibitem [{\citenamefont {Vagnozzi}(2017)}]{Vagnozzi:2017ilo}%
  \BibitemOpen
  \bibfield  {author} {\bibinfo {author} {\bibfnamefont {S.}~\bibnamefont
  {Vagnozzi}},\ }\href {\doibase 10.1088/1361-6382/aa838b} {\bibfield
  {journal} {\bibinfo  {journal} {Class. Quant. Grav.}\ }\textbf {\bibinfo
  {volume} {34}},\ \bibinfo {pages} {185006} (\bibinfo {year} {2017})},\
  \Eprint {http://arxiv.org/abs/1708.00603} {arXiv:1708.00603 [gr-qc]}
  \BibitemShut {NoStop}%
\bibitem [{\citenamefont {Dutta}\ \emph {et~al.}(2018)\citenamefont {Dutta},
  \citenamefont {Khyllep}, \citenamefont {Saridakis}, \citenamefont
  {Tamanini},\ and\ \citenamefont {Vagnozzi}}]{Dutta:2017fjw}%
  \BibitemOpen
  \bibfield  {author} {\bibinfo {author} {\bibfnamefont {J.}~\bibnamefont
  {Dutta}}, \bibinfo {author} {\bibfnamefont {W.}~\bibnamefont {Khyllep}},
  \bibinfo {author} {\bibfnamefont {E.~N.}\ \bibnamefont {Saridakis}}, \bibinfo
  {author} {\bibfnamefont {N.}~\bibnamefont {Tamanini}}, \ and\ \bibinfo
  {author} {\bibfnamefont {S.}~\bibnamefont {Vagnozzi}},\ }\href {\doibase
  10.1088/1475-7516/2018/02/041} {\bibfield  {journal} {\bibinfo  {journal}
  {JCAP}\ }\textbf {\bibinfo {volume} {1802}},\ \bibinfo {pages} {041}
  (\bibinfo {year} {2018})},\ \Eprint {http://arxiv.org/abs/1711.07290}
  {arXiv:1711.07290 [gr-qc]} \BibitemShut {NoStop}%
\bibitem [{\citenamefont {Casalino}\ \emph {et~al.}(2018)\citenamefont
  {Casalino}, \citenamefont {Rinaldi}, \citenamefont {Sebastiani},\ and\
  \citenamefont {Vagnozzi}}]{Casalino:2018tcd}%
  \BibitemOpen
  \bibfield  {author} {\bibinfo {author} {\bibfnamefont {A.}~\bibnamefont
  {Casalino}}, \bibinfo {author} {\bibfnamefont {M.}~\bibnamefont {Rinaldi}},
  \bibinfo {author} {\bibfnamefont {L.}~\bibnamefont {Sebastiani}}, \ and\
  \bibinfo {author} {\bibfnamefont {S.}~\bibnamefont {Vagnozzi}},\ }\href
  {\doibase 10.1016/j.dark.2018.10.001} {\bibfield  {journal} {\bibinfo
  {journal} {Phys. Dark Univ.}\ }\textbf {\bibinfo {volume} {22}},\ \bibinfo
  {pages} {108} (\bibinfo {year} {2018})},\ \Eprint
  {http://arxiv.org/abs/1803.02620} {arXiv:1803.02620 [gr-qc]} \BibitemShut
  {NoStop}%
\bibitem [{\citenamefont {Bambi}\ and\ \citenamefont
  {Freese}(2009)}]{Bambi:2008jg}%
  \BibitemOpen
  \bibfield  {author} {\bibinfo {author} {\bibfnamefont {C.}~\bibnamefont
  {Bambi}}\ and\ \bibinfo {author} {\bibfnamefont {K.}~\bibnamefont {Freese}},\
  }\href {\doibase 10.1103/PhysRevD.79.043002} {\bibfield  {journal} {\bibinfo
  {journal} {Phys. Rev.}\ }\textbf {\bibinfo {volume} {D79}},\ \bibinfo {pages}
  {043002} (\bibinfo {year} {2009})},\ \Eprint {http://arxiv.org/abs/0812.1328}
  {arXiv:0812.1328 [astro-ph]} \BibitemShut {NoStop}%
\bibitem [{\citenamefont {Amarilla}\ \emph {et~al.}(2010)\citenamefont
  {Amarilla}, \citenamefont {Eiroa},\ and\ \citenamefont
  {Giribet}}]{Amarilla:2010zq}%
  \BibitemOpen
  \bibfield  {author} {\bibinfo {author} {\bibfnamefont {L.}~\bibnamefont
  {Amarilla}}, \bibinfo {author} {\bibfnamefont {E.~F.}\ \bibnamefont {Eiroa}},
  \ and\ \bibinfo {author} {\bibfnamefont {G.}~\bibnamefont {Giribet}},\ }\href
  {\doibase 10.1103/PhysRevD.81.124045} {\bibfield  {journal} {\bibinfo
  {journal} {Phys. Rev.}\ }\textbf {\bibinfo {volume} {D81}},\ \bibinfo {pages}
  {124045} (\bibinfo {year} {2010})},\ \Eprint {http://arxiv.org/abs/1005.0607}
  {arXiv:1005.0607 [gr-qc]} \BibitemShut {NoStop}%
\bibitem [{\citenamefont {Amarilla}\ and\ \citenamefont
  {Eiroa}(2012)}]{Amarilla:2011fx}%
  \BibitemOpen
  \bibfield  {author} {\bibinfo {author} {\bibfnamefont {L.}~\bibnamefont
  {Amarilla}}\ and\ \bibinfo {author} {\bibfnamefont {E.~F.}\ \bibnamefont
  {Eiroa}},\ }\href {\doibase 10.1103/PhysRevD.85.064019} {\bibfield  {journal}
  {\bibinfo  {journal} {Phys. Rev.}\ }\textbf {\bibinfo {volume} {D85}},\
  \bibinfo {pages} {064019} (\bibinfo {year} {2012})},\ \Eprint
  {http://arxiv.org/abs/1112.6349} {arXiv:1112.6349 [gr-qc]} \BibitemShut
  {NoStop}%
\bibitem [{\citenamefont {Amarilla}\ and\ \citenamefont
  {Eiroa}(2013)}]{Amarilla:2013sj}%
  \BibitemOpen
  \bibfield  {author} {\bibinfo {author} {\bibfnamefont {L.}~\bibnamefont
  {Amarilla}}\ and\ \bibinfo {author} {\bibfnamefont {E.~F.}\ \bibnamefont
  {Eiroa}},\ }\href {\doibase 10.1103/PhysRevD.87.044057} {\bibfield  {journal}
  {\bibinfo  {journal} {Phys. Rev.}\ }\textbf {\bibinfo {volume} {D87}},\
  \bibinfo {pages} {044057} (\bibinfo {year} {2013})},\ \Eprint
  {http://arxiv.org/abs/1301.0532} {arXiv:1301.0532 [gr-qc]} \BibitemShut
  {NoStop}%
\bibitem [{\citenamefont {Nedkova}\ \emph {et~al.}(2013)\citenamefont
  {Nedkova}, \citenamefont {Tinchev},\ and\ \citenamefont
  {Yazadjiev}}]{Nedkova:2013msa}%
  \BibitemOpen
  \bibfield  {author} {\bibinfo {author} {\bibfnamefont {P.~G.}\ \bibnamefont
  {Nedkova}}, \bibinfo {author} {\bibfnamefont {V.~K.}\ \bibnamefont
  {Tinchev}}, \ and\ \bibinfo {author} {\bibfnamefont {S.~S.}\ \bibnamefont
  {Yazadjiev}},\ }\href {\doibase 10.1103/PhysRevD.88.124019} {\bibfield
  {journal} {\bibinfo  {journal} {Phys. Rev.}\ }\textbf {\bibinfo {volume}
  {D88}},\ \bibinfo {pages} {124019} (\bibinfo {year} {2013})},\ \Eprint
  {http://arxiv.org/abs/1307.7647} {arXiv:1307.7647 [gr-qc]} \BibitemShut
  {NoStop}%
\bibitem [{\citenamefont {Tinchev}\ and\ \citenamefont
  {Yazadjiev}(2014)}]{Tinchev:2013nba}%
  \BibitemOpen
  \bibfield  {author} {\bibinfo {author} {\bibfnamefont {V.~K.}\ \bibnamefont
  {Tinchev}}\ and\ \bibinfo {author} {\bibfnamefont {S.~S.}\ \bibnamefont
  {Yazadjiev}},\ }\href {\doibase 10.1142/S0218271814500606} {\bibfield
  {journal} {\bibinfo  {journal} {Int. J. Mod. Phys.}\ }\textbf {\bibinfo
  {volume} {D23}},\ \bibinfo {pages} {1450060} (\bibinfo {year} {2014})},\
  \Eprint {http://arxiv.org/abs/1311.1353} {arXiv:1311.1353 [gr-qc]}
  \BibitemShut {NoStop}%
\bibitem [{\citenamefont {Wei}\ and\ \citenamefont {Liu}(2013)}]{Wei:2013kza}%
  \BibitemOpen
  \bibfield  {author} {\bibinfo {author} {\bibfnamefont {S.-W.}\ \bibnamefont
  {Wei}}\ and\ \bibinfo {author} {\bibfnamefont {Y.-X.}\ \bibnamefont {Liu}},\
  }\href {\doibase 10.1088/1475-7516/2013/11/063} {\bibfield  {journal}
  {\bibinfo  {journal} {JCAP}\ }\textbf {\bibinfo {volume} {1311}},\ \bibinfo
  {pages} {063} (\bibinfo {year} {2013})},\ \Eprint
  {http://arxiv.org/abs/1311.4251} {arXiv:1311.4251 [gr-qc]} \BibitemShut
  {NoStop}%
\bibitem [{\citenamefont {Grenzebach}\ \emph {et~al.}(2014)\citenamefont
  {Grenzebach}, \citenamefont {Perlick},\ and\ \citenamefont
  {L{\"a}mmerzahl}}]{Grenzebach:2014fha}%
  \BibitemOpen
  \bibfield  {author} {\bibinfo {author} {\bibfnamefont {A.}~\bibnamefont
  {Grenzebach}}, \bibinfo {author} {\bibfnamefont {V.}~\bibnamefont {Perlick}},
  \ and\ \bibinfo {author} {\bibfnamefont {C.}~\bibnamefont {L{\"a}mmerzahl}},\
  }\href {\doibase 10.1103/PhysRevD.89.124004} {\bibfield  {journal} {\bibinfo
  {journal} {Phys. Rev.}\ }\textbf {\bibinfo {volume} {D89}},\ \bibinfo {pages}
  {124004} (\bibinfo {year} {2014})},\ \Eprint {http://arxiv.org/abs/1403.5234}
  {arXiv:1403.5234 [gr-qc]} \BibitemShut {NoStop}%
\bibitem [{\citenamefont {Papnoi}\ \emph {et~al.}(2014)\citenamefont {Papnoi},
  \citenamefont {Atamurotov}, \citenamefont {Ghosh},\ and\ \citenamefont
  {Ahmedov}}]{Papnoi:2014aaa}%
  \BibitemOpen
  \bibfield  {author} {\bibinfo {author} {\bibfnamefont {U.}~\bibnamefont
  {Papnoi}}, \bibinfo {author} {\bibfnamefont {F.}~\bibnamefont {Atamurotov}},
  \bibinfo {author} {\bibfnamefont {S.~G.}\ \bibnamefont {Ghosh}}, \ and\
  \bibinfo {author} {\bibfnamefont {B.}~\bibnamefont {Ahmedov}},\ }\href
  {\doibase 10.1103/PhysRevD.90.024073} {\bibfield  {journal} {\bibinfo
  {journal} {Phys. Rev.}\ }\textbf {\bibinfo {volume} {D90}},\ \bibinfo {pages}
  {024073} (\bibinfo {year} {2014})},\ \Eprint {http://arxiv.org/abs/1407.0834}
  {arXiv:1407.0834 [gr-qc]} \BibitemShut {NoStop}%
\bibitem [{\citenamefont {Sakai}\ \emph {et~al.}(2014)\citenamefont {Sakai},
  \citenamefont {Saida},\ and\ \citenamefont {Tamaki}}]{Sakai:2014pga}%
  \BibitemOpen
  \bibfield  {author} {\bibinfo {author} {\bibfnamefont {N.}~\bibnamefont
  {Sakai}}, \bibinfo {author} {\bibfnamefont {H.}~\bibnamefont {Saida}}, \ and\
  \bibinfo {author} {\bibfnamefont {T.}~\bibnamefont {Tamaki}},\ }\href
  {\doibase 10.1103/PhysRevD.90.104013} {\bibfield  {journal} {\bibinfo
  {journal} {Phys. Rev.}\ }\textbf {\bibinfo {volume} {D90}},\ \bibinfo {pages}
  {104013} (\bibinfo {year} {2014})},\ \Eprint {http://arxiv.org/abs/1408.6929}
  {arXiv:1408.6929 [gr-qc]} \BibitemShut {NoStop}%
\bibitem [{\citenamefont {Wei}\ \emph {et~al.}(2015)\citenamefont {Wei},
  \citenamefont {Cheng}, \citenamefont {Zhong},\ and\ \citenamefont
  {Zhou}}]{Wei:2015dua}%
  \BibitemOpen
  \bibfield  {author} {\bibinfo {author} {\bibfnamefont {S.-W.}\ \bibnamefont
  {Wei}}, \bibinfo {author} {\bibfnamefont {P.}~\bibnamefont {Cheng}}, \bibinfo
  {author} {\bibfnamefont {Y.}~\bibnamefont {Zhong}}, \ and\ \bibinfo {author}
  {\bibfnamefont {X.-N.}\ \bibnamefont {Zhou}},\ }\href {\doibase
  10.1088/1475-7516/2015/08/004} {\bibfield  {journal} {\bibinfo  {journal}
  {JCAP}\ }\textbf {\bibinfo {volume} {1508}},\ \bibinfo {pages} {004}
  (\bibinfo {year} {2015})},\ \Eprint {http://arxiv.org/abs/1501.06298}
  {arXiv:1501.06298 [gr-qc]} \BibitemShut {NoStop}%
\bibitem [{\citenamefont {Moffat}(2015)}]{Moffat:2015kva}%
  \BibitemOpen
  \bibfield  {author} {\bibinfo {author} {\bibfnamefont {J.~W.}\ \bibnamefont
  {Moffat}},\ }\href {\doibase 10.1140/epjc/s10052-015-3352-6} {\bibfield
  {journal} {\bibinfo  {journal} {Eur. Phys. J.}\ }\textbf {\bibinfo {volume}
  {C75}},\ \bibinfo {pages} {130} (\bibinfo {year} {2015})},\ \Eprint
  {http://arxiv.org/abs/1502.01677} {arXiv:1502.01677 [gr-qc]} \BibitemShut
  {NoStop}%
\bibitem [{\citenamefont {Ghasemi-Nodehi}\ \emph {et~al.}(2015)\citenamefont
  {Ghasemi-Nodehi}, \citenamefont {Li},\ and\ \citenamefont
  {Bambi}}]{Ghasemi-Nodehi:2015raa}%
  \BibitemOpen
  \bibfield  {author} {\bibinfo {author} {\bibfnamefont {M.}~\bibnamefont
  {Ghasemi-Nodehi}}, \bibinfo {author} {\bibfnamefont {Z.}~\bibnamefont {Li}},
  \ and\ \bibinfo {author} {\bibfnamefont {C.}~\bibnamefont {Bambi}},\ }\href
  {\doibase 10.1140/epjc/s10052-015-3539-x} {\bibfield  {journal} {\bibinfo
  {journal} {Eur. Phys. J.}\ }\textbf {\bibinfo {volume} {C75}},\ \bibinfo
  {pages} {315} (\bibinfo {year} {2015})},\ \Eprint
  {http://arxiv.org/abs/1506.02627} {arXiv:1506.02627 [gr-qc]} \BibitemShut
  {NoStop}%
\bibitem [{\citenamefont {Atamurotov}\ \emph {et~al.}(2016)\citenamefont
  {Atamurotov}, \citenamefont {Ghosh},\ and\ \citenamefont
  {Ahmedov}}]{Atamurotov:2015xfa}%
  \BibitemOpen
  \bibfield  {author} {\bibinfo {author} {\bibfnamefont {F.}~\bibnamefont
  {Atamurotov}}, \bibinfo {author} {\bibfnamefont {S.~G.}\ \bibnamefont
  {Ghosh}}, \ and\ \bibinfo {author} {\bibfnamefont {B.}~\bibnamefont
  {Ahmedov}},\ }\href {\doibase 10.1140/epjc/s10052-016-4122-9} {\bibfield
  {journal} {\bibinfo  {journal} {Eur. Phys. J.}\ }\textbf {\bibinfo {volume}
  {C76}},\ \bibinfo {pages} {273} (\bibinfo {year} {2016})},\ \Eprint
  {http://arxiv.org/abs/1506.03690} {arXiv:1506.03690 [gr-qc]} \BibitemShut
  {NoStop}%
\bibitem [{\citenamefont {Cunha}\ \emph {et~al.}(2015)\citenamefont {Cunha},
  \citenamefont {Herdeiro}, \citenamefont {Radu},\ and\ \citenamefont
  {Runarsson}}]{Cunha:2015yba}%
  \BibitemOpen
  \bibfield  {author} {\bibinfo {author} {\bibfnamefont {P.~V.~P.}\
  \bibnamefont {Cunha}}, \bibinfo {author} {\bibfnamefont {C.~A.~R.}\
  \bibnamefont {Herdeiro}}, \bibinfo {author} {\bibfnamefont {E.}~\bibnamefont
  {Radu}}, \ and\ \bibinfo {author} {\bibfnamefont {H.~F.}\ \bibnamefont
  {Runarsson}},\ }\href {\doibase 10.1103/PhysRevLett.115.211102} {\bibfield
  {journal} {\bibinfo  {journal} {Phys. Rev. Lett.}\ }\textbf {\bibinfo
  {volume} {115}},\ \bibinfo {pages} {211102} (\bibinfo {year} {2015})},\
  \Eprint {http://arxiv.org/abs/1509.00021} {arXiv:1509.00021 [gr-qc]}
  \BibitemShut {NoStop}%
\bibitem [{\citenamefont {Amir}\ and\ \citenamefont
  {Ghosh}(2016)}]{Amir:2016cen}%
  \BibitemOpen
  \bibfield  {author} {\bibinfo {author} {\bibfnamefont {M.}~\bibnamefont
  {Amir}}\ and\ \bibinfo {author} {\bibfnamefont {S.~G.}\ \bibnamefont
  {Ghosh}},\ }\href {\doibase 10.1103/PhysRevD.94.024054} {\bibfield  {journal}
  {\bibinfo  {journal} {Phys. Rev.}\ }\textbf {\bibinfo {volume} {D94}},\
  \bibinfo {pages} {024054} (\bibinfo {year} {2016})},\ \Eprint
  {http://arxiv.org/abs/1603.06382} {arXiv:1603.06382 [gr-qc]} \BibitemShut
  {NoStop}%
\bibitem [{\citenamefont {Dastan}\ \emph {et~al.}(2016)\citenamefont {Dastan},
  \citenamefont {Saffari},\ and\ \citenamefont {Soroushfar}}]{Dastan:2016vhb}%
  \BibitemOpen
  \bibfield  {author} {\bibinfo {author} {\bibfnamefont {S.}~\bibnamefont
  {Dastan}}, \bibinfo {author} {\bibfnamefont {R.}~\bibnamefont {Saffari}}, \
  and\ \bibinfo {author} {\bibfnamefont {S.}~\bibnamefont {Soroushfar}},\
  }\href@noop {} {\  (\bibinfo {year} {2016})},\ \Eprint
  {http://arxiv.org/abs/1606.06994} {arXiv:1606.06994 [gr-qc]} \BibitemShut
  {NoStop}%
\bibitem [{\citenamefont {Tretyakova}\ and\ \citenamefont
  {Adyev}(2016)}]{Tretyakova:2016ale}%
  \BibitemOpen
  \bibfield  {author} {\bibinfo {author} {\bibfnamefont {D.~A.}\ \bibnamefont
  {Tretyakova}}\ and\ \bibinfo {author} {\bibfnamefont {T.~M.}\ \bibnamefont
  {Adyev}},\ }\href@noop {} {\  (\bibinfo {year} {2016})},\ \Eprint
  {http://arxiv.org/abs/1610.07300} {arXiv:1610.07300 [gr-qc]} \BibitemShut
  {NoStop}%
\bibitem [{\citenamefont {Mureika}\ and\ \citenamefont
  {Varieschi}(2017)}]{Mureika:2016efo}%
  \BibitemOpen
  \bibfield  {author} {\bibinfo {author} {\bibfnamefont {J.~R.}\ \bibnamefont
  {Mureika}}\ and\ \bibinfo {author} {\bibfnamefont {G.~U.}\ \bibnamefont
  {Varieschi}},\ }\href {\doibase 10.1139/cjp-2017-0241} {\bibfield  {journal}
  {\bibinfo  {journal} {Can. J. Phys.}\ }\textbf {\bibinfo {volume} {95}},\
  \bibinfo {pages} {1299} (\bibinfo {year} {2017})},\ \Eprint
  {http://arxiv.org/abs/1611.00399} {arXiv:1611.00399 [gr-qc]} \BibitemShut
  {NoStop}%
\bibitem [{\citenamefont {Sharif}\ and\ \citenamefont
  {Iftikhar}(2016)}]{Sharif:2016znp}%
  \BibitemOpen
  \bibfield  {author} {\bibinfo {author} {\bibfnamefont {M.}~\bibnamefont
  {Sharif}}\ and\ \bibinfo {author} {\bibfnamefont {S.}~\bibnamefont
  {Iftikhar}},\ }\href {\doibase 10.1140/epjc/s10052-016-4472-3} {\bibfield
  {journal} {\bibinfo  {journal} {Eur. Phys. J.}\ }\textbf {\bibinfo {volume}
  {C76}},\ \bibinfo {pages} {630} (\bibinfo {year} {2016})},\ \Eprint
  {http://arxiv.org/abs/1611.00611} {arXiv:1611.00611 [gr-qc]} \BibitemShut
  {NoStop}%
\bibitem [{\citenamefont {Alhamzawi}(2017)}]{Alhamzawi:2017iyn}%
  \BibitemOpen
  \bibfield  {author} {\bibinfo {author} {\bibfnamefont {A.}~\bibnamefont
  {Alhamzawi}},\ }\href {\doibase 10.1142/S0218271817501565} {\bibfield
  {journal} {\bibinfo  {journal} {Int. J. Mod. Phys.}\ }\textbf {\bibinfo
  {volume} {D26}},\ \bibinfo {pages} {1750156} (\bibinfo {year}
  {2017})}\BibitemShut {NoStop}%
\bibitem [{\citenamefont {Cunha}\ \emph {et~al.}(2017)\citenamefont {Cunha},
  \citenamefont {Herdeiro}, \citenamefont {Kleihaus}, \citenamefont {Kunz},\
  and\ \citenamefont {Radu}}]{Cunha:2016wzk}%
  \BibitemOpen
  \bibfield  {author} {\bibinfo {author} {\bibfnamefont {P.~V.~P.}\
  \bibnamefont {Cunha}}, \bibinfo {author} {\bibfnamefont {C.~A.~R.}\
  \bibnamefont {Herdeiro}}, \bibinfo {author} {\bibfnamefont {B.}~\bibnamefont
  {Kleihaus}}, \bibinfo {author} {\bibfnamefont {J.}~\bibnamefont {Kunz}}, \
  and\ \bibinfo {author} {\bibfnamefont {E.}~\bibnamefont {Radu}},\ }\href
  {\doibase 10.1016/j.physletb.2017.03.020} {\bibfield  {journal} {\bibinfo
  {journal} {Phys. Lett.}\ }\textbf {\bibinfo {volume} {B768}},\ \bibinfo
  {pages} {373} (\bibinfo {year} {2017})},\ \Eprint
  {http://arxiv.org/abs/1701.00079} {arXiv:1701.00079 [gr-qc]} \BibitemShut
  {NoStop}%
\bibitem [{\citenamefont {Singh}\ and\ \citenamefont
  {Ghosh}(2018)}]{Singh:2017vfr}%
  \BibitemOpen
  \bibfield  {author} {\bibinfo {author} {\bibfnamefont {B.~P.}\ \bibnamefont
  {Singh}}\ and\ \bibinfo {author} {\bibfnamefont {S.~G.}\ \bibnamefont
  {Ghosh}},\ }\href {\doibase 10.1016/j.aop.2018.05.010} {\bibfield  {journal}
  {\bibinfo  {journal} {Annals Phys.}\ }\textbf {\bibinfo {volume} {395}},\
  \bibinfo {pages} {127} (\bibinfo {year} {2018})},\ \Eprint
  {http://arxiv.org/abs/1707.07125} {arXiv:1707.07125 [gr-qc]} \BibitemShut
  {NoStop}%
\bibitem [{\citenamefont {Tsukamoto}(2018)}]{Tsukamoto:2017fxq}%
  \BibitemOpen
  \bibfield  {author} {\bibinfo {author} {\bibfnamefont {N.}~\bibnamefont
  {Tsukamoto}},\ }\href {\doibase 10.1103/PhysRevD.97.064021} {\bibfield
  {journal} {\bibinfo  {journal} {Phys. Rev.}\ }\textbf {\bibinfo {volume}
  {D97}},\ \bibinfo {pages} {064021} (\bibinfo {year} {2018})},\ \Eprint
  {http://arxiv.org/abs/1708.07427} {arXiv:1708.07427 [gr-qc]} \BibitemShut
  {NoStop}%
\bibitem [{\citenamefont {Eiroa}\ and\ \citenamefont
  {Sendra}(2018)}]{Eiroa:2017uuq}%
  \BibitemOpen
  \bibfield  {author} {\bibinfo {author} {\bibfnamefont {E.~F.}\ \bibnamefont
  {Eiroa}}\ and\ \bibinfo {author} {\bibfnamefont {C.~M.}\ \bibnamefont
  {Sendra}},\ }\href {\doibase 10.1140/epjc/s10052-018-5586-6} {\bibfield
  {journal} {\bibinfo  {journal} {Eur. Phys. J.}\ }\textbf {\bibinfo {volume}
  {C78}},\ \bibinfo {pages} {91} (\bibinfo {year} {2018})},\ \Eprint
  {http://arxiv.org/abs/1711.08380} {arXiv:1711.08380 [gr-qc]} \BibitemShut
  {NoStop}%
\bibitem [{\citenamefont {Kumar}\ \emph {et~al.}(2017)\citenamefont {Kumar},
  \citenamefont {Singh}, \citenamefont {Ali},\ and\ \citenamefont
  {Ghosh}}]{Kumar:2017vuh}%
  \BibitemOpen
  \bibfield  {author} {\bibinfo {author} {\bibfnamefont {R.}~\bibnamefont
  {Kumar}}, \bibinfo {author} {\bibfnamefont {B.~P.}\ \bibnamefont {Singh}},
  \bibinfo {author} {\bibfnamefont {M.~S.}\ \bibnamefont {Ali}}, \ and\
  \bibinfo {author} {\bibfnamefont {S.~G.}\ \bibnamefont {Ghosh}},\ }\href@noop
  {} {\  (\bibinfo {year} {2017})},\ \Eprint {http://arxiv.org/abs/1712.09793}
  {arXiv:1712.09793 [gr-qc]} \BibitemShut {NoStop}%
\bibitem [{\citenamefont {Hennigar}\ \emph {et~al.}(2018)\citenamefont
  {Hennigar}, \citenamefont {Poshteh},\ and\ \citenamefont
  {Mann}}]{Hennigar:2018hza}%
  \BibitemOpen
  \bibfield  {author} {\bibinfo {author} {\bibfnamefont {R.~A.}\ \bibnamefont
  {Hennigar}}, \bibinfo {author} {\bibfnamefont {M.~B.~J.}\ \bibnamefont
  {Poshteh}}, \ and\ \bibinfo {author} {\bibfnamefont {R.~B.}\ \bibnamefont
  {Mann}},\ }\href {\doibase 10.1103/PhysRevD.97.064041} {\bibfield  {journal}
  {\bibinfo  {journal} {Phys. Rev.}\ }\textbf {\bibinfo {volume} {D97}},\
  \bibinfo {pages} {064041} (\bibinfo {year} {2018})},\ \Eprint
  {http://arxiv.org/abs/1801.03223} {arXiv:1801.03223 [gr-qc]} \BibitemShut
  {NoStop}%
\bibitem [{\citenamefont {Vetsov}\ \emph {et~al.}(2018)\citenamefont {Vetsov},
  \citenamefont {Gyulchev},\ and\ \citenamefont {Yazadjiev}}]{Vetsov:2018mld}%
  \BibitemOpen
  \bibfield  {author} {\bibinfo {author} {\bibfnamefont {T.}~\bibnamefont
  {Vetsov}}, \bibinfo {author} {\bibfnamefont {G.}~\bibnamefont {Gyulchev}}, \
  and\ \bibinfo {author} {\bibfnamefont {S.}~\bibnamefont {Yazadjiev}},\
  }\href@noop {} {\  (\bibinfo {year} {2018})},\ \Eprint
  {http://arxiv.org/abs/1801.04592} {arXiv:1801.04592 [gr-qc]} \BibitemShut
  {NoStop}%
\bibitem [{\citenamefont {Shaikh}\ \emph {et~al.}(2019)\citenamefont {Shaikh},
  \citenamefont {Kocherlakota}, \citenamefont {Narayan},\ and\ \citenamefont
  {Joshi}}]{Shaikh:2018lcc}%
  \BibitemOpen
  \bibfield  {author} {\bibinfo {author} {\bibfnamefont {R.}~\bibnamefont
  {Shaikh}}, \bibinfo {author} {\bibfnamefont {P.}~\bibnamefont
  {Kocherlakota}}, \bibinfo {author} {\bibfnamefont {R.}~\bibnamefont
  {Narayan}}, \ and\ \bibinfo {author} {\bibfnamefont {P.~S.}\ \bibnamefont
  {Joshi}},\ }\href {\doibase 10.1093/mnras/sty2624} {\bibfield  {journal}
  {\bibinfo  {journal} {Mon. Not. Roy. Astron. Soc.}\ }\textbf {\bibinfo
  {volume} {482}},\ \bibinfo {pages} {52} (\bibinfo {year} {2019})},\ \Eprint
  {http://arxiv.org/abs/1802.08060} {arXiv:1802.08060 [astro-ph.HE]}
  \BibitemShut {NoStop}%
\bibitem [{\citenamefont {Shaikh}(2018)}]{Shaikh:2018kfv}%
  \BibitemOpen
  \bibfield  {author} {\bibinfo {author} {\bibfnamefont {R.}~\bibnamefont
  {Shaikh}},\ }\href {\doibase 10.1103/PhysRevD.98.024044} {\bibfield
  {journal} {\bibinfo  {journal} {Phys. Rev.}\ }\textbf {\bibinfo {volume}
  {D98}},\ \bibinfo {pages} {024044} (\bibinfo {year} {2018})},\ \Eprint
  {http://arxiv.org/abs/1803.11422} {arXiv:1803.11422 [gr-qc]} \BibitemShut
  {NoStop}%
\bibitem [{\citenamefont {Mizuno}\ \emph {et~al.}(2018)\citenamefont {Mizuno},
  \citenamefont {Younsi}, \citenamefont {Fromm}, \citenamefont {Porth},
  \citenamefont {De~Laurentis}, \citenamefont {Olivares}, \citenamefont
  {Falcke}, \citenamefont {Kramer},\ and\ \citenamefont
  {Rezzolla}}]{Mizuno:2018lxz}%
  \BibitemOpen
  \bibfield  {author} {\bibinfo {author} {\bibfnamefont {Y.}~\bibnamefont
  {Mizuno}}, \bibinfo {author} {\bibfnamefont {Z.}~\bibnamefont {Younsi}},
  \bibinfo {author} {\bibfnamefont {C.~M.}\ \bibnamefont {Fromm}}, \bibinfo
  {author} {\bibfnamefont {O.}~\bibnamefont {Porth}}, \bibinfo {author}
  {\bibfnamefont {M.}~\bibnamefont {De~Laurentis}}, \bibinfo {author}
  {\bibfnamefont {H.}~\bibnamefont {Olivares}}, \bibinfo {author}
  {\bibfnamefont {H.}~\bibnamefont {Falcke}}, \bibinfo {author} {\bibfnamefont
  {M.}~\bibnamefont {Kramer}}, \ and\ \bibinfo {author} {\bibfnamefont
  {L.}~\bibnamefont {Rezzolla}},\ }\href {\doibase 10.1038/s41550-018-0449-5}
  {\bibfield  {journal} {\bibinfo  {journal} {Nat. Astron.}\ }\textbf {\bibinfo
  {volume} {2}},\ \bibinfo {pages} {585} (\bibinfo {year} {2018})},\ \Eprint
  {http://arxiv.org/abs/1804.05812} {arXiv:1804.05812 [astro-ph.GA]}
  \BibitemShut {NoStop}%
\bibitem [{\citenamefont {Amir}\ \emph {et~al.}(2019)\citenamefont {Amir},
  \citenamefont {Jusufi}, \citenamefont {Banerjee},\ and\ \citenamefont
  {Hansraj}}]{Amir:2018pcu}%
  \BibitemOpen
  \bibfield  {author} {\bibinfo {author} {\bibfnamefont {M.}~\bibnamefont
  {Amir}}, \bibinfo {author} {\bibfnamefont {K.}~\bibnamefont {Jusufi}},
  \bibinfo {author} {\bibfnamefont {A.}~\bibnamefont {Banerjee}}, \ and\
  \bibinfo {author} {\bibfnamefont {S.}~\bibnamefont {Hansraj}},\ }\href
  {\doibase 10.1088/1361-6382/ab42be} {\bibfield  {journal} {\bibinfo
  {journal} {Class. Quant. Grav.}\ }\textbf {\bibinfo {volume} {36}},\ \bibinfo
  {pages} {215007} (\bibinfo {year} {2019})},\ \Eprint
  {http://arxiv.org/abs/1806.07782} {arXiv:1806.07782 [gr-qc]} \BibitemShut
  {NoStop}%
\bibitem [{\citenamefont {Ovgun}\ \emph {et~al.}(2018)\citenamefont {Ovgun},
  \citenamefont {Sakalli},\ and\ \citenamefont {Saavedra}}]{Ovgun:2018tua}%
  \BibitemOpen
  \bibfield  {author} {\bibinfo {author} {\bibfnamefont {A.}~\bibnamefont
  {Ovgun}}, \bibinfo {author} {\bibfnamefont {I.}~\bibnamefont {Sakalli}}, \
  and\ \bibinfo {author} {\bibfnamefont {J.}~\bibnamefont {Saavedra}},\ }\href
  {\doibase 10.1088/1475-7516/2018/10/041} {\bibfield  {journal} {\bibinfo
  {journal} {JCAP}\ }\textbf {\bibinfo {volume} {1810}},\ \bibinfo {pages}
  {041} (\bibinfo {year} {2018})},\ \Eprint {http://arxiv.org/abs/1807.00388}
  {arXiv:1807.00388 [gr-qc]} \BibitemShut {NoStop}%
\bibitem [{\citenamefont {Ayzenberg}\ and\ \citenamefont
  {Yunes}(2018)}]{Ayzenberg:2018jip}%
  \BibitemOpen
  \bibfield  {author} {\bibinfo {author} {\bibfnamefont {D.}~\bibnamefont
  {Ayzenberg}}\ and\ \bibinfo {author} {\bibfnamefont {N.}~\bibnamefont
  {Yunes}},\ }\href {\doibase 10.1088/1361-6382/aae87b} {\bibfield  {journal}
  {\bibinfo  {journal} {Class. Quant. Grav.}\ }\textbf {\bibinfo {volume}
  {35}},\ \bibinfo {pages} {235002} (\bibinfo {year} {2018})},\ \Eprint
  {http://arxiv.org/abs/1807.08422} {arXiv:1807.08422 [gr-qc]} \BibitemShut
  {NoStop}%
\bibitem [{\citenamefont {Okounkova}\ \emph {et~al.}(2019)\citenamefont
  {Okounkova}, \citenamefont {Scheel},\ and\ \citenamefont
  {Teukolsky}}]{Okounkova:2018abo}%
  \BibitemOpen
  \bibfield  {author} {\bibinfo {author} {\bibfnamefont {M.}~\bibnamefont
  {Okounkova}}, \bibinfo {author} {\bibfnamefont {M.~A.}\ \bibnamefont
  {Scheel}}, \ and\ \bibinfo {author} {\bibfnamefont {S.~A.}\ \bibnamefont
  {Teukolsky}},\ }\href {\doibase 10.1088/1361-6382/aafcdf} {\bibfield
  {journal} {\bibinfo  {journal} {Class. Quant. Grav.}\ }\textbf {\bibinfo
  {volume} {36}},\ \bibinfo {pages} {054001} (\bibinfo {year} {2019})},\
  \Eprint {http://arxiv.org/abs/1810.05306} {arXiv:1810.05306 [gr-qc]}
  \BibitemShut {NoStop}%
\bibitem [{\citenamefont {Wang}\ \emph
  {et~al.}(2019{\natexlab{c}})\citenamefont {Wang}, \citenamefont {Xu},\ and\
  \citenamefont {Wei}}]{Wang:2018prk}%
  \BibitemOpen
  \bibfield  {author} {\bibinfo {author} {\bibfnamefont {H.-M.}\ \bibnamefont
  {Wang}}, \bibinfo {author} {\bibfnamefont {Y.-M.}\ \bibnamefont {Xu}}, \ and\
  \bibinfo {author} {\bibfnamefont {S.-W.}\ \bibnamefont {Wei}},\ }\href
  {\doibase 10.1088/1475-7516/2019/03/046} {\bibfield  {journal} {\bibinfo
  {journal} {JCAP}\ }\textbf {\bibinfo {volume} {1903}},\ \bibinfo {pages}
  {046} (\bibinfo {year} {2019}{\natexlab{c}})},\ \Eprint
  {http://arxiv.org/abs/1810.12767} {arXiv:1810.12767 [gr-qc]} \BibitemShut
  {NoStop}%
\bibitem [{\citenamefont {Haroon}\ \emph
  {et~al.}(2019{\natexlab{a}})\citenamefont {Haroon}, \citenamefont {Jusufi},\
  and\ \citenamefont {Jamil}}]{Haroon:2019new}%
  \BibitemOpen
  \bibfield  {author} {\bibinfo {author} {\bibfnamefont {S.}~\bibnamefont
  {Haroon}}, \bibinfo {author} {\bibfnamefont {K.}~\bibnamefont {Jusufi}}, \
  and\ \bibinfo {author} {\bibfnamefont {M.}~\bibnamefont {Jamil}},\
  }\href@noop {} {\  (\bibinfo {year} {2019}{\natexlab{a}})},\ \Eprint
  {http://arxiv.org/abs/1904.00711} {arXiv:1904.00711 [gr-qc]} \BibitemShut
  {NoStop}%
\bibitem [{\citenamefont {Kumar}\ \emph
  {et~al.}(2019{\natexlab{b}})\citenamefont {Kumar}, \citenamefont {Singh},\
  and\ \citenamefont {Ghosh}}]{Kumar:2019ohr}%
  \BibitemOpen
  \bibfield  {author} {\bibinfo {author} {\bibfnamefont {R.}~\bibnamefont
  {Kumar}}, \bibinfo {author} {\bibfnamefont {B.~P.}\ \bibnamefont {Singh}}, \
  and\ \bibinfo {author} {\bibfnamefont {S.~G.}\ \bibnamefont {Ghosh}},\
  }\href@noop {} {\  (\bibinfo {year} {2019}{\natexlab{b}})},\ \Eprint
  {http://arxiv.org/abs/1904.07652} {arXiv:1904.07652 [gr-qc]} \BibitemShut
  {NoStop}%
\bibitem [{\citenamefont {Ovgun}\ \emph {et~al.}(2019)\citenamefont {Ovgun},
  \citenamefont {Sakalli}, \citenamefont {Saavedra},\ and\ \citenamefont
  {Leiva}}]{Ovgun:2019jdo}%
  \BibitemOpen
  \bibfield  {author} {\bibinfo {author} {\bibfnamefont {A.}~\bibnamefont
  {Ovgun}}, \bibinfo {author} {\bibfnamefont {I.}~\bibnamefont {Sakalli}},
  \bibinfo {author} {\bibfnamefont {J.}~\bibnamefont {Saavedra}}, \ and\
  \bibinfo {author} {\bibfnamefont {C.}~\bibnamefont {Leiva}},\ }\href@noop {}
  {\  (\bibinfo {year} {2019})},\ \Eprint {http://arxiv.org/abs/1906.05954}
  {arXiv:1906.05954 [hep-th]} \BibitemShut {NoStop}%
\bibitem [{\citenamefont {Das}\ \emph {et~al.}(2019)\citenamefont {Das},
  \citenamefont {Saha},\ and\ \citenamefont {Gangopadhyay}}]{Das:2019sty}%
  \BibitemOpen
  \bibfield  {author} {\bibinfo {author} {\bibfnamefont {A.}~\bibnamefont
  {Das}}, \bibinfo {author} {\bibfnamefont {A.}~\bibnamefont {Saha}}, \ and\
  \bibinfo {author} {\bibfnamefont {S.}~\bibnamefont {Gangopadhyay}},\
  }\href@noop {} {\  (\bibinfo {year} {2019})},\ \Eprint
  {http://arxiv.org/abs/1909.01988} {arXiv:1909.01988 [gr-qc]} \BibitemShut
  {NoStop}%
\bibitem [{\citenamefont {Amarilla}\ and\ \citenamefont
  {Eiroa}(2017)}]{Amarilla:2015pgp}%
  \BibitemOpen
  \bibfield  {author} {\bibinfo {author} {\bibfnamefont {L.}~\bibnamefont
  {Amarilla}}\ and\ \bibinfo {author} {\bibfnamefont {E.~F.}\ \bibnamefont
  {Eiroa}},\ }in\ \href {\doibase 10.1142/9789813226609_0459} {\emph {\bibinfo
  {booktitle} {{Proceedings, 14th Marcel Grossmann Meeting on Recent
  Developments in Theoretical and Experimental General Relativity,
  Astrophysics, and Relativistic Field Theories (MG14) (In 4 Volumes): Rome,
  Italy, July 12-18, 2015}}}},\ Vol.~\bibinfo {volume} {4}\ (\bibinfo {year}
  {2017})\ pp.\ \bibinfo {pages} {3543--3548},\ \Eprint
  {http://arxiv.org/abs/1512.08956} {arXiv:1512.08956 [gr-qc]} \BibitemShut
  {NoStop}%
\bibitem [{\citenamefont {Xu}\ \emph {et~al.}(2018)\citenamefont {Xu},
  \citenamefont {Hou}, \citenamefont {Gong},\ and\ \citenamefont
  {Wang}}]{Xu:2018wow}%
  \BibitemOpen
  \bibfield  {author} {\bibinfo {author} {\bibfnamefont {Z.}~\bibnamefont
  {Xu}}, \bibinfo {author} {\bibfnamefont {X.}~\bibnamefont {Hou}}, \bibinfo
  {author} {\bibfnamefont {X.}~\bibnamefont {Gong}}, \ and\ \bibinfo {author}
  {\bibfnamefont {J.}~\bibnamefont {Wang}},\ }\href {\doibase
  10.1088/1475-7516/2018/09/038} {\bibfield  {journal} {\bibinfo  {journal}
  {JCAP}\ }\textbf {\bibinfo {volume} {1809}},\ \bibinfo {pages} {038}
  (\bibinfo {year} {2018})},\ \Eprint {http://arxiv.org/abs/1803.00767}
  {arXiv:1803.00767 [gr-qc]} \BibitemShut {NoStop}%
\bibitem [{\citenamefont {Hou}\ \emph {et~al.}(2018{\natexlab{a}})\citenamefont
  {Hou}, \citenamefont {Xu}, \citenamefont {Zhou},\ and\ \citenamefont
  {Wang}}]{Hou:2018bar}%
  \BibitemOpen
  \bibfield  {author} {\bibinfo {author} {\bibfnamefont {X.}~\bibnamefont
  {Hou}}, \bibinfo {author} {\bibfnamefont {Z.}~\bibnamefont {Xu}}, \bibinfo
  {author} {\bibfnamefont {M.}~\bibnamefont {Zhou}}, \ and\ \bibinfo {author}
  {\bibfnamefont {J.}~\bibnamefont {Wang}},\ }\href {\doibase
  10.1088/1475-7516/2018/07/015} {\bibfield  {journal} {\bibinfo  {journal}
  {JCAP}\ }\textbf {\bibinfo {volume} {1807}},\ \bibinfo {pages} {015}
  (\bibinfo {year} {2018}{\natexlab{a}})},\ \Eprint
  {http://arxiv.org/abs/1804.08110} {arXiv:1804.08110 [gr-qc]} \BibitemShut
  {NoStop}%
\bibitem [{\citenamefont {Haroon}\ \emph
  {et~al.}(2019{\natexlab{b}})\citenamefont {Haroon}, \citenamefont {Jamil},
  \citenamefont {Jusufi}, \citenamefont {Lin},\ and\ \citenamefont
  {Mann}}]{Haroon:2018ryd}%
  \BibitemOpen
  \bibfield  {author} {\bibinfo {author} {\bibfnamefont {S.}~\bibnamefont
  {Haroon}}, \bibinfo {author} {\bibfnamefont {M.}~\bibnamefont {Jamil}},
  \bibinfo {author} {\bibfnamefont {K.}~\bibnamefont {Jusufi}}, \bibinfo
  {author} {\bibfnamefont {K.}~\bibnamefont {Lin}}, \ and\ \bibinfo {author}
  {\bibfnamefont {R.~B.}\ \bibnamefont {Mann}},\ }\href {\doibase
  10.1103/PhysRevD.99.044015} {\bibfield  {journal} {\bibinfo  {journal} {Phys.
  Rev.}\ }\textbf {\bibinfo {volume} {D99}},\ \bibinfo {pages} {044015}
  (\bibinfo {year} {2019}{\natexlab{b}})},\ \Eprint
  {http://arxiv.org/abs/1810.04103} {arXiv:1810.04103 [gr-qc]} \BibitemShut
  {NoStop}%
\bibitem [{\citenamefont {Hou}\ \emph {et~al.}(2018{\natexlab{b}})\citenamefont
  {Hou}, \citenamefont {Xu},\ and\ \citenamefont {Wang}}]{Hou:2018avu}%
  \BibitemOpen
  \bibfield  {author} {\bibinfo {author} {\bibfnamefont {X.}~\bibnamefont
  {Hou}}, \bibinfo {author} {\bibfnamefont {Z.}~\bibnamefont {Xu}}, \ and\
  \bibinfo {author} {\bibfnamefont {J.}~\bibnamefont {Wang}},\ }\href {\doibase
  10.1088/1475-7516/2018/12/040} {\bibfield  {journal} {\bibinfo  {journal}
  {JCAP}\ }\textbf {\bibinfo {volume} {1812}},\ \bibinfo {pages} {040}
  (\bibinfo {year} {2018}{\natexlab{b}})},\ \Eprint
  {http://arxiv.org/abs/1810.06381} {arXiv:1810.06381 [gr-qc]} \BibitemShut
  {NoStop}%
\bibitem [{\citenamefont {Konoplya}(2019)}]{Konoplya:2019sns}%
  \BibitemOpen
  \bibfield  {author} {\bibinfo {author} {\bibfnamefont {R.~A.}\ \bibnamefont
  {Konoplya}},\ }\href {\doibase 10.1016/j.physletb.2019.05.043} {\bibfield
  {journal} {\bibinfo  {journal} {Phys. Lett.}\ }\textbf {\bibinfo {volume}
  {B795}},\ \bibinfo {pages} {1} (\bibinfo {year} {2019})},\ \Eprint
  {http://arxiv.org/abs/1905.00064} {arXiv:1905.00064 [gr-qc]} \BibitemShut
  {NoStop}%
\bibitem [{\citenamefont {Boyle}\ \emph {et~al.}(1988)\citenamefont {Boyle},
  \citenamefont {Shanks},\ and\ \citenamefont {Peterson}}]{Boyle:1988zz}%
  \BibitemOpen
  \bibfield  {author} {\bibinfo {author} {\bibfnamefont {B.~J.}\ \bibnamefont
  {Boyle}}, \bibinfo {author} {\bibfnamefont {T.}~\bibnamefont {Shanks}}, \
  and\ \bibinfo {author} {\bibfnamefont {B.~A.}\ \bibnamefont {Peterson}},\
  }\href@noop {} {\bibfield  {journal} {\bibinfo  {journal} {Mon. Not. Roy.
  Astron. Soc.}\ }\textbf {\bibinfo {volume} {235}},\ \bibinfo {pages} {935}
  (\bibinfo {year} {1988})}\BibitemShut {NoStop}%
\bibitem [{\citenamefont {Wyithe}\ and\ \citenamefont
  {Loeb}(2002)}]{Wyithe:2002ij}%
  \BibitemOpen
  \bibfield  {author} {\bibinfo {author} {\bibfnamefont {J.~S.~B.}\
  \bibnamefont {Wyithe}}\ and\ \bibinfo {author} {\bibfnamefont
  {A.}~\bibnamefont {Loeb}},\ }\href {\doibase 10.1086/344249} {\bibfield
  {journal} {\bibinfo  {journal} {Astrophys. J.}\ }\textbf {\bibinfo {volume}
  {581}},\ \bibinfo {pages} {886} (\bibinfo {year} {2002})},\ \Eprint
  {http://arxiv.org/abs/astro-ph/0206154} {arXiv:astro-ph/0206154 [astro-ph]}
  \BibitemShut {NoStop}%
\bibitem [{\citenamefont {Silverman}\ \emph {et~al.}(2008)\citenamefont
  {Silverman} \emph {et~al.}}]{Silverman:2007qa}%
  \BibitemOpen
  \bibfield  {author} {\bibinfo {author} {\bibfnamefont {J.~D.}\ \bibnamefont
  {Silverman}} \emph {et~al.},\ }\href {\doibase 10.1086/529572} {\bibfield
  {journal} {\bibinfo  {journal} {Astrophys. J.}\ }\textbf {\bibinfo {volume}
  {679}},\ \bibinfo {pages} {118} (\bibinfo {year} {2008})},\ \Eprint
  {http://arxiv.org/abs/0710.2461} {arXiv:0710.2461 [astro-ph]} \BibitemShut
  {NoStop}%
\bibitem [{\citenamefont {Fiore}\ \emph {et~al.}(2012)\citenamefont {Fiore}
  \emph {et~al.}}]{Fiore:2011iv}%
  \BibitemOpen
  \bibfield  {author} {\bibinfo {author} {\bibfnamefont {F.}~\bibnamefont
  {Fiore}} \emph {et~al.},\ }\href {\doibase 10.1051/0004-6361/201117581}
  {\bibfield  {journal} {\bibinfo  {journal} {Astron. Astrophys.}\ }\textbf
  {\bibinfo {volume} {537}},\ \bibinfo {pages} {A16} (\bibinfo {year}
  {2012})},\ \Eprint {http://arxiv.org/abs/1109.2888} {arXiv:1109.2888
  [astro-ph.CO]} \BibitemShut {NoStop}%
\bibitem [{\citenamefont {Georgakakis}\ \emph {et~al.}(2015)\citenamefont
  {Georgakakis} \emph {et~al.}}]{Georgakakis:2015rfa}%
  \BibitemOpen
  \bibfield  {author} {\bibinfo {author} {\bibfnamefont {A.}~\bibnamefont
  {Georgakakis}} \emph {et~al.},\ }\href {\doibase 10.1093/mnras/stv1703}
  {\bibfield  {journal} {\bibinfo  {journal} {Mon. Not. Roy. Astron. Soc.}\
  }\textbf {\bibinfo {volume} {453}},\ \bibinfo {pages} {1946} (\bibinfo {year}
  {2015})},\ \Eprint {http://arxiv.org/abs/1507.07558} {arXiv:1507.07558
  [astro-ph.HE]} \BibitemShut {NoStop}%
\bibitem [{\citenamefont {Kulkarni}\ \emph {et~al.}(2019)\citenamefont
  {Kulkarni}, \citenamefont {Worseck},\ and\ \citenamefont
  {Hennawi}}]{Kulkarni:2018ebj}%
  \BibitemOpen
  \bibfield  {author} {\bibinfo {author} {\bibfnamefont {G.}~\bibnamefont
  {Kulkarni}}, \bibinfo {author} {\bibfnamefont {G.}~\bibnamefont {Worseck}}, \
  and\ \bibinfo {author} {\bibfnamefont {J.~F.}\ \bibnamefont {Hennawi}},\
  }\href {\doibase 10.1093/mnras/stz1493} {\bibfield  {journal} {\bibinfo
  {journal} {Mon. Not. Roy. Astron. Soc.}\ }\textbf {\bibinfo {volume} {488}},\
  \bibinfo {pages} {1035} (\bibinfo {year} {2019})},\ \Eprint
  {http://arxiv.org/abs/1807.09774} {arXiv:1807.09774 [astro-ph.GA]}
  \BibitemShut {NoStop}%
\end{thebibliography}%
\end{document}